\documentclass[prd,tightenlines,showpacs,groupedaddress,
	superscriptaddress,11pt,nofootinbib,floatfix,
	preprintnumbers,letterpaper]{revtex4}
\usepackage{hyperref,amssymb,amsmath,graphicx,xcolor,cancel}
\usepackage{bm}

\newcommand{\lsim}{\raisebox{-0.7ex}{$\stackrel{\textstyle <}{\sim}$ }}

\def\si{^1 \hskip -0.03in S _0}
\def\siii{^3 \hskip -0.025in S _1}
\def\diii{^3 \hskip -0.03in D _1}

\def\pislash{{\pi\hskip-0.55em /}}

\def\pionmass{806~{\rm MeV} }

\def\mun{-1.972\left({\tiny \begin{array}{l}
			+0.027\\-0.033
		\end{array}}\right) (0.059) }
\def\mup{3.17\left({\tiny \begin{array}{l}
		+0.10\\-0.09
	\end{array}}\right)(0.09) }
\def\mud{1.41\left({\tiny \begin{array}{l}
			+0.28\\-0.25
		\end{array}}\right)(0.04) }
\def\muHethree{-2.28\left({\tiny \begin{array}{l}
			+0.59\\-1.04
		\end{array}}\right)(0.07) }
\def\mutriton{3.32\left({\tiny \begin{array}{l}
			+0.79\\-0.59
		\end{array}}\right)(0.10) }

\def\betan{0.198\left({\tiny \begin{array}{l}
		+0.009\\-0.011
		\end{array}}\right)(0.010)}
\def\betann{0.296\left({\tiny \begin{array}{l}
		+0.019\\-0.018
		\end{array}}\right)(0.015)}
\def\betap{0.83\left({\tiny \begin{array}{l}
		+0.10\\-0.07
		\end{array}}\right)(0.04)}
\def\betapp{0.84\left({\tiny \begin{array}{l}
		+0.41\\-0.36
		\end{array}}\right)(0.04)}
\def\betadpm{0.70\left({\tiny \begin{array}{l}
		+0.24\\-0.23
		\end{array}}\right)(0.04)}
\def\betaHethree{0.85\left({\tiny \begin{array}{l}
		+0.34\\-0.32
		\end{array}}\right)(0.04)}
\def\betatriton{0.40\left({\tiny \begin{array}{l}
			+0.27\\-0.27
		\end{array}}\right)(0.02)}
\def\betaHefour{0.54\left({\tiny \begin{array}{l}
			+0.32\\-0.31
		\end{array}}\right)(0.03)}

\def\dbetann{0.1}
\def\dbetannCORR{-0.070\left({\tiny \begin{array}{l}
			+0.006\\-0.009
		\end{array}}\right)(0.004) }
\def\dbetapp{ -0.82 \left({\tiny \begin{array}{l}
			+0.42\\-0.37
		\end{array}}\right)(0.04) }

\def\betapPhysNoUnit{5.22\left({\tiny \begin{array}{l}
		+0.66\\-0.45
		\end{array}}\right)(0.23)}
\def\betanPhysNoUnit{1.253\left({\tiny \begin{array}{l}
		+0.056\\-0.067
		\end{array}}\right)(0.055)}
\def\betannPhysNoUnit{1.872\left({\tiny \begin{array}{l}
		+0.121\\-0.113
		\end{array}}\right)(0.082)}
\def\betappPhysNoUnit{5.31\left({\tiny \begin{array}{l}
		+2.59\\-2.27
		\end{array}}\right)(0.23)}
\def\betadpmPhysNoUnit{4.4\left({\tiny \begin{array}{l}
		+1.6\\-1.5
		\end{array}}\right)(0.2)}
\def\betaHethreePhysNoUnit{5.4\left({\tiny \begin{array}{l}
			+2.2\\-2.1
		\end{array}}\right)(0.2)}
\def\betatritonPhysNoUnit{2.6(1.7)(0.1)}
\def\betaHefourPhysNoUnit{3.4\left({\tiny \begin{array}{l}
			+2.0\\-1.9
		\end{array}}\right)(0.2)}

\def\betapPhys{\betapPhysNoUnit \times 10^{-4}\ {\rm fm}^3}
\def\betanPhys{\betanPhysNoUnit \times 10^{-4}\ {\rm fm}^3}
\def\betannPhys{\betannPhysNoUnit \times 10^{-4}\ {\rm fm}^3}

\def\betadpmPhys{\betadpmPhysNoUnit \times 10^{-4}\ {\rm fm}^3}
\def\betaHethreePhys{\betaHethreePhysNoUnit \times 10^{-4}\ {\rm fm}^3}
\def\betatritonPhys{\betatritonPhysNoUnit \times 10^{-4}\ {\rm fm}^3}
\def\betaHefourPhys{\betaHefourPhysNoUnit \times 10^{-4}\ {\rm fm}^3}

\def\kappaplusLonebar{ 2.74\left({\tiny \begin{array}{l}
			+0.07\\-0.05
		\end{array}}\right)(0.07)\ {\rm nNM}}
\def\Lonebar{  0.207\left({\tiny \begin{array}{l}
			+0.020\\-0.020
		\end{array}}\right)(0.006)\ {\rm nNM}}

\begin{document}

\begin{figure}[!t]
  \vskip -1.1cm \leftline{
    \includegraphics[width=3.0 cm]{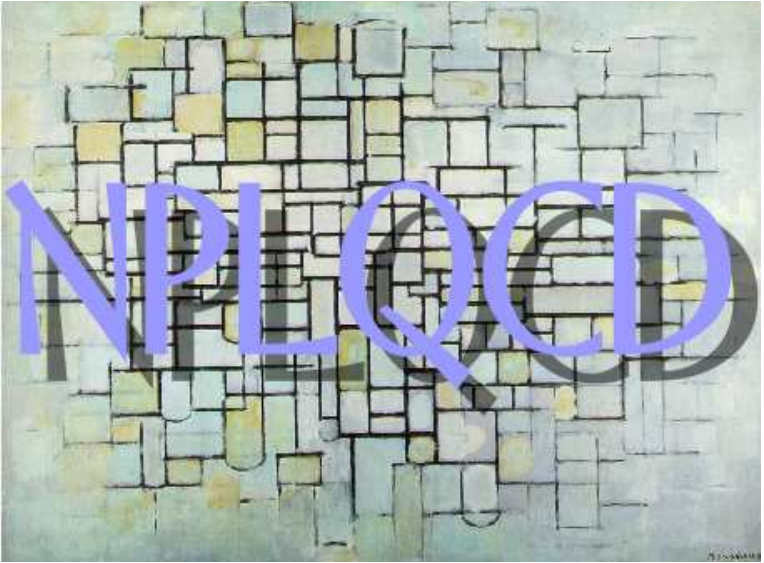}} \vskip
  -0.5cm
\end{figure}

\title{The Magnetic Structure of Light Nuclei from Lattice QCD}

\author{Emmanuel~Chang} 
\affiliation{Institute for Nuclear Theory, University of Washington, Seattle, WA 98195-1550, USA}

\author{William Detmold} \affiliation{ Center for Theoretical Physics,
  Massachusetts Institute of Technology, Cambridge, MA 02139, USA}
 
\author{Kostas~Orginos} \affiliation{Department of Physics, College of
  William and Mary, Williamsburg, VA 23187-8795, USA}
\affiliation{Jefferson Laboratory, 12000 Jefferson Avenue, Newport
  News, VA 23606, USA}

\author{Assumpta~Parre\~no} \affiliation{Departament d'Estructura i
  Constituents de la Mat\`eria.  Institut de Ci\`encies del Cosmos
  (ICC), Universitat de Barcelona, Mart\'{\i} Franqu\`es 1,
  E08028-Spain}

\author{Martin J. Savage}
\affiliation{Institute for Nuclear Theory, University of Washington, Seattle, WA 98195-1550, USA}

\author{Brian C. Tiburzi} \affiliation{ Department of Physics, The City
  College of New York, New York, NY 10031, USA } \affiliation{Graduate
  School and University Center, The City University of New York, New
  York, NY 10016, USA } \affiliation{RIKEN BNL Research Center,
  Brookhaven National Laboratory, Upton, NY 11973, USA }

\author{Silas~R.~Beane} \affiliation{Department of Physics, University
	of Washington, Box 351560, Seattle, WA 98195, USA}

\collaboration{NPLQCD Collaboration}

\date{\today}

\preprint{INT-PUB-15-011}
\preprint{NT@UW-15-03}
\preprint{ICC@UB-15-016}
\preprint{MIT-CTP-4667}

\pacs{11.15.Ha, 
  12.38.Gc, 
  13.40.Gp 
}

\begin{abstract}
  Lattice QCD with background magnetic fields
  is used to calculate the  magnetic moments and magnetic polarizabilities of the
  nucleons and of light nuclei with $A\le4$, along with the cross-section for the $M1$ transition
  $np\rightarrow d\gamma$, 
  at the flavor SU(3)-symmetric point where the pion mass is  $m_\pi\sim \pionmass$. 
   These magnetic properties are extracted from nucleon and nuclear energies in
  six  uniform  magnetic fields of varying strengths. 
  The magnetic moments are presented in a recent Letter \cite{Beane:2014ora}. 
  For the charged states, the extraction of the polarizability requires careful
  treatment of Landau levels, which enter  non-trivially  
  in the method that is employed. 
   The nucleon polarizabilities are found to be of similar magnitude 
  to their physical values, with $\beta_p=\betapPhys$ and $\beta_n=\betanPhys$,
  exhibiting a significant isovector component.  
  The dineutron is bound at these heavy quark masses and its magnetic polarizability, $\beta_{nn}=\betannPhys$  differs significantly 
  from twice that of the neutron. 
  A linear combination of deuteron scalar and tensor polarizabilities is determined by the energies  of 
  the $j_z=\pm 1$ deuteron states, and is found to be  
  $\beta_{d,\pm 1}=\betadpmPhys$. 
  The magnetic polarizabilities of  the three-nucleon and four-nucleon systems are found to be 
   positive and similar in size to those of the proton, 
  $\beta_{^{3}\rm He}=\betaHethreePhys$,
  $\beta_{^{3}\rm H}=\betatritonPhys$, 
  $\beta_{^{4}\rm  He}=\betaHefourPhys$. 
   Mixing between the $j_z=0$ deuteron state  and the  spin-singlet
  $np$ state induced by the background magnetic  field is used to 
  extract the short-distance two-nucleon counterterm,  ${\bar L}_1$, of the pionless
  effective theory for $NN$ systems  
  (equivalent to the meson-exchange current contribution in nuclear potential models),
  that dictates the cross-section for the $np\to d\gamma$ process near threshold. 
  Combined with previous determinations of NN scattering parameters, this enables an ab initio determination of the threshold cross-section at these unphysical masses. 
\end{abstract}

\maketitle

\newpage

\section{Introduction}
\label{sec:intro}

The charge, magnetic moment and magnetic polarizability of a composite system describe its  linear and
quadratic response to a uniform, time-independent  magnetic field. 
These properties are determined by the distribution of the constituents of the system and by the
currents induced by the   field. As such, measurements of the magnetic properties of the nucleons and nuclei provide important
information about their internal structure.  
Furthermore, these quantities also serve to parameterize the cross-section for low-energy Compton scattering on such
targets.  
Historically, the magnetic moments of nucleons and light nuclei
provided some of the first indications of substructure, 
and by now they are well known.
The primary focus of this article is on the magnetic polarizabilities 
and the cross-section for the radiative capture process
$np\rightarrow d\gamma$ at low energies 
which is dominated by the $M1$ multipole. 
While the magnetic
polarizability of the proton\footnote{
	The experimental polarizability reported here is defined with Born terms subtracted; the total ${\cal O}(B^2)$ shift in the energy of the proton  is larger by $(16\pi M^3)^{-1}=0.15 \times 10^{-4} {\rm fm}^3$ \cite{Lee:2014iha}.}, $\beta_p^{\rm exp} = (3.15\pm 0.35 \pm
0.20 \pm 0.30) \times 10^{-4}~{\rm fm}^3$, is well determined
experimentally~\cite{Federspiel:1991yd,Zieger:1992jq,MacGibbon:1995in,Beringer:1900zz,Myers:2014ace,Lensky:2014efa},
the magnetic polarizability of the neutron $\beta_n^{\rm exp} =
(3.65\pm 1.25\pm 0.20\pm 0.80)\times 10^{-4}~{\rm fm}^3$ remains quite
uncertain~\cite{Kossert:2002ws,Lundin:2002jy,Beringer:1900zz,Phillips:2009af,Myers:2014ace}.
This uncertainty is largely a consequence of the lack of a free
neutron target; the neutron polarizability must be determined from
that of light nuclei, primarily the deuteron (see Ref.~\cite{Phillips:2009af} for a recent summary).  
The smallness of the nucleon
polarizabilities, compared with their spatial volumes, $\sim 1\ {\rm fm}^3$,
indicates that they are quite magnetically rigid, with the spins and
currents of their constituents influenced only at a modest level by 
external fields.  The uncertainty in $\beta_n^{\rm exp}$ is large enough so that
a significant isovector polarizability remains a possibility. For a
recent review, see Ref.~\cite{Holstein:2013kia}.

From a theoretical standpoint, the leading contributions to the
nucleon magnetic polarizabilities result from both pion-loop effects
and $\Delta$-resonance pole
contributions~\cite{Butler:1992ci,Butler:1992pn,Broniowski:1992vj,Cohen:1992uy,Pascalutsa:2002pi,Holstein:2013kia}.
The $\Delta$-pole contribution is 
${\cal O}(e^2 /[ M_N^2 (M_\Delta -M_N)] )$ and is considerably larger than the
experimentally determined polarizabilities.  At the physical
light-quark masses, significant cancellations between the paramagnetic
($\Delta$-pole) and diamagnetic (loop) contributions are in effect.
As the various contributions depend differently on the
light-quark masses (the leading pion-loop contributions scales as
$\sim m_\pi^{-1}$ while the  $\Delta$-pole
contribution is only weakly dependent on the quark masses), it is
expected that  $\beta_p$ and $\beta_n$ will vary reasonably rapidly
near the physical point.\footnote{Recent chiral effective field theory 
	calculations support this expectation \cite{GriesshammerPC}.}
Because it is relatively mass independent, the $\Delta$-pole contribution provides an 
estimate of the expected size for polarizabilities at any quark mass and it will be used to assess the naturalness of 
the polarizabilities extracted from the present Lattice QCD (LQCD) calculations.

The leading contributions to the magnetic polarizabilities of light
nuclei are determined by the nucleon charges, magnetic moments and
polarizabilities.  At a sub-leading level, the forces
between nucleons, both those responsible for nuclear binding and 
meson-exchange currents (MECs), 
modify these one-body contributions
and produce short-distance contributions that are unrelated to the
electromagnetic interactions of single nucleons.  
Accounting for this in a consistent manner is non-trivial and requires a 
well-controlled power-counting.
This has been investigated in detail for
the deuteron~\cite{Chen:1998vi} and light nuclei~\cite{Griesshammer:2012we}.  
Experimentally, the polarizabilities of
the deuteron have been measured through Compton scattering 
(leading to extractions of the  neutron magnetic polarizability) 
and further measurements
will be performed with increased precision at the HI$\gamma$S facility~\cite{Weller:2009zza}, 
MAX-Lab at Lund~\cite{MAXLab} and at MAMI in Mainz~\cite{Downie:2011mm}. 
Plans for a new generation of 
Compton scattering experiments on other light nuclei are also being developed 
(see, for example, Refs.~\cite{Myers:2012xw,Myers:2014qhi,Myers:2014qaa}).

The radiative capture process, $np \to d\gamma$, and the inverse 
processes of deuteron  electro- and photo-disintegration, 
$\gamma^{(\ast)}d \to np$, are important in early universe cosmology and have led to critical insights into the 
interactions between nucleons, and in particular, interactions 
with  photons. 
At very low energies, the $M1$ magnetic multipole transition amplitude
is dominant and is determined primarily by the isovector nucleon magnetic 
moment. 
The short distance contributions (equivalently, MECs) are sub-leading and
modify the cross-sections 
at ${\cal O}(10\%)$ \cite{Riska:1972zz,Hockert:1974qt} and are well determined by experiment in this particular case.
Given the phenomenological importance of this  and other related processes, 
it is important to understand these contributions from first principles.

In this work, we present lattice QCD (LQCD) calculations of the
magnetic moments and polarizabilities of the proton, neutron and
$s$-shell nuclei up to atomic number $A=4$. 
Further, the $j_z=I_z=0$ $np$ systems are used 
to investigate the short-distance two-nucleon  
contributions to the cross-section for  $np \to d\gamma$. 
The methods and calculations that are presented 
are an extension of those used to calculate the magnetic moments of light nuclei in
Ref.~\cite{Beane:2014ora} and the result for the capture cross-section have been recently highlighted in Ref.~\cite{Beane:2015yha}. 
 In
Section~\ref{sec:methods}, the methodology of using
background magnetic fields in LQCD calculations to determine hadron magnetic
moments and polarizabilities and magnetic transition amplitudes is presented. 
Section~\ref{sec:results} discusses the
results of our calculations, by first discussing the general analysis
methods used to extract the magnetic properties from energy shifts before turning to a discussion of each of the nucleons and nuclei that are
studied.  
Our results are summarized in Section~\ref{sec:end}, and opportunities  for possible extensions of this work are also outlined.
Finally, Appendix~\ref{app:CPcors} is dedicated to defining the formalism underpinning the LQCD methodology used in the present calculations.  Explicit examples centered on the behavior of point-like charged particles are explored.

\section{Methodology}
\label{sec:methods}

In order to calculate the magnetic structure of  nucleons and light nuclei, and the low-energy cross-section
for $np\rightarrow d\gamma$, 
it is sufficient to perform LQCD calculations of these systems in uniform time-independent  magnetic fields.
In 
sufficiently weak 
background fields, the quantities of interest can be extracted directly from the energy eigenvalues of the 
spin states of these systems.

\subsection{Background Magnetic Fields}

In this work, lattice QCD calculations are performed using one ensemble of 
QCD gauge-field configurations with $N_f=3$ degenerate dynamical flavors of light quarks 
on a 
$L^3\times T = 
(32^3\times 48) \,  a^4$
discretized spacetime. 
This ensemble was generated using
a L\"uscher-Weisz gauge action \cite{Luscher:1984xn} and a tadpole-improved 
clover-fermion action \cite{Sheikholeslami:1985ij}
at a  coupling corresponding to a lattice spacing of 
$a=0.110(1)~{\rm fm}$ \cite{MeinelPC}
(further details of this ensemble can be found in
Refs.~\cite{Beane:2012vq,Beane:2013br}).  
All three light-quark masses in this ensemble
are equal to that of the physical strange quark, producing a pion of
mass $m_\pi \sim\pionmass$. 
In the present set of calculations, ${\cal O}(10^3)$ gauge-field configurations separated by $10$ Hybrid Monte-Carlo
trajectories were used from this ensemble.

Background electromagnetic fields have been used extensively 
in  previous calculations of the  electromagnetic properties of single hadrons, such
as the magnetic moments of the lowest-lying
baryons~\cite{Bernard:1982yu,Martinelli:1982cb,Lee:2005ds,
  Lee:2005dq,Detmold:2006vu,Aubin:2008qp,Detmold:2009dx,Detmold:2010ts,Primer:2013pva,Beane:2014ora},
 the polarizabilities of 
mesons~\cite{Fiebig:1988en,Christensen:2004ca,Lee:2005dq,Detmold:2009dx,Lujan:2014kia}
and the electric polarizabilities of baryons
\cite{Lujan:2014kia,Freeman:2014kka}. 
In addition, the magnetic moments of light nuclei have been recently calculated in 
Ref.~\cite{Beane:2014ora}. 
In order for the quark fields,
with electric charges $q_q$ ($q_u=+\frac{2}{3}$ and $q_{d,s}=-\frac{1}{3}$ for the up-, down- and strange-quarks, respectively), 
to encounter uniform magnetic flux throughout the lattice,
the field strength is quantized according to 
the condition~\cite{'tHooft:1979uj}
\begin{eqnarray}
  e {\bf B} & = & 
  {6\pi\over L^2} \tilde n {\bf e_z}
  \ ,\quad \tilde n\in{\mathbb{Z}}\,,
\end{eqnarray}
where $e$ is the magnitude of the electron charge and we consider explicitly fields in the $x_3\equiv z$ direction as indicated by the unit vector ${\bf e_z}$.  
Neglecting the electric charges of the sea-quarks,
a quantized,
time-independent and  uniform background magnetic field oriented in the positive $x_3$-direction can
be implemented by multiplying the QCD gauge link variables by $U_Q(1)$ link
fields, $U^{(Q)}_\mu(x)$, of the form
\begin{eqnarray}
 U^{(Q)}_1(x) & =& \left\{ \begin{array}{lcr}  1 &  {\rm for} 
 &  x_1 \ne  L -a \\
 \exp\left[-i q_q \tilde n {6\pi  x_2 \over L} \right]
 & {\rm for}& x_1 = L-a
 \end{array}\right. ,
   \nonumber\\
 U^{(Q)}_2(x) & = & 
  \exp\left[i q_q \tilde n {6\pi  a \, x_1 \over L^2} \right]\,,
  \nonumber\\
  U^{(Q)}_3(x)& = & 1
  \ , 
  \nonumber \\
  U_4^{(Q)}(x)& =& 1
  \ ,
  \label{eq:bkdgfield}
\end{eqnarray}
where $\tilde n$ is an integer
satisfying 
$| \tilde{n} | \leq \frac{1}{4} L^2 / a^2$ 
(in Euclidean space,  $x_4$ corresponds to the Wick-rotated time coordinate).  
In physical units, the
background magnetic fields used in this
work are quantized in units of 
$e {\bf B}  \sim 0.059 \ \tilde n\ {\rm GeV}^2 {\bf e_z}$;
 in comparison to the nucleon mass, the dimensionless
ratio $e |{\bf B}|/M_N^2 \sim 0.013$ for the smallest field suggesting the deformations 
arising from the magnetic field are perturbatively small compared to QCD effects for $|\tilde{n}|\, \lsim 10$.  
As $m_u=m_d=m_s$ in these calculations, the up-quark propagator 
in the $\tilde n=1$ field is the same as the down- or strange-quark propagator in the 
$\tilde n=-2$ field, with similar relations for the other magnetic field strengths.
To optimize the re-use of light-quark propagators in the calculation,
different quark electric charges were implemented by using a different magnetic field (with the same charge).
Consequently, 
 the $U_Q(1)$ fields  that are used in this work correspond to
$\tilde n=0,+1,-2,+3,+4,-6,+12$, 
corresponding to magnetic fields of 
$e {\bf  B} \sim  0, 0.06, -0.12, 0.18, 0.24, -0.36, 0.71~{\rm GeV}^2 {\bf e_z}$,
respectively.

In the presence of a time-independent and  uniform magnetic field, 
the energy eigenstates of a structureless charged particle with definite angular momentum along the field direction
are described by Landau levels and plane waves, 
rather than by three-momentum plane waves alone.
One of the subtle finite-volume (FV) effects introduced into this calculation is the loss of translational invariance in the interaction of 
charged particles with the background gauge field. 
We give a brief description of this problem, 
and relegate the more technical aspects of the discussion to  Appendix~\ref{app:CPcors}. 
For the implementation of the magnetic field using the links in Eq.~(\ref{eq:bkdgfield}), 
the lack of translational invariance is made obvious by  the Wilson loops,
\begin{eqnarray}
W_1(x_2) & = & 
\prod\limits_{j=0}^{L/a -1}\ U_1^{(Q)} (x+j a \hat x_1) 
\ =\ 
   \exp\left[-i q_q \tilde n {6\pi  x_2 \over L} \right]
   \ \ ,
   \nonumber\\
   W_2(x_1) & = & 
\prod\limits_{j=0}^{L/a -1}\ U_2^{(Q)} (x+j a \hat x_2) 
\ =\ 
   \exp\left[i q_q \tilde n {6\pi  x_1 \over L} \right]
\ \ \ ,
\label{eq:wilsonloops}
\end{eqnarray}
which wrap around the $x_1$ and $x_2$ directions of the lattice geometry, respectively. These exhibit explicit spatial dependence. 
Further, there are additional effects for charged-particle correlation functions
arising from their gauge dependence.

Because of the lack of translational invariance, 
the links employed in Eq.~(\ref{eq:bkdgfield}) 
define a spatial origin 
${\bf x} = {\bf 0}$, 
where the gauge potential vanishes, 
${\bf A}({\bf x})={\bf 0}$. 
In performing the present calculations, 
the source points for the correlation functions are not restricted to 
${\bf x} = {\bf 0}$ but instead are randomly distributed within the lattice volume, approximately restores translational invariance.
In the case of charged-particle correlation functions, 
this averaging leads to non-trivial effects, 
because the overlap of a given hadronic operator onto the various Landau levels depends on the source location. 
This is explicated in  Appendix~\ref{app:CPcors}, 
where additional methods of restoring translational invariance are discussed in the context of a structureless charged particle.

Post-multiplication of  $U_Q(1)$ links onto the QCD gauge links
omits the effects of the external magnetic field on the gluonic degrees
of freedom through the fermion determinant. The present calculations therefore 
correspond to a partially-quenched 
theory in which the sea quark charges are set to zero while the valence 
quark charges assume their physical values~\cite{Savage:2001dy,Chen:2001yi,Beane:2002vq,Detmold:2006vu}.  
For a SU(3) symmetric
choice of quark masses, as used herein, this does not affect the magnetic moments 
or the $np \to d \gamma$ transition
(linear responses to the field)  because the $N_f=3$
charge matrix is traceless \cite{Detmold:2006vu,Beane:2014ora} and 
couplings to sea quarks explicitly cancel at this order 
(indeed the isovector nature of the $np \to d \gamma$ transition make it insensitive to disconnected contributions even away from the SU(3) symmetric point).  
However,  the magnetic
polarizabilities receive contributions from terms in which the
two photons associated with the magnetic field  interact
with zero, one or two sea quark loops. 
The terms involving zero and two sea quark loops are correctly implemented,
however the square of the light-quark
electric-charge matrix is not traceless and  terms involving the
two photons interacting with one sea quark loop will contribute to isoscalar quantities for any non-zero charge matrix. 
In the present work, these terms are omitted for computational 
expediency.\footnote{
   Several approaches to these terms have been considered recently~\cite{Lujan:2014kia,Freeman:2014kka} 
   and may be investigated  in future studies of nuclei,  although  significant computational resources are required.
  }  
Generally,
it has been found that the related disconnected contributions to analogous single-hadron observables are small for the vector current \cite{Deka:2013zha,Green:2015wqa}, and we expect that this behavior
persists in nuclei. 
It is important to remember that  this systematic
ambiguity is restricted to the case of the isoscalar polarizabilities,
and that the isovector and isotensor combinations, such as
$\beta_{p} - \beta_{n}$, will be correctly calculated for the SU(3) symmetric case.

\subsection{Interpolating Operators and Contractions}

In order to probe the energy eigenstates of the systems under consideration, 
interpolating operators are constructed
with the desired quantum numbers. 
In principle, any choice of operator that has a  overlap onto a given eigenstate
is acceptable.
However, poor choices will have small overlaps onto the  state of interest
and hence will 
give rise to ``noisy'' signals with significant contamination from other states with the same quantum numbers.
For a vanishing background magnetic field, the energy eigenstates are also momentum 
eigenstates, and in order to access the ground state energy, it is simplest to 
choose interpolating operators that project onto states with zero three-momentum. 
In this work, this approach is followed
for both the neutron and proton (despite the proton carrying a positive electric charge).
The proton correlation functions  are of the form
\begin{eqnarray}
\label{eq:corr_prot}
C^{(P,S)}_p(t; {\bf x}_i) = \langle 0 | \widetilde{\cal O}^{(P,S)}_p(t;{\bf 0}) \overline{\cal O}^{(S)}_p({\bf x}_i,0) | 0 \rangle_{\bf B}\,,
\end{eqnarray}
with interpolating operators that are given by
\begin{eqnarray}
{\cal O}^{(S)}_p({\bf x},t)
& = & \epsilon^{ijk} [\tilde u_+^i({\bf x},t) C \gamma_5 \tilde d_+^j({\bf x},t)] \tilde u_+^k({\bf x},t)\,, 
\nonumber\\
\widetilde {\cal O}^{(P)}_p(t;{\bf p}) 
& = & \sum_{{\bf x}} e^{i\bf{p}\cdot{\bf x}}\epsilon^{ijk} [u_+^i({\bf x},t) C \gamma_5 d_+^j({\bf x},t)] u_+^k({\bf x},t) \,, 
\nonumber\\
\widetilde {\cal O}^{(S)}_p(t;{\bf p}) 
& = & \sum_{{\bf x}} e^{i\bf{p}\cdot{\bf x}}\epsilon^{ijk} [\tilde u_+^i({\bf x},t) C \gamma_5 \tilde d_+^j({\bf x},t)] 
\tilde u_+^k({\bf x},t)
\ =\ 
\sum_{{\bf x}} e^{i\bf{p}\cdot{\bf x}} \  {\cal O}^{(S)}_p({\bf x},t)
 \,,
\label{eq:ops}
\end{eqnarray}
where $\langle\ldots\rangle_{{\bf B}}$ indicates ensemble averaging with respect to 
QCD in the presence of the $U_Q(1)$ links corresponding to a uniform background magnetic field
${\bf B} = B {\bf e_z}$,
and  where the spin indices of the operators, carried by the third quark, are suppressed.
In Eq.~(\ref{eq:ops}), $\tilde q(x)$ corresponds to a quark field of flavor $q=u,d$ that has been 
smeared \cite{Albanese:1987ds} in the spatial directions using a Gaussian form, while $q(x)$ corresponds to a local field. Additionally, the subscript $_+$ on the quark fields implies that they are explicitly projected onto positive energy modes, that is $u^i_+({\bf x},t)= (1+\gamma_4)u^i({\bf x},t)$. 
The superscript $(P,S)$ on an interpolating operator (and hence the correlation function) indicates a point or smeared form, respectively,
$C=i\gamma_4\gamma_2$ is the charge conjugation matrix, and 
$\overline{\cal O}^{(S)}_p = {\cal O}^{(S)\dagger }_p\gamma_4$. 
Neutron correlation functions are constructed from those of the proton by interchanging $u\leftrightarrow d$.
The correlation functions with the quantum numbers of nuclei,
constructed using the methods discussed in detail in Refs.~\cite{Detmold:2012eu,Beane:2012vq}, 
are built recursively using sink-projected nucleon ``blocks'' that involve either smeared or local fields. 
For the present calculations,  zero momentum states  are built from zero-momentum blocks,
although more complicated constructions can also be considered.

As suggested in Ref.~\cite{Tiburzi:2012ks}, in order to study the proton in a magnetic field, it 
would be more appropriate  to use interpolating operators that project onto the lowest-lying Landau level,
rather than projecting onto three-momentum plane-waves.
This would enhance the overlap of the interpolating operator onto the lowest, close-to-Landau 
energy eigenstate and suppress the overlap onto higher states.
However, it is unclear how to generalize such a  framework to nuclei  constructed from nucleon blocks. 
While  single-hadron blocks provide
a good basis for the construction of correlation functions of nuclei in the absence of background fields, 
this will not necessarily be the case in a magnetic field.  
The problem is exemplified by $^3$He, 
a $j ={1\over 2}$ nucleus
comprised of two protons and a neutron.  
Assuming a compact state
(which it has been shown to be at the quark masses used in this work
through calculations in multiple lattice volumes~\cite{Beane:2012vq}), 
the wavefunction of the nucleus is a Landau
level determined by its electric charge of $Q=2e$, 
while that of the proton is a Landau level determined by $Q=e$, 
and that of the neutron is a momentum eigenstate. 
Proton blocks could be constructed by projection onto a given Landau level at the sink, 
leading to presumably improved signals for the proton correlation functions. 
However, combining Landau-projected proton blocks and momentum-projected neutron blocks
will not necessarily produce interpolating operators that couple well to the
$^3$He Landau levels. 
For larger nuclei, this problem becomes even more severe. 
There are  interesting directions that could be pursued in this regard, 
for example
constructing single hadron blocks that are projected onto the Landau levels 
of the ``target'' nucleus. 
However, {\it a priori} it is difficult to   ascertain how efficacious 
such approaches will be. 
In the current work the same interpolating 
operators used for zero-field calculations are used for calculations in the presence of background magnetic fields.
Although  these are not  ideal  interpolating operators, 
 they are not orthogonal to the expected eigenstates, and the results extracted from this
naive approach  serve as a benchmark for more sophisticated methods 
that can be explored in future investigations.

The correlation functions investigated in this work with the quantum numbers of nuclei  are of the 
form
\begin{eqnarray}
C_{h;j_z}(t; {\bf B}) & = & 
\langle 0 | \tilde f_{h;j_z}\left[\widetilde{\cal O}^{(S,P)}_{p,n}(t;\vec{0})\right]
f_{h;j_z}\left[\overline{\cal O}^{(S)}_{p,n}(t;{\bf x}_i)\right] |0 \rangle_{{\bf B}}
\, .
\label{eq:nuccorr}
\end{eqnarray}
The exact source and sink interpolating functions, 
$f_{h;j_z}$ and $\tilde f_{h;j_z}$,
depend on the quantum numbers of the nucleus, 
and are defined implicitly through the recursive procedures of Ref.~\cite{Detmold:2012eu}. 
For nuclei with non-zero total spin $j$,  the $z$-component of spin, $j_z$, is made explicit 
as each magnetic sub-state is studied individually.
On each QCD gauge configuration, up- and down-quark propagators are generated for each of the
seven magnetic field strengths from 48 uniformly distributed Gaussian-smeared sources. 
The position of the first source was randomly chosen and the remaining sources were placed on a regular, three-dimensional grid  relative to that location. 
The sources locations were selected after the background magnetic field was applied and hence there was no correlation between 
the source location and the position of the zero of the vector potential.
As the calculations from the different source locations on each gauge field are averaged over,
the dependence of $C_{h;j_z}(t; {\bf B})$ on the source location ${\bf  x}_i$ is suppressed.
This location averaging effectively projects 
the source interpolating operator onto zero momentum and is discussed in detail in Appendix \ref{app:CPcors}.
In most cases, two  correlation functions are  constructed for each 
nuclear state using the smeared and point sink interpolators, 
although for larger nuclei there are more possibilities than are calculated.

\subsection{Magnetic Field Strength Dependence of Energies}
\label{sec:fieldstrength}

In a uniform background magnetic field, the
energy eigenvalues of a hadron, $h$, either a nucleon or nucleus, with spin $j\leq1$ polarized in the $z$-direction, 
with magnetic quantum number $j_z$, 
are of the form
\begin{eqnarray}
  E_{h;j_z}({\bf B}) & = & 
  \sqrt{M_h^2 + P_\parallel^2 + (2 n_L +1) | Q_h e {\bf B}|}
   - {\bm \mu}_h\cdot{\bf B}
  - 2\pi \beta_h^{(M0)} |{\bf B}|^2 
  - 2\pi \beta^{(M2)}_h \langle \hat T_{ij} B_i B_j \rangle  + \ldots
  \,, \nonumber \\
  \label{eq:Eshift}
\end{eqnarray}
where $M_h$ is the mass of the hadron, 
$P_\parallel$ is its momentum parallel to the magnetic field,
$Q_h$ is its charge in units of $e$,
and $n_L$ is the
quantum number of the Landau level that it occupies.  
For a nucleon or nucleus with spin $j\ge {1\over 2}$, there is a contribution from the
magnetic moment, ${\bm \mu}_h$, that is linear in the magnetic field.
The magnetic polarizability is conveniently decomposed into
multipoles, with $\beta_h\equiv\beta^{(M0)}_h$ denoting the scalar polarizability and
$\beta^{(M2)}_h$ denoting the tensor polarizability 
(the latter contributes for hadrons with $j\ge 1$).  
$\hat T_{ij}$ is a traceless symmetric tensor operator which, when written in terms of angular momentum generators, 
is of the form
\begin{eqnarray}
\hat T_{ij} & = & {1\over 2}\left[\ \hat J_i \hat J_j +  \hat J_j \hat J_i - {2\over 3} \delta_{ij} \hat J^2\ \right]
\,,
\label{eq:Tijdef}
\end{eqnarray}
and 
$\langle ... \rangle$ in Eq.~(\ref{eq:Eshift}) denotes its expectation 
value.\footnote{ 
For a magnetic field aligned in the z-direction,
it follows that 
  $ \langle \hat T_{ij} B_i B_j \rangle = \langle \hat T_{zz} B^2 \rangle =  \left( j_z^2-{1\over 3} j(j+1) \right) B^2 $.
   This vanishes for both the $j=0$ and $j={1\over 2}$ states, and takes the values
   $ \langle \hat T_{ij} B_i B_j \rangle = {1\over 3}B^2$ for  $j=1, j_z=\pm 1$ states  and 
   $ \langle \hat T_{ij} B_i B_j \rangle = -{2\over 3}B^2$ for  $j=1, j_z=0$ states.
    }  
Note that the polarizabilities defined here represent the full quadratic response to the field and 
differ from other conventions used in the literature where Born terms are explicitly removed 
(for a discussion, see e.g. Ref.~\cite{Birse:2000ve}).
The ellipses denote contributions that involve three or more powers of the
magnetic field.  
The spin-averaged energy eigenvalues project 
onto the scalar contributions,
\begin{eqnarray}
  \langle E_h({\bf B}) \rangle  & \equiv & 
  {1\over 2j+1}\sum_{j_z=-j}^j E_{h;j_z}({\bf B}) 
  =  \sqrt{M_h^2 + P_\parallel^2 + (2 n_L +1) | Q_h e {\bf B}|}  
  \ -\  2\pi \beta_h^{(M0)} |{\bf B}|^2 \ +\  ...
  \ ,
  \label{eq:avEshift}
\end{eqnarray}
where the ellipsis denotes contributions of ${\cal O}(|{\bf B}|^4$) and higher.
For spin-$j$ states, the energy difference between $j_z=\pm j$ 
isolates the magnetic moment at lowest order in the expansion. 
Other combinations of the energy eigenvalues of the individual  spin components
can be formed to isolate  higher multipoles.

\section{Results}
\label{sec:results}

\subsection{Extraction of Energy Levels}

With the  background magnetic  field given in Eq.~(\ref{eq:bkdgfield}),  
well-defined energy levels exist for each value of the magnetic field strength. 
In order to determine the magnetic polarizabilities, 
energy eigenvalues are determined from the appropriate correlation functions, the 
$C_{h;j_z}(t; {\bf B})$ defined in Eq.~(\ref{eq:nuccorr}).
\begin{figure}[!ht]
  \centering
  \includegraphics[width=\columnwidth]{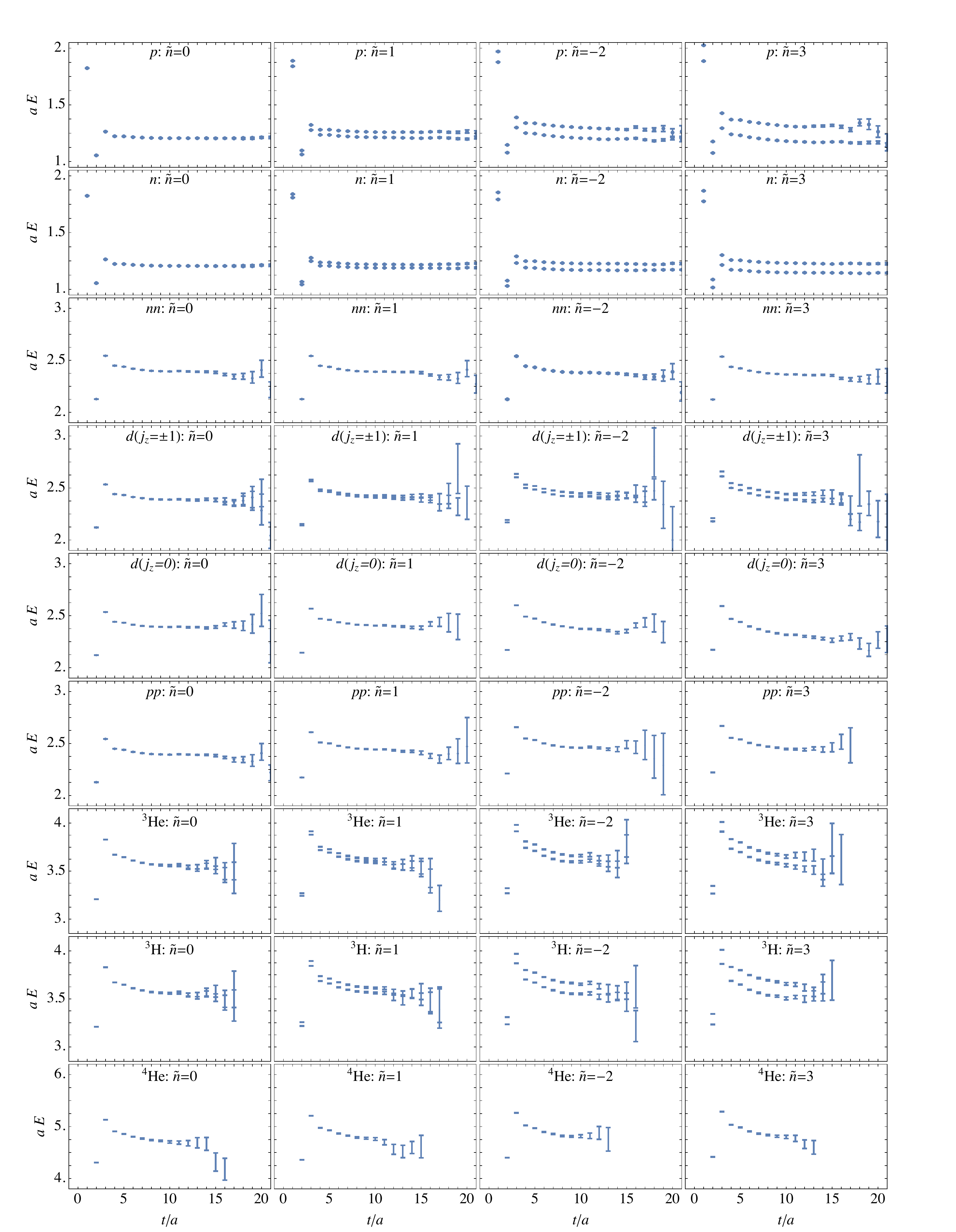}\ \
  \caption{
  EMPs obtained from the smeared-smeared 
  nucleon and nuclear 
  correlation functions for the lowest-four magnetic field strengths. 
  The $j_z=\pm j$ states are shown for each hadron.
  }
  \label{fig:corrs}
\end{figure}
The individual correlation functions associated with each state in each magnetic field
are examined, and the time intervals over which  they are consistent with
single exponential behavior are determined.
Effective mass plots (EMPs) associated with 
representative correlation functions obtained in the  magnetic fields with 
$\tilde n=0,1,-2,3$  are shown  in Fig.~\ref{fig:corrs}.
Having identified these time intervals, the main analysis focuses on ratios 
of these correlation functions constructed as
\begin{eqnarray}
  R_{h;j_z}(t;{\bf B}) \ = \ 
  { C_{h;j_z}(t; {\bf B}) \over C_{h;j_z}(t; {\bf B}=0) }
  & \stackrel{t\to\infty}{\longrightarrow} & 
  Z_{h;j_z}({\bf B})\ e^{- \delta E_{h;j_z}({\bf B}) t }
  \ \ \ ,
  \label{eq:ratcorr}
\end{eqnarray}
where $\delta E_{h;j_z}({\bf B}) = E_{h;j_z}({\bf B}) - E_{h;j_z}({\bf 0})$
is the energy difference induced by the magnetic field,
and $Z_{h;j_z}({\bf B})$ is a
time-independent, but field-dependent, overlap factor.
To be specific,  
$h=p,\ n,\ nn,\ d,\ pp,\ ^{3}{\rm He},\ ^{3}{\rm H}$ and $^{4}{\rm He}$ 
are considered in all magnetic substates.
It is advantageous to work with ratios of correlation functions
because fluctuations present in the  energies  extracted
from  individual correlation functions cancel to a significant degree in the ratio.\footnote{ 
Note that such ratios are formed after averaging over an ensemble of measurements.
} 
The energy shifts are extracted from these ratios 
in the time regions in which the 
contributing
individual correlation functions  show single state dominance
by either directly fitting the ratio or, alternatively,
by fitting the effective mass resulting from the double ratio
\begin{eqnarray}
  \delta E_{h;j_z}({\bf B};t,t_J)
  & = & 
  {1\over t_J} \ \log\left({ R_{h;j_z}(t;{\bf B})\over R_{h;j_z}(t+t_J;{\bf B})}\right)
  \stackrel{t\to\infty}{\longrightarrow} 
  \delta E_{h;j_z}({\bf B})   \, ,
  \label{eq:EMPratcorr}
\end{eqnarray}
where $t_J$ represents a temporal offset that can be chosen to optimize energy extractions. 
The fits are performed using the correlated $\chi^2$-minimization procedure, 
with the covariance matrix  determined using  jackknife or bootstrap resampling. 
A systematic fitting uncertainty is estimated by performing fits over multiple fitting intervals 
within the  region of single-exponential dominance for a given system.
The primary analysis in this work
is based on 
fitting the ratios of correlation functions obtained from
binning  source-averaged measurements into 
$N_b=100$ blocked samples 
and  generating $N_B=200$ bootstrap samples from these blocked samples. 
Alternate analyses are also undertaken in which differences include varying
the statistical procedures and also performing constant fits to effective mass functions formed from different 
values of $t_J$ and other possible variations. Consistent results are obtained.

In order to extract the magnetic polarizabilities,
ratios of the correlation functions associated with the maximally-stretched spin states
can be formed such that the leading magnetic-moment contributions cancel,
\begin{eqnarray}
  \overline R_{h}(t;{\bf B}) \ = \ 
  { C_{h;j_z=j}(t;{\bf B}) \over C_{h;j_z=j}(t;{\bf 0}) }{ C_{h;j_z=-j}(t;{\bf B}) \over C_{h;j_z=-j}(t;{\bf 0}) }
  & \rightarrow & 
  \tilde Z\ e^{-2\delta \langle E_h({\bf B})\rangle t }
  \ \ \ ,
  \label{eq:ratratcorr}
\end{eqnarray}
where 
$ \delta \langle E_h({\bf B})\rangle= \langle E_h({\bf  B})\rangle -M_h$ 
is the  spin-averaged energy shift. 
  Similarly, the spin difference between maximal $j_z=\pm j$ states eliminates 
  the spin-independent terms, leaving only the magnetic-moment contribution and ${\cal O}(|{\bf B}|^3)$ and higher terms. 
  This has been used to extract magnetic moments in Refs.~\cite{Primer:2013pva,Beane:2014ora} using the ratio
  \begin{eqnarray}
  \Delta R_{h}(t;{\bf B}) \ = \ 
  { C_{h;j_z=j}(t;{\bf B}) \over C_{h;j_z=j}(t;{\bf 0}) }{ C_{h;j_z=-j}(t;{\bf 0})  \over C_{h;j_z=-j}(t;{\bf B})}
  & \rightarrow & 
  \Delta \tilde Z\ e^{- (E_{h;j}({\bf B})- E_{h;-j}({\bf B})) t }
  \,.
  \label{eq:polratratcorr}
\end{eqnarray}
In the present work, 
the individual spin states are used to extract the magnetic moments and polarizabilities in a coupled fit as the latter quantities are the primary target of this study. However,  the magnetic moments have also been extracted from the spin-difference ratios,
Eq.~(\ref{eq:polratratcorr}) \cite{Beane:2014ora},
and that approach is  
found to lead to consistent, but more precise  results and remains the preferred method for extracting the  magnetic moments.
Given the more complicated nature of the fits we perform here in order to 
obtain sensitivity to the polarizabilities, it is unsurprising that the uncertainties on the lower order terms are  larger. 
In what follows, we present the magnetic moments that result from the coupled fits for completeness, 
but use the previous fits to spin-difference ratios as the best estimates of these quantities.

\begin{figure}[!th]
  \centering
  \includegraphics[width=\columnwidth]{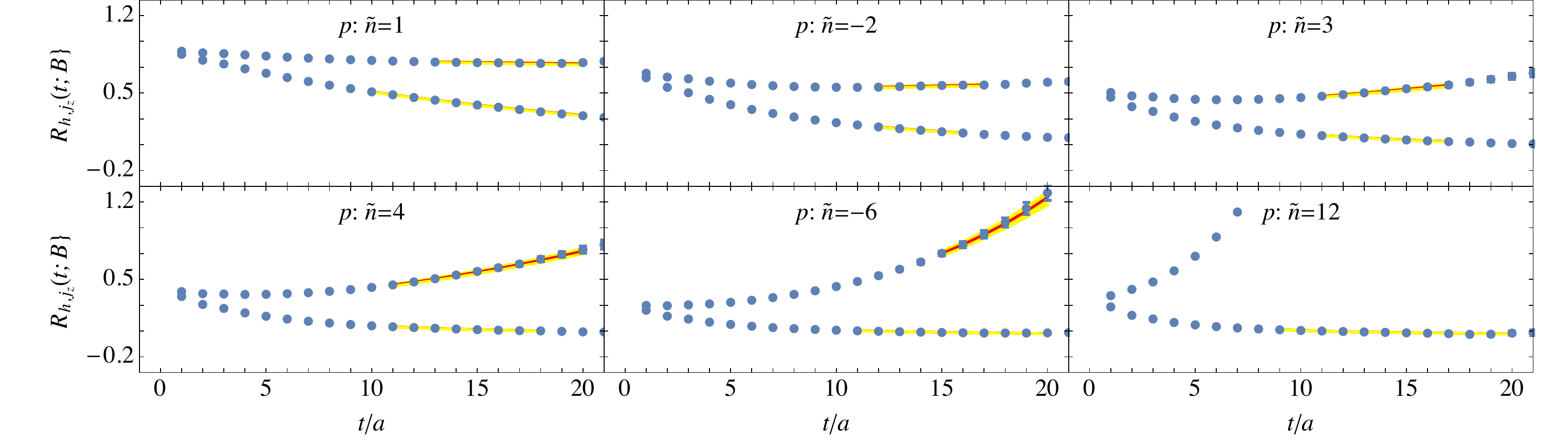}\\ 
    \includegraphics[width=\columnwidth]{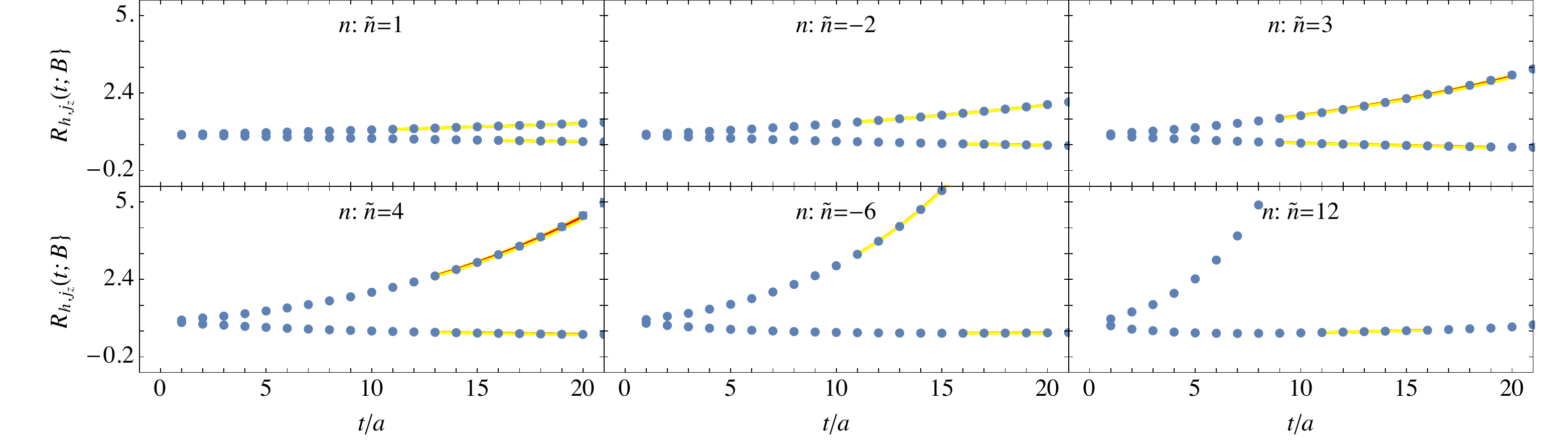}\\ 
      \includegraphics[width=\columnwidth]{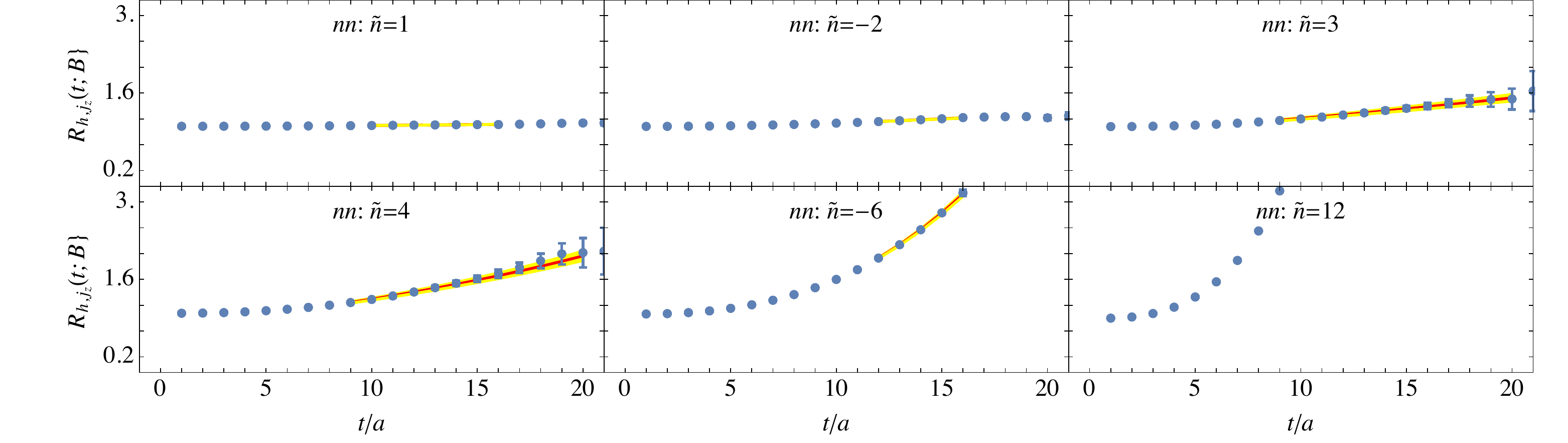}\\ 
        \includegraphics[width=\columnwidth]{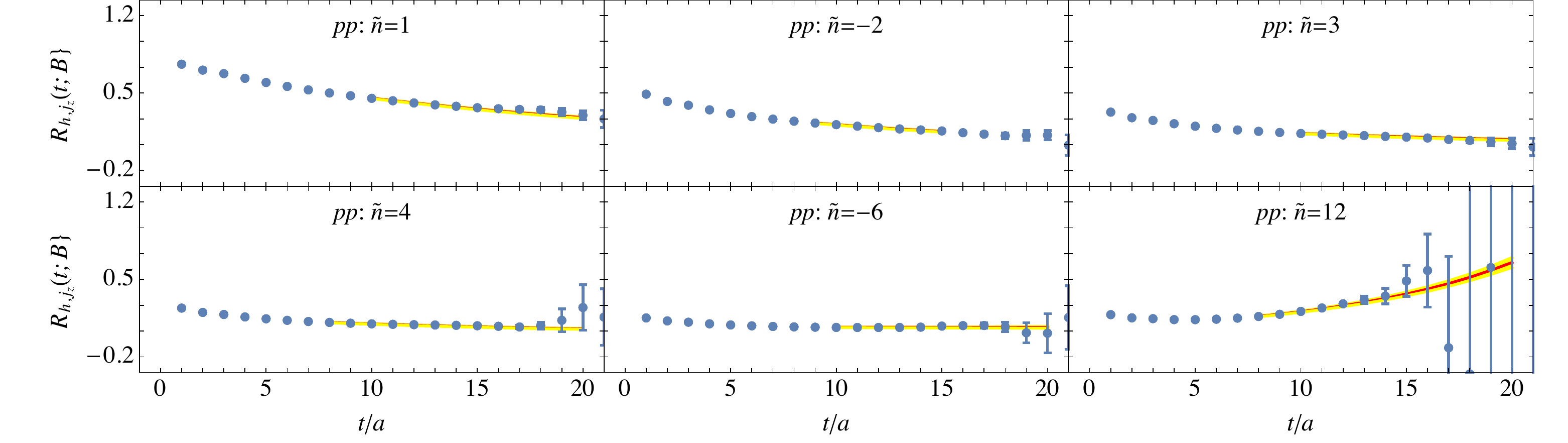}
  \caption{
  The ratio of correlation functions  associated with  the $p$, $n$, $nn$ and $pp$ systems. 
  Results are shown for all six field strengths for the smeared-smeared correlators and for both $|j_z|=j$ states for 
  states with $j>0$. 
  The shaded bands correspond to the statistical uncertainties of the given fit.}
  \label{fig:corrratio1}
\end{figure}

\begin{figure}[!th]
  \centering
  \includegraphics[width=\columnwidth]{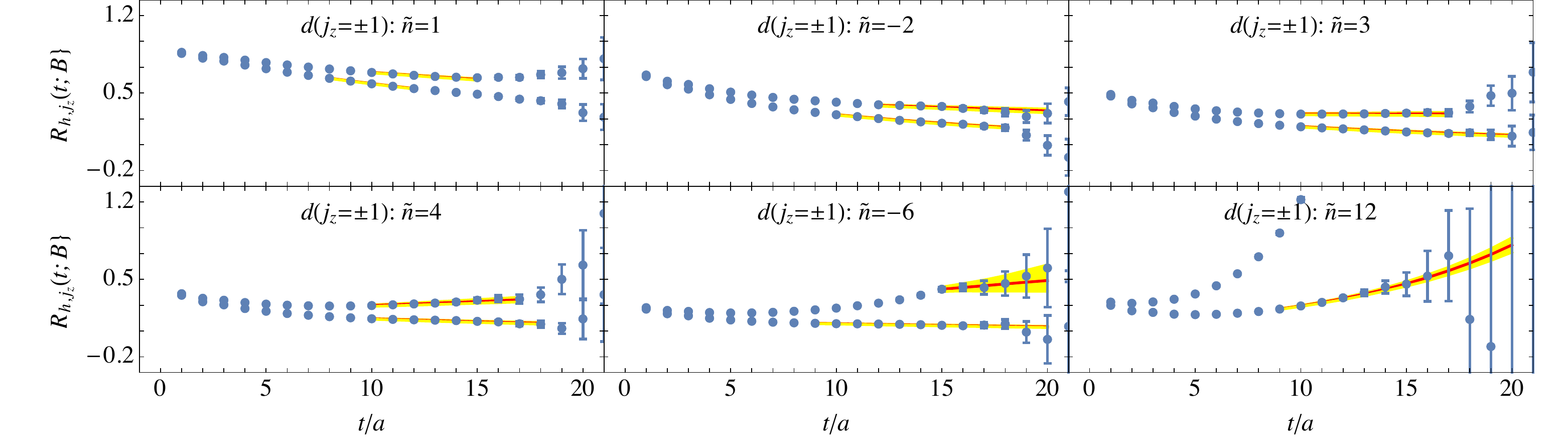}\\ 
    \includegraphics[width=\columnwidth]{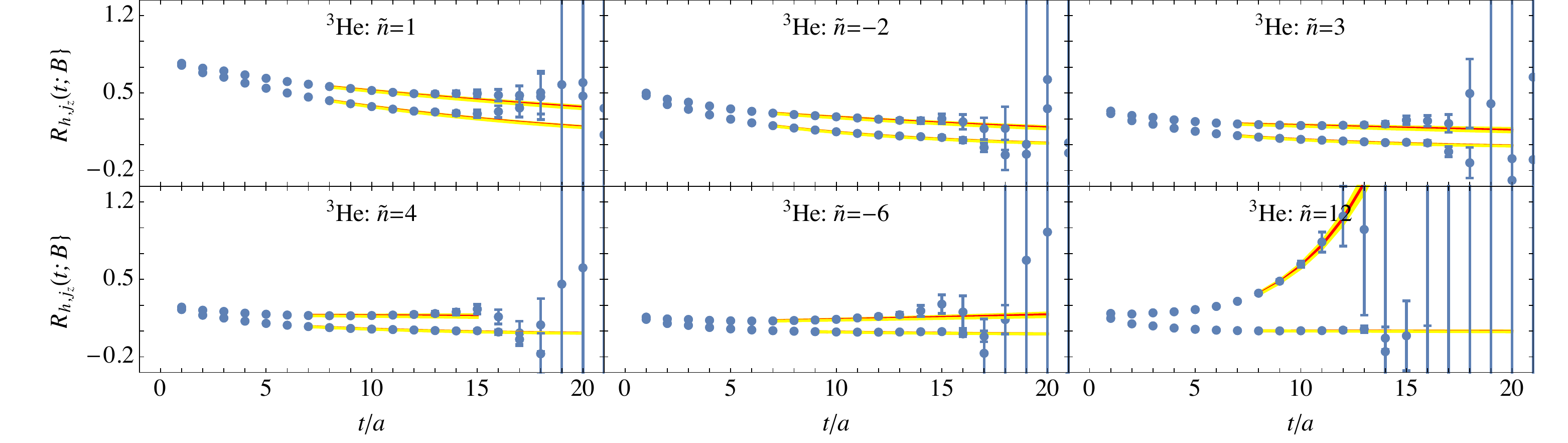}\\ 
      \includegraphics[width=\columnwidth]{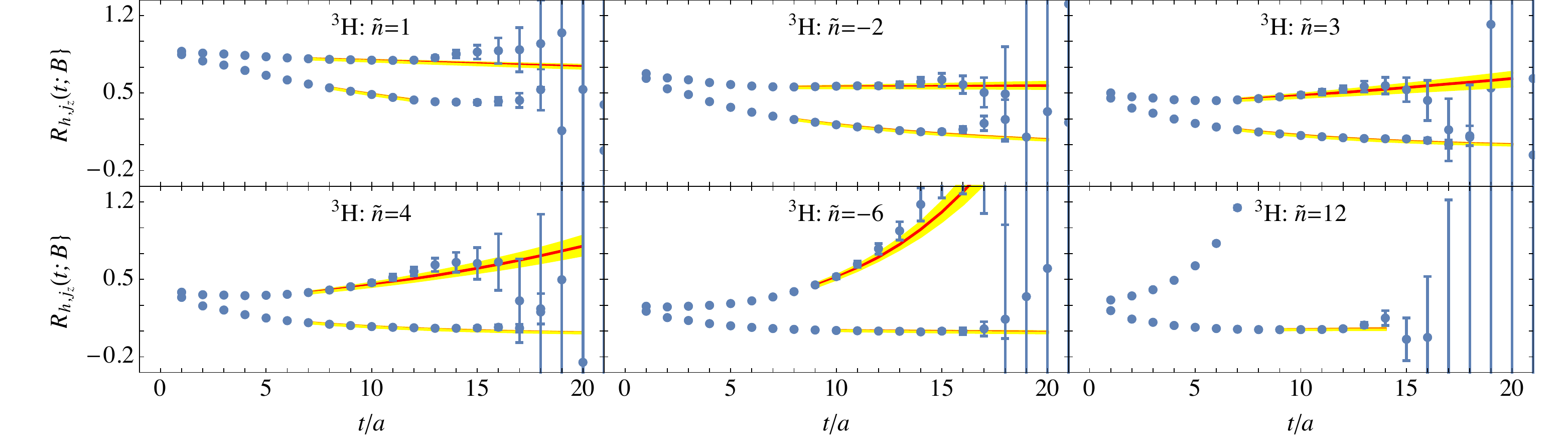}\\ 
        \includegraphics[width=\columnwidth]{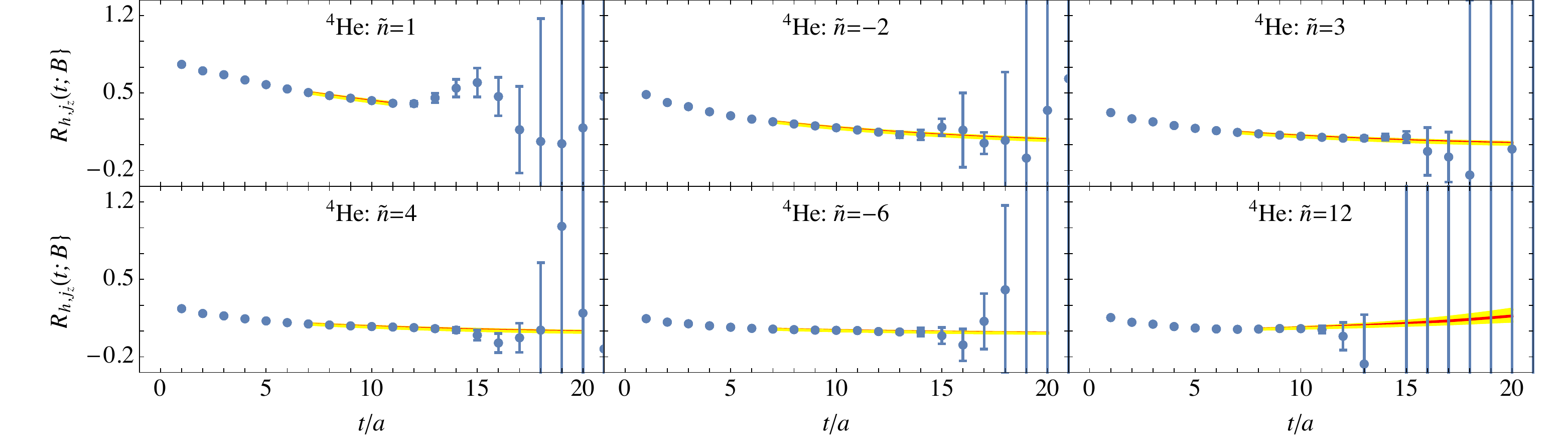}
  \caption{ The ratio of correlation functions  associated with the  $d$, $^3$He, $^3$H and  $^4$He states. 
  Results are shown for all six field strengths for the smeared-smeared correlators and for both $|j_z|=j$ states for 
  states with $j>0$. The shaded bands correspond to the statistical uncertainties of the given fit.}
  \label{fig:corrratio2}
\end{figure}

Figures~\ref{fig:corrratio1} and \ref{fig:corrratio2} show the 
ratios of correlation functions used in the extraction of energy shifts for each magnetic field strength and spin component. 
Results from all six nonzero magnetic field strengths are shown.
In each case, the associated single exponential fit to the ratio of correlation functions is shown, along with the associated 
statistical uncertainty. 
Fits over all time ranges
in $[t_{\rm min},t_{\rm max}]$ are considered,
where $t_{\rm max}=24$ and $t_{\rm min}$ is set by requiring consistency with single exponential 
behavior of the individual correlators that form a given ratio. 
The central fit is identified as that 
over the time range with $t_{\rm max}-t_{\rm min}>3$ with the lowest correlated $\chi^2/d.o.f$. 
The standard deviation of all fits over subsets of time ranges in the interval  $[t_{\rm min}-1,t_{\rm max}]$
that have a $\chi^2$ within one unit of the minimum $\chi^2$
($\chi^2\rightarrow \chi_{\rm min}^2 + 1$)
is taken as the fitting systematic uncertainty.
The extracted energy shifts are tabulated in Table \ref{tab:energies}.
\begin{table}[!t]
  \begin{ruledtabular}
    \begin{tabular}{ c r | c c c c c c}
   \multicolumn{2}{c|}{State} &\multicolumn{6}{c}{$a\ \delta E_{h;j_z}(\tilde n)$} \\
   \hline
    $h$ & $j_z$ & $\tilde n=1$ & $\tilde n=-2$ & $\tilde n=3$ & $\tilde n=4$ & $\tilde n=-6$ & $\tilde n=12$   \\
    \hline\hline
    $p$ & $\frac{1}{2}$ & \footnotesize{0.0032(11)(17)} & \footnotesize{0.0839(24)(0)} & \footnotesize{$-$0.0324(22)(19)} & \footnotesize{$-$0.0581(26)(12)} & \footnotesize{0.1288(51)(65)} & \footnotesize{$-$0.2495(17)(13)} 
    \\
    $p$ & $-\frac{1}{2}$ & \footnotesize{0.05372(63)(68)} & \footnotesize{$-$0.0073(16)(7)} & \footnotesize{0.1035(34)(23)} & \footnotesize{0.1087(37)(42)} & \footnotesize{$-$0.1045(57)(41)} & \footnotesize{0.142(4)(16)} 
    \\
    \hline
    $n$ & $\frac{1}{2}$ & \footnotesize{0.01297(32)(19)} & \footnotesize{$-$0.03741(12)(8)} & \footnotesize{0.0249(12)(19)} & \footnotesize{0.0184(21)(12)} & \footnotesize{$-$0.12694(35)(27)} & \footnotesize{$-$0.02318(91)(64)} 
    \\
    $n$ & $-\frac{1}{2}$ & \footnotesize{$-$0.01711(7)(17)} & \footnotesize{0.01749(75)(46)} & \footnotesize{$-$0.0584(10)(11)} & \footnotesize{$-$0.0831(17)(6)} & \footnotesize{$-$0.0027(23)(24)} & \footnotesize{$-$0.24212(63)(25)} 
    \\
    \hline
    $nn$ & 0 & \footnotesize{$-$0.00321(15)(32)} & \footnotesize{$-$0.0146(8)(12)} & \footnotesize{$-$0.0285(31)(21)} & \footnotesize{$-$0.0488(32)(36)} & \footnotesize{$-$0.1147(24)(99)} & \footnotesize{$-$0.2793(27)(26)} 
    \\
    \hline
    $d$ & 1 & \footnotesize{0.0190(16)(74)} & \footnotesize{0.0588(34)(44)} & \footnotesize{$-$0.0009(54)(31)} & \footnotesize{$-$0.0262(72)(61)} & \footnotesize{0.033(7)(15)} & \footnotesize{$-$0.337(22)(19)} 
    \\
    $d$ & -1 & \footnotesize{0.0398(8)(33)} & \footnotesize{0.0169(59)(79)} & \footnotesize{0.0523(58)(62)} & \footnotesize{0.041(8)(13)} & \footnotesize{$-$0.039(47)(28)} & \footnotesize{$-$0.114(7)(20)} 
    \\
    \hline
    $pp$ & 0 & \footnotesize{0.0490(19)(82)} & \footnotesize{0.0679(36)(56)} & \footnotesize{0.0536(55)(88)} & \footnotesize{0.062(5)(12)} & \footnotesize{0.001(11)(34)} & \footnotesize{$-$0.114(5)(15)} 
    \\
    \hline
    $^3$He & $\frac{1}{2}$ & \footnotesize{0.067(3)(24)} & \footnotesize{0.0408(38)(53)} & \footnotesize{0.123(8)(10)} & \footnotesize{0.126(8)(22)} & \footnotesize{$-$0.028(7)(27)} & \footnotesize{0.01(1)(69)} 
    \\
    $^3$He & $-\frac{1}{2}$ & \footnotesize{0.034(3)(16)} & \footnotesize{0.112(4)(13)} & \footnotesize{0.023(7)(16)} & \footnotesize{0.0045(85)(94)} & \footnotesize{0.112(15)(77)} & \footnotesize{$-$0.262(11)(89)} 
    \\
    \hline
    $^3$H & $\frac{1}{2}$ & \footnotesize{0.007(1)(16)} & \footnotesize{0.100(4)(19)} & \footnotesize{$-$0.027(7)(25)} & \footnotesize{$-$0.058(7)(39)} & \footnotesize{0.06(2)(11)} & \footnotesize{$-$0.31(8)(12)} 
    \\
    $^3$H & $-\frac{1}{2}$ & \footnotesize{0.058(2)(27)} & \footnotesize{$-$0.0025(35)(67)} & \footnotesize{0.125(7)(29)} & \footnotesize{0.138(8)(38)} & \footnotesize{$-$0.152(11)(35)} & \footnotesize{$-$0.02(2)(70)} 
    \\
    \hline
    $^4$He & 0 & \footnotesize{0.056(4)(69)} & \footnotesize{0.086(7)(30)} & \footnotesize{0.090(14)(35)} & \footnotesize{0.098(15)(74)} & \footnotesize{0.09(1)(11)} & \footnotesize{$-$0.10(3)(62)} \\
    \end{tabular}
  \end{ruledtabular}
  \caption{
   Extracted energy shifts of the nucleons and nuclei, where the first uncertainty is statistical and the second corresponds to the systematic uncertainty obtained from variation of fitting ranges.
   The $j_z=0$ deuteron state is studied separately.
}
  \label{tab:energies}
\end{table}

\begin{figure}[!th]
  \centering
  \includegraphics[width=\columnwidth]{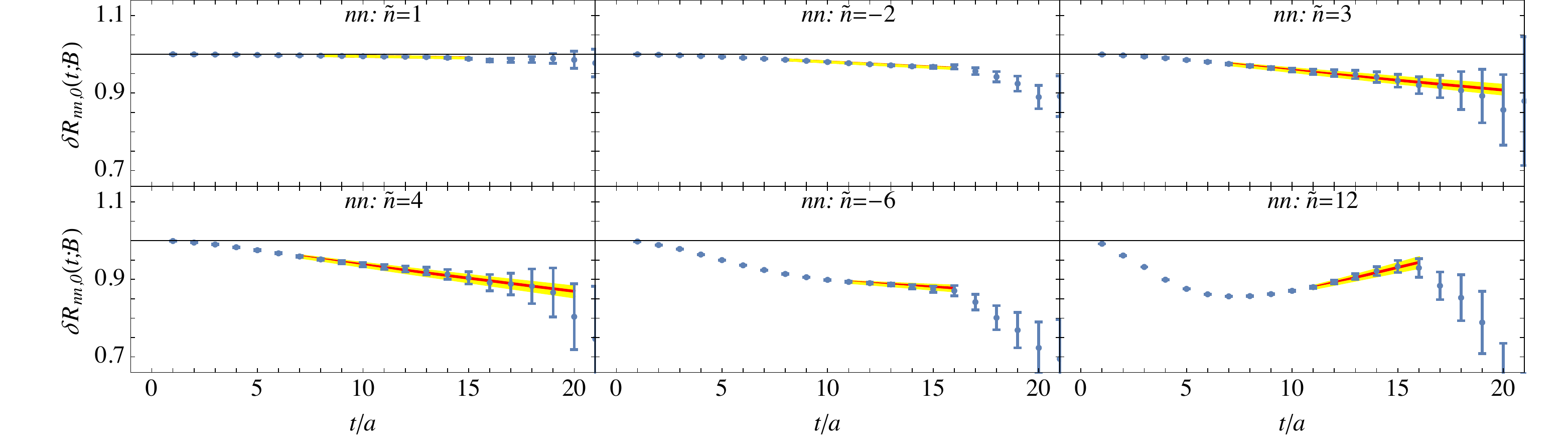}\\ 
    \includegraphics[width=\columnwidth]{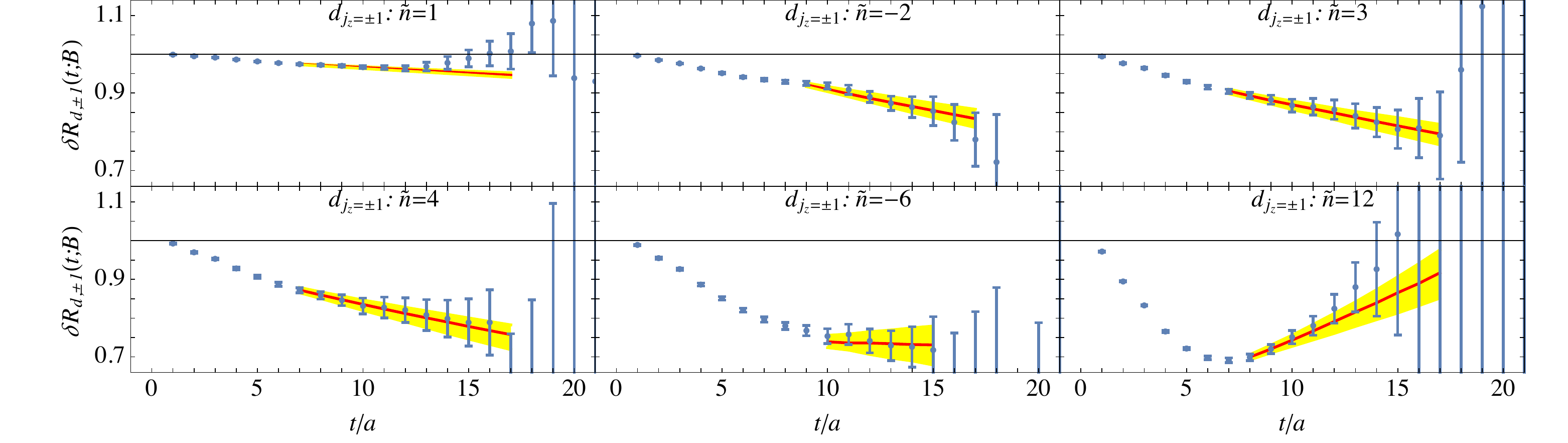}
    \includegraphics[width=\columnwidth]{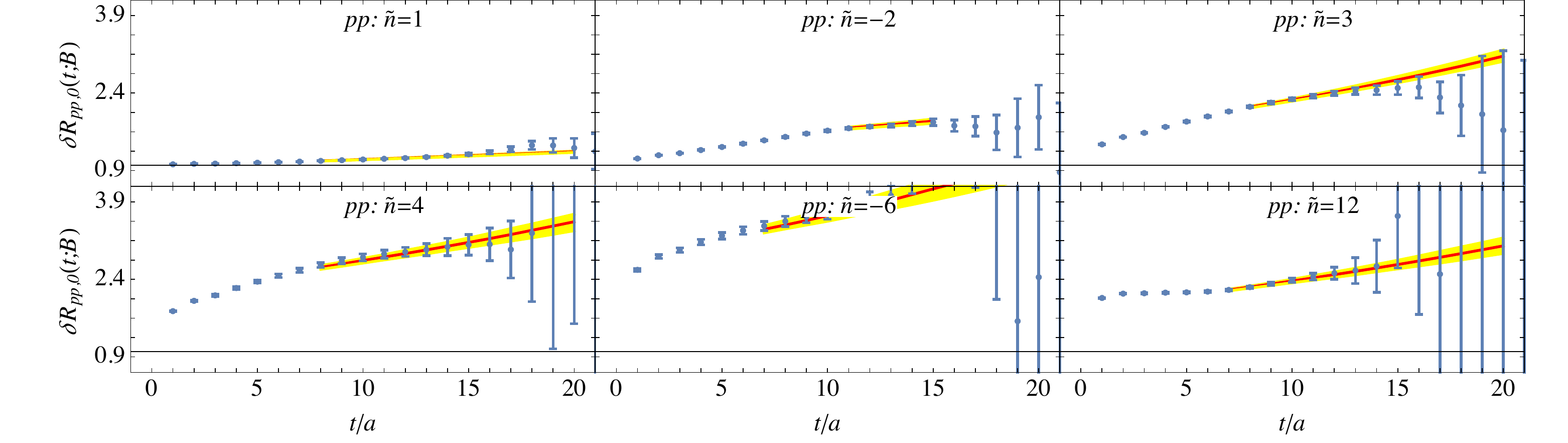}
    \caption{
The ratios of correlation functions defined in Eq.~(\ref{eq:ratcorrAminus}) for the $nn$, $d (j_z=\pm1)$ and $pp$ systems. 
Results are shown for all six magnetic fields for the smeared-smeared correlation functions. 
The shaded bands correspond to the statistical uncertainties of the given fit.
The deuteron spin states are averaged for simplicity. }
  \label{fig:deltaNNcorr}
\end{figure}

The 
correlation functions associated with the nucleons and nuclei are highly correlated,
and therefore  differences between the energies of two given 
states can be more accurately determined  
than those of each state individually.
Of particular interest is the difference between the 
magnetic properties of a nucleus and that of its constituent nucleons. 
To highlight these differences,  further ratios are constructed,
\begin{eqnarray}
  \delta R_{{\cal A},j_z}(t;{\bf B}) \ = \ 
  { R_{{\cal A}}(t; {\bf B}) \over 
  \prod\limits_{h\in {\cal A}}R_{h}(t; {\bf B}) }
  & \stackrel{t\to\infty}{\longrightarrow} & 
  \delta Z_{{\cal A}}({\bf B})\exp\left[{- \left(\delta E_{{\cal A}}({\bf B}) - \sum_{h\in{\cal A}} \delta E_{h}({\bf B}) \right) t }\right]
  \ \ \ ,
  \label{eq:ratcorrAminus}
\end{eqnarray}
where the nucleus ${\cal A}$ contains a set of nucleons, $h$,
 and the spin indices have been suppressed  for brevity.
Figure~\ref{fig:deltaNNcorr} shows this ratio of correlation functions for 
the $nn$, $j_z=\pm1$ deuteron, and $pp$ systems.

As discussed previously, the momentum-projected 
interpolating operators are not expected to 
provide particularly good overlap onto the 
low-energy eigenstates of the 
proton and 
charged nuclei in magnetic fields, 
which are expected to more closely resemble Landau  wavefunctions. 
Indeed,  the  interpolating operators are found to overlap
most strongly with states {\it other} than the lowest Landau level, as will be discussed below and in detail in Appendix \ref{app:CPcors}. 
In Fig.~\ref{fig:Zfactor},
 the ratio of overlap factors of the extracted state at nonzero and zero 
background field strengths are shown. 
For the neutral states, the overlap is only weakly dependent 
on the field strength, but for charged states, the overlap rapidly 
decreases with increasing magnetic field strength. 
This indicates that 
care must be taken in interpreting the extracted states. 
It does not mean that unrelated states 
(those that are not continuously related as a function of the magnetic field)
are being extracted for different field strengths, 
but instead the overlap onto the given state is decreasing.\footnote{
Similar effects have been seen in constructing multi-pion correlation functions from 
combinations of pion interpolating operators built with differing momenta.
Exponentially smaller overlaps
were observed,  although a consistent energy could be extracted~\cite{Zhifeng,Shi:2011mr}. 
In background electric fields, the overlap factors of momentum projected interpolating operators 
were also seen to decrease significantly with 
the applied field strength~\cite{Detmold:2009dx,Detmold:2010ts}.}
\begin{figure}[!ht]
  \centering
  \includegraphics[width=\columnwidth]{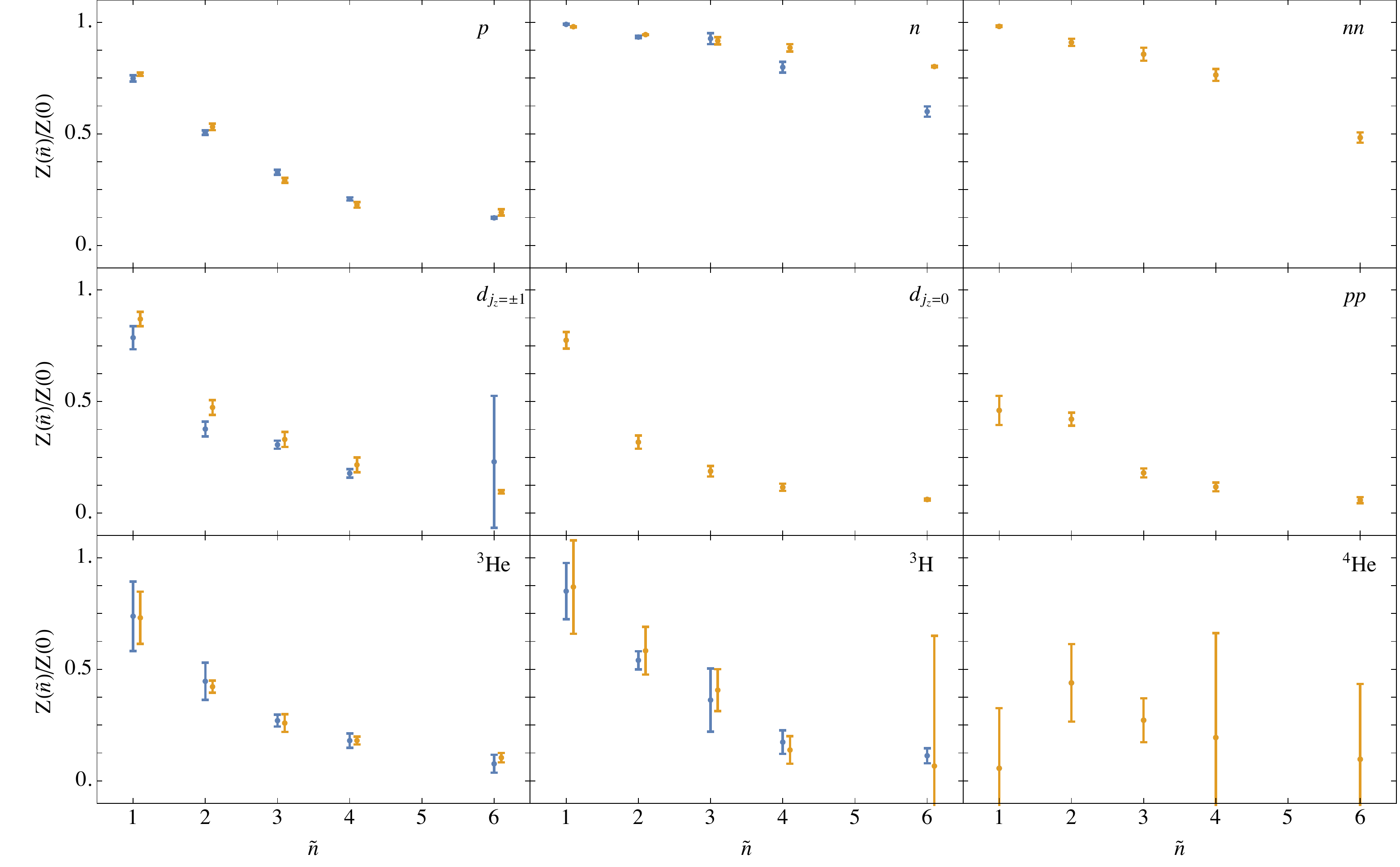}\ \
  \caption{
  The magnetic field strength dependence of the ground state overlap factors, 
  defined in Eq.~(\protect\ref{eq:ratcorr}).
  For $j_z\ne0$  states, both spin states are shown as the different colored points offset slightly for clarity.
   }
  \label{fig:Zfactor}
\end{figure}

\subsection{The Coupled  $j_z=I_z=0$ Two-Nucleon Channel}
\label{sec:Jz0corrs}

The $j_z=I_z=0$ channel is special in that the presence of the
magnetic field breaks  isospin symmetry through the charge matrix
and also introduces explicit spin dependence to the low-energy effective Hamiltonian. 
Consequently, the background magnetic field couples the 
$j_z=0$ deuteron and  $I_z=0$ dinucleon states,
and 
a more complicated analysis is required to extract the essential physics.
The energy eigenvalues of this system result from
diagonalizing a $2\times 2$ truncated Hamiltonian in the basis formed from the
$\si$ and the $\siii$ states.\footnote{
 For the small magnetic field strengths considered in this work, 
 the gap to excitations is significant, and such excitations can be neglected. 
 Additionally, 
 due to the tensor interaction,
 the $j=1$ state involves both $S$-wave and $D$-wave contributions. 
The $\siii$-$\diii$ coupled channels are truncated down to the $\siii$ channel  
because the deuteron and dineutron are close in energy in the
absence of a magnetic field, and the deuteron is predominantly $S$-wave
(at least at the physical pion mass). 
For a more detailed discussion of the deuteron in a FV, see Refs.~\cite{Beane:2010em,Briceno:2013bda}.
} 
For non-vanishing magnetic fields, the off-diagonal elements of this truncated Hamiltonian
receive contributions from magnetic transitions between the $\siii$
and $\si$ channels induced by the nucleon isovector magnetic moment and 
short-distance two-nucleon interactions with the magnetic field resulting from
meson-exchange currents  in the context of potential models 
or local two-nucleon current couplings in effective field theory (EFT).
As the nucleon isovector
magnetic moment is large, the energy splittings between these states are significant for the magnetic field strengths employed.

The pionless EFT (EFT($\pislash$)) can be used to describe the low-energy strong and electroweak 
interactions of two-nucleon, three-nucleon and multi-nucleon systems~\cite{Kaplan:1998sz,Chen:1999tn}.  
It provides a systematic way to include the gauge-invariant electroweak interactions that are not related to strong-interaction  
$S$-matrix elements through local multi-nucleon operators.
While conventionally formulated in terms of four-nucleon interactions with insertions of derivatives (as well as higher body interactions), EFT($\pislash$) can be fruitfully
defined in terms of dibaryon fields, permitting a dramatic simplification in calculations beyond leading order (LO) in the 
expansion~\cite{Beane:2000fi}.
The  Lagrange density describing the two-nucleon
electromagnetic interactions at LO and next-to-leading order (NLO) in EFT($\pislash$) 
using  dibaryon fields is~\cite{Beane:2000fi}
\begin{eqnarray}
{\cal L} &=&
{e\over 2 M_N} N^\dagger \left[ \kappa_0 + \kappa_1 \tau^3 \right] { \mathbf{\Sigma}}\cdot {\bf B} N
\ -\ 
{e\over M_N} \left( \kappa_0 - { \tilde l_2\over r_3}\right) i \epsilon_{ijk} t_i^\dagger t_j B_k
\nonumber \\
 &&\qquad +\ 
 {e\over M_N} {l_1\over\sqrt{r_1 r_3}} \left[ t_j^\dagger s_3 B_j \ +\ {\rm h.c.}\ \right],
\label{eq:traniL}
\end{eqnarray}
where $t_i$ are the SO(3) vector components of the $\siii$ dibaryon field and $s_3$ is the $I_z=0$, $\si$ dibaryon field, $M_N$ is the nucleon mass  and $\mathbf{\Sigma}$ is the spin operator.
The effective ranges in the $\si$ and $\siii$ channels are denoted as $r_1$ and $r_3$, respectively while
$\kappa_0$ and $\kappa_1$ are the isoscalar and isovector magnetic moments of the nucleon.
The NLO interactions, described by dibaryon operators coupled to the magnetic field,
are accompanied by the coefficients $l_{1,2}$ in Eq.~(\ref{eq:traniL}).\footnote{In this expression, $l_2$ has been replaced by 
$\tilde l_2 - r_3 \kappa_0$ to make explicit the deviation of the  deuteron magnetic moment,
$\mu_d =  {e\over M}
 \left( \kappa_0 + {\gamma_0\over 1-\gamma_0 r_3} \tilde l_2\right)$,
from the single nucleon contribution.
}

In Ref.~\cite{Detmold:2004qn}, it was recognized that LQCD calculations employing background magnetic fields could be used to extract the deuteron magnetic moment, 
and the rate for low-energy $np\rightarrow d\gamma$ radiative capture, by determining the energy eigenvalues of the
two-nucleon systems~\cite{Detmold:2004qn,Meyer:2012wk}.  
The deuteron magnetic moment is extracted from the energies of the $j_z=\pm 1$ states in the background fields, while the 
$np\rightarrow d\gamma$ radiative capture cross-section is determined from the nucleon isovector magnetic 
moment and the value of $l_1$ determined from the 
energies of the two $j_z=0$ states in the coupled $\si$--$\siii$ $np$ sector. 
This latter combination is probed
through the determinant condition~\cite{Detmold:2004qn}
\begin{eqnarray}
 \left[ p\cot\delta_1-{S_++S_-\over 2\pi L} \right] 
\left[ p\cot\delta_3-{S_++S_-\over 2\pi L} \right] 
&=&
\left[ {|e  {\bf B}|l_1\over 2} + {S_+-S_-\over 2\pi L} \right]^2
\ \ ,
\label{det_rel}
\end{eqnarray}
where $\delta_{1,3}$ are the phase-shifts in the $\si$ and $\siii$ channels, respectively.
Solutions to this
equation correspond to the energy eigenvalues of the system, 
with the functions $S_\pm$  given by
\begin{equation}
S_\pm\equiv S\left(\frac{L^2}{4\pi^2}(p^2\pm |e {\bf B}| \kappa_1)\right)\,,
\end{equation}
where 
\begin{equation}
S(\eta)=\sum_{\bf n\ne0}^{|{\bf n}|<\Lambda}\frac{1}{|{\bf n}|^2-\eta} -4\pi\Lambda
\end{equation}
is the three-dimensional Riemann-zeta function associated with the $A_1^+$ 
irreducible representation of the cubic group~\cite{Luscher:1986pf,Luscher:1990ux,Beane:2003da}.

At the quark masses used in these calculations, the deuteron and bound dineutron are approximately degenerate~\cite{Beane:2012vq}, 
and have  scattering lengths, $a_{1,3}$, and effective ranges, $r_{1,3}$, 
that are numerically close to each other 
($a_1\sim a_3=a$ and $r_1\sim r_3=r$)~\cite{Beane:2013br} and hence $\delta_1\sim\delta_3=\delta$.\footnote{
The difference in binding energies  is $\Delta_{\siii,\si}=E_{\si}-E_{\siii} = 5.8(1.4)$ MeV \cite{Beane:2012vq}; 
provided the difference in energies is small compared to the shifts induced by the magnetic field, 
it can be neglected. If it cannot be neglected, the determinant condition must be solved numerically.
}
Because of this,  Eq.~(\ref{det_rel}) simplifies to
\begin{eqnarray}
p \cot\delta & = & \frac{1}{\pi L}S_\pm \pm {|e {\bf B}| l_1\over 2} \,,
\end{eqnarray}
where both signs should be taken together for the two solutions.
Expanding this for small $|e {\bf B}|$,  the shifts in the energies of the two eigenstates are  
\begin{eqnarray}
\Delta E_{\siii,\si} & = & 
\mp Z_d^2  \left(\kappa_1 + \gamma_0 l_1 \right) \ {|e {\bf B}|\over M}
\ +\ ...
\ =\ 
\mp \left( \kappa_1 + \overline{L}_1 \right) \ {|e {\bf B}|\over M}
\ +\ ...
\ \ \ ,
\label{eq:npEsplit}
\end{eqnarray}
where  $Z_d = 1/\sqrt{1-\gamma_0 r}$ is the square-root of the residue of the deuteron propagator at the 
pole and the ellipsis denotes terms that are higher order in the strength of the magnetic field.
In Eq.~(\ref{eq:npEsplit}), 
the deviations of the energy shifts from their naive single-particle values are defined using
\begin{eqnarray}
\overline{L}_1 
& = & 
\gamma_0 Z_d^2 
 \left( l_1 + r \kappa_1 \right)
\, .
\label{eq:Lonedef}
\end{eqnarray}

To numerically study this coupled system, it proves useful
to first construct
the correlation matrix
\begin{eqnarray} {\bf C}(t; {\bf B}) & = & \left(
    \begin{array}{cc}
      C_{\siii,\siii}(t; {\bf B}) & C_{\siii,\si}(t; {\bf B}) \\
      C_{\si,\siii}(t; {\bf B}) & C_{\si,\si}(t; {\bf B}) 
    \end{array}
  \right)
  \, ,
  \label{eq:cormat}
\end{eqnarray}
where the matrix elements 
$C_{A,B}(t;{\bf B})$ 
are generated from source and sink operators associated with the
$A,B\in \{\si, \siii\}$ channels (which are orthogonal in the absence of the magnetic field). 
The generalized eigenvalue problem,  defined by this correlation matrix, 
can be solved to extract the (diagonalized) principal correlation functions~\cite{Luscher:1990ck},
energies and energy differences.
That is, solutions of the system
\begin{equation}
[ {\bf C}(t_0; {\bf B}) ]^{-1/2}  {\bf C}(t; {\bf B}) [ {\bf C}(t_0; {\bf B}) ]^{-1/2}  v = \lambda(t; {\bf B}) v
\label{eq:gevp}
\end{equation}
are sought,
where the eigenvalues are the principal correlation functions
$\lambda_\pm (t; {\bf B}) = \exp[-(\bar E\pm \Delta E_{\siii,\si} )t]$ with
average energy $\bar E$ and energy difference $ \Delta E_{\siii,\si} $.
The parameter $t_0$ can be chosen to stabilize the extraction but has little numerical effect in the current results.
To extract the response to a background magnetic field, the ratio of the principle correlation functions
\begin{equation}
R_{\siii,\si}(t;{\bf B}) = {\lambda_+ (t; {\bf B}) \over \lambda_- (t; {\bf B})} \stackrel{t\to\infty}{\longrightarrow} \hat Z \exp\left[2\ \Delta E_{\siii,\si} t  \right]
\,,
\label{eq:R3s11s0}
\end{equation}
permits a refined determination of the energy difference $\Delta E_{\siii,\si} $,
significantly reducing correlated fluctuations,
where $\hat Z$ is a $t$-independent constant.

\begin{figure}[!th]
  \centering
  \includegraphics[width=\columnwidth]{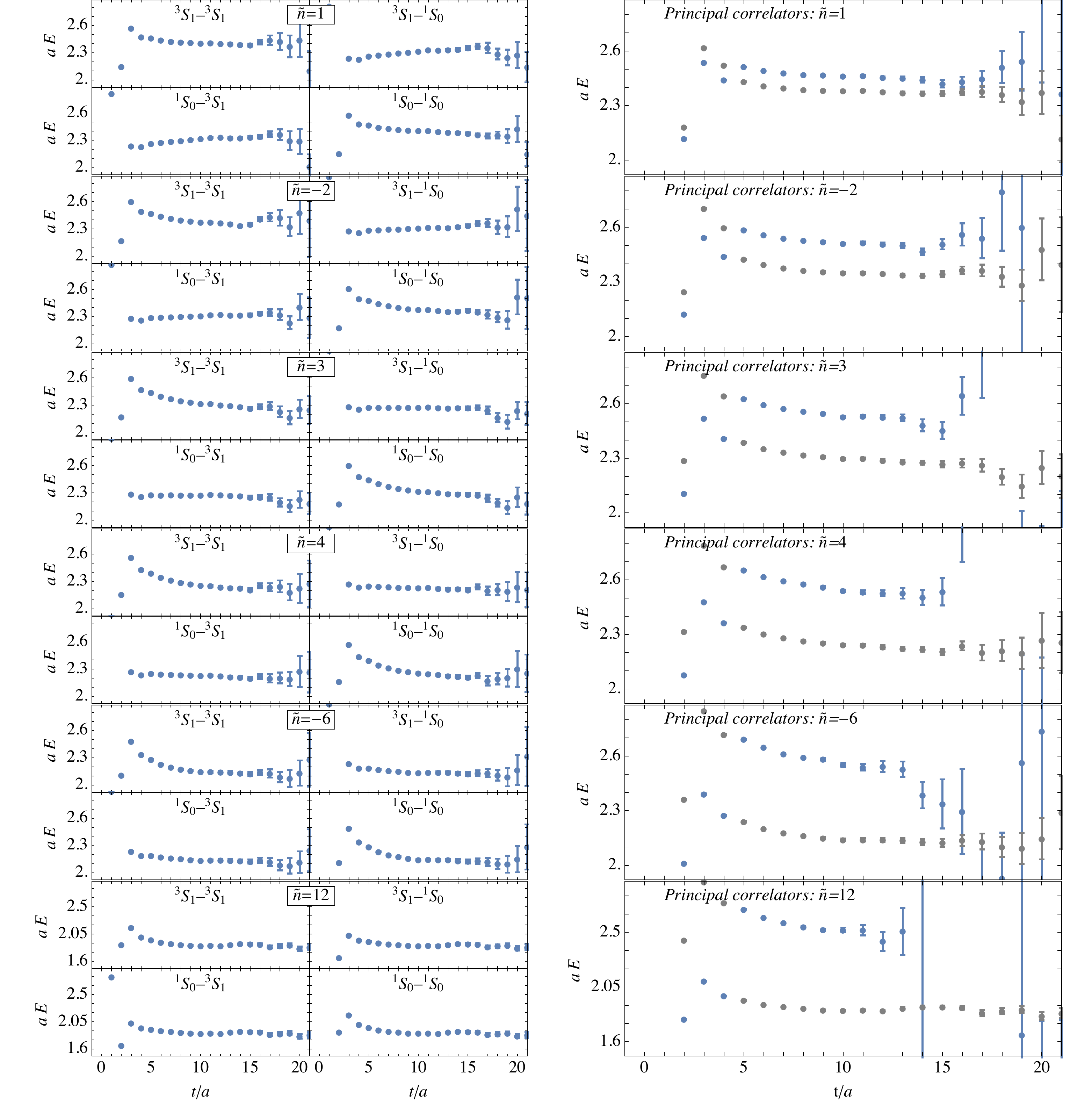} 
\caption{
Results from nucleon-nucleon smeared-smeared correlation functions in the mixed $j_z=I_z=0$ sector.
The left panels show the effective masses of the elements of the $2\times2$ matrix of correlation functions,
with each quartet of plots corresponding to a different magnetic field strength. 
In the right panels, the EMPs of the principal correlation functions resulting from solving the associated generalized 
eigenvalue problem are shown. 
}
  \label{fig:corrJzeq0}
\end{figure}

\begin{figure}[!th]
  \centering
  \includegraphics[width=\columnwidth]{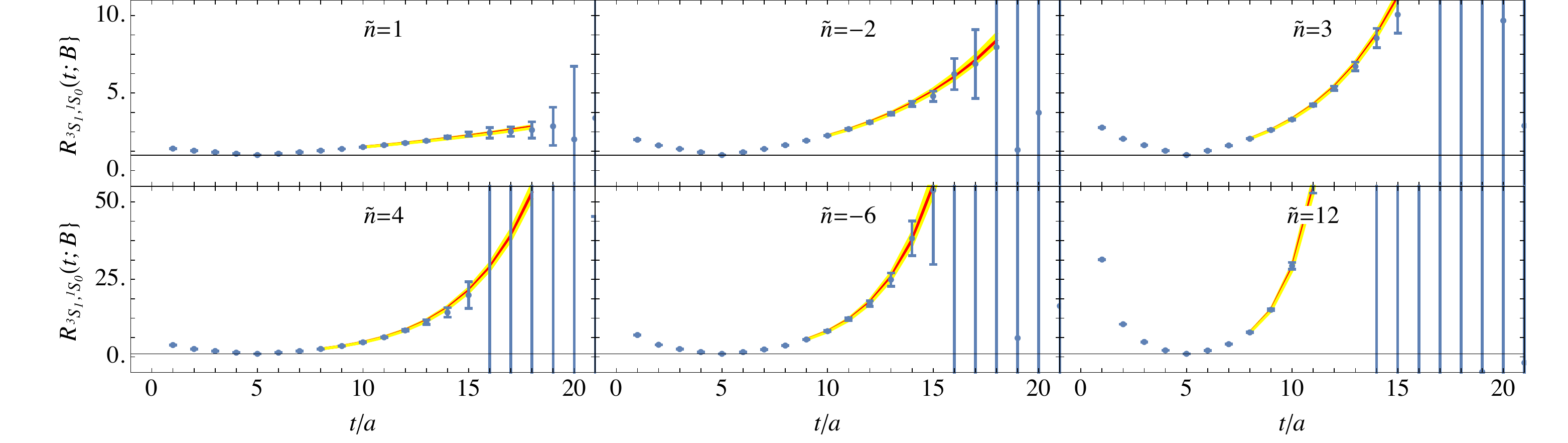} 
\caption{
The ratio of correlation functions, $R_{\siii,\si}(t;{\bf B})$ determined from the principal correlation functions for $t_0=5$ 
for all six magnetic field strengths used in this work. 
Fits to the correlation functions are also shown with  uncertainties represented by the shaded region.
}
  \label{fig:corrratioJzeq0}
\end{figure}

Figure~\ref{fig:corrJzeq0} shows the EMPs of the original correlation functions 
of the coupled-channel system in Eq.~(\ref{eq:cormat})
according to their source and sink type. 
This figure also shows the EMPs constructed from the principal correlation functions 
that are determined by solving the generalized eigenvalue problem, Eq.~(\ref{eq:gevp}), 
for $t_0=5$. 
The diagonalization of the matrix of correlation functions in Eq.~(\ref{eq:cormat})
is particularly effective in this case because the states are orthogonal in the limit of vanishing magnetic field. 
In most cases, plateau behavior is visible in both principal correlation functions, 
indicating that  the lowest two eigenvalues of the system can be extracted. 
Given this, focus is placed on the ratios $R_{\siii,\si}(t;{\bf B})$ in the region where the principal correlation functions 
are consistent with single exponential behavior. 
Figure~\ref{fig:corrratioJzeq0} shows the ratios for all magnetic field strengths along with the associated single exponential fits. 
Analysis of these ratios in the coupled system is performed with the same methods used to analyze the ratios in the unmixed channels.

\begin{figure}[!ht]
  \centering
  \includegraphics[width=\columnwidth]{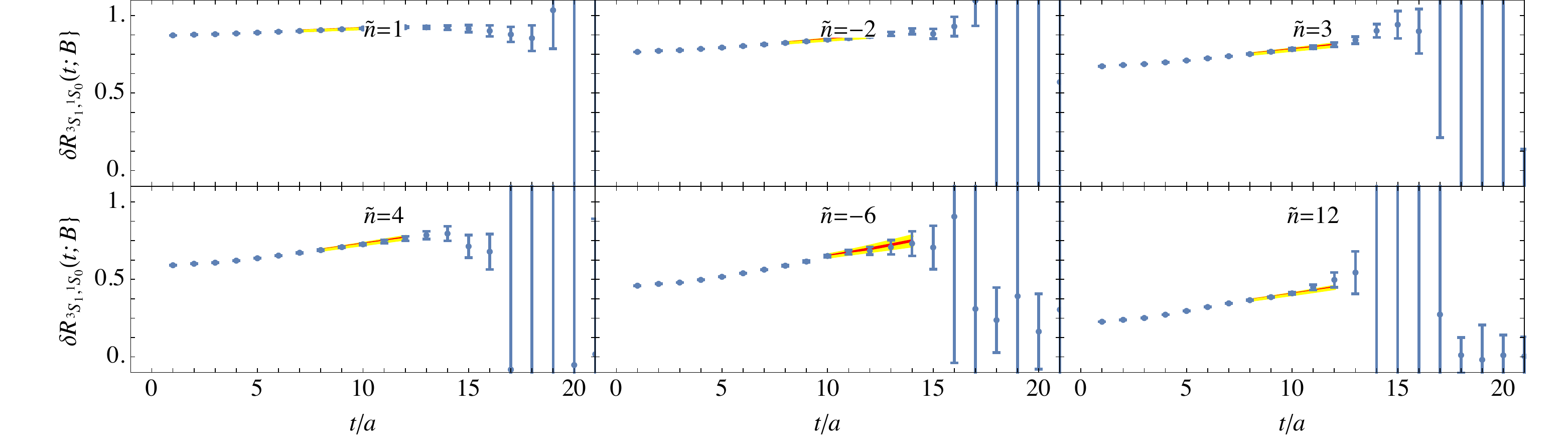} 
\caption{
The ratio $\delta R_{\siii,\si}(t;{\bf B})$ computed from the principal correlation functions with $t_0=5$,   
divided by the appropriate isovector combination of the spin differences of the single nucleon correlation functions are shown
for all six magnetic field strengths used in this work. 
Fits to these ratios, and the associated  uncertainties (the bands), are also  shown.
}
  \label{fig:corrratioratioJzeq0}
\end{figure}

As in Eq.~(\ref{eq:ratcorrAminus}), 
the calculated correlation functions associated with nucleons and nuclei share 
the same quantum fluctuations, to a large degree.   
This makes it possible to determine differences 
between properties of the $np$ system and those of a free neutron and proton with more precision than the individual 
properties. 
In the current context,  the ratio
\begin{equation}
\delta R_{\siii,\si}(t;{\bf B}) = \frac{R_{\siii,\si}(t;{\bf B})}{\Delta R_p(t;{\bf B})/\Delta R_n(t;{\bf B})}\,,
\label{eq:deltaR3s11s0}
\end{equation}
decays with a characteristic exponent 
$2\Delta E_{\siii,\si}({\bf B})- (E_{p,\uparrow} - E_{p,\downarrow}) +(E_{n,\uparrow} - E_{n,\downarrow}) =  2 |e {\bf B}| \overline L_1 /M +{\cal O}(|{\bf B}|^3)$, 
permitting direct access to  deviations from single nucleon physics,
where the 
$\Delta R_h(t;{\bf B})$ are given in Eq.~(\ref{eq:ratratcorr}).
Figure~\ref{fig:corrratioratioJzeq0} shows these ratios for each field strength,
from which the energy shifts can be extracted with remarkable  precision.

\subsection{Magnetic Field Strength Dependence: General Strategies}
\label{sec:Bfits}

Having extracted the energies and energy differences 
as a function of the magnetic field strength, the
remaining task is to use them to determine the magnetic
properties of the nucleons and nuclei  through fits to the expected forms shown in 
Eq.~(\ref{eq:Eshift}). 
The fits and  extracted properties of each nucleon and nucleus are presented
individually in the following subsection.
Here, the general features of the analysis,
and the difficulties  encountered in confronting Landau levels, are discussed.

In dimensionless units, the form used for the fits  to the ground states (${\bf B}= B {\bf e_z}$ and $P_\parallel=0$) is 
\begin{eqnarray}
a\ \delta E_{h;j_z} 
&=&  \sqrt{a^2 M_h^2 + (2n_L+1)Q_h a^2|e\ {\bf B}|} - a M_h 
- \frac{2 e}{ a M_N} \hat \mu_h j_z a^2 |e\ {\bf B}| 
\nonumber \\
&& 
\qquad 
- \frac{2\pi}{a^3 M_N^2(M_\Delta-M_N)} \left[\hat \beta_h 
+ \hat \beta_h^{(2)}  ( j_z^2 - {1\over 3} j(j+1) )  \right]\left(a^2 |e\ {\bf B}|\right)^2
\nonumber \\
&& 
\qquad +j_z \hat \gamma_h \left(a^2 |e\ {\bf B}|\right)^3
+ \hat \delta_h \left(a^2 |e\ {\bf B}|\right)^4\,,
\label{eq:Bdependlatt}
\end{eqnarray}
where the  fit parameters are 
\begin{eqnarray}
&& 
n_L, \qquad 
\hat \mu_h = \mu_h {M_N\over 2e}, \ \ 
\hat\gamma_h  ,\ \ 
\hat\delta_h  \,
\nonumber\\
&&
\hat\beta_h = \frac{M_N^2(M_\Delta-M_N)}{e^2} \beta_h^{(M0)}  ,\ \ 
\hat\beta_h^{(2)} =  \frac{M_N^2(M_\Delta-M_N)}{e^2} \beta_h^{(M2)} \,,
\label{eq:imlessvardef}
\end{eqnarray}
and $a^2|e\ {\bf B}|=\frac{6\pi | \tilde n|}{L^2}$ is the dimensionless field strength.
This extends to higher orders in $|e\ {\bf B}|$ than the form given in Eq.~(\ref{eq:Eshift}), 
providing for estimates of fitting systematic uncertainties in the extraction of the magnetic moments and polarizabilities resulting from
the  choice of fit form. 
The hadron masses, $a M_h$, 
 are taken from our previous studies on this ensemble of gauge-field configurations~\cite{Beane:2012vq} and are known precisely ($a M_\Delta= 1.3321(10)(19)$ on this ensemble). 
For uncharged states, $n_L$ does not enter the fit, 
and for states with $j=0$, the parameters $\hat\mu$, $\hat\beta_h^{(2)}$ and $\hat\gamma$ are absent. 
As in Ref.~\cite{Beane:2014ora}, the extracted magnetic moments are expressed in terms of ``natural nuclear magnetons''  (nNM)
defined with respect to the nucleon mass at the quark masses used in the calculation. 
The  polarizabilities are given in terms of the natural dimensionless polarizability $e^2/M_N^2(M_\Delta - M_N)$
(given the expected dominance of the $\Delta$-resonance, this is the appropriate scale for the magnetic polarizabilities),
but are also presented in physical units in the conclusion.
A physical interpretation of the higher order parameters is  not provided,
and they are used only to control the systematic 
uncertainties in the magnetic moments and polarizabilities
extracted from the fit.

In performing fits, the same bootstrap sets of extractions of the energy shifts are used at each magnetic field strength in order
to exploit the correlations between them. 
The ensemble averages of the energy shifts are used to obtain the central values of fit parameters describing 
the magnetic field strength dependence.
An ensemble of fits to the bootstrap data set is
used to obtain the associated statistical uncertainties. 
To propagate the
systematic uncertainties from the fits to the ratios of correlation functions into the field dependence analysis, 
the bootstrap sets of energy 
shifts are spread away from their mean by the ratio of the quadrature-combined statistical and systematic uncertainties to
the statistical uncertainty. That is, for an energy variable $E$ with an ensemble of extracted bootstrap values $\{E_i\}$, mean value $\overline{E}=\frac{1}{N_B} \sum\limits_{b=1}^{N_B} 
E_b$, and statistical and fitting systematic uncertainties $\delta E_{\rm stat}$ and $\delta E_{\rm sys}$, the spreaded bootstrap 
ensemble values $\{\widetilde E_i\}$ are taken to be 
\begin{equation}
\widetilde E_i= \overline{E} + \frac{\sqrt{\delta E_{\rm stat}^2+\delta E_{\rm sys}^2}}{\delta E_{\rm stat}} (E_i-\overline{E})\,,
\end{equation}
and it is these quantities that are used in subsequent analyses.
The highly correlated nature of the results obtained 
at different field strengths 
(that is, results obtained at different magnetic field strengths on a given configuration have correlated statistical fluctuations)
makes this approach essential.

An important aspect of the present analysis is to address 
the range of magnetic field strengths for which the fit forms in Eq.~(\ref{eq:Bdependlatt}) 
describe the energy shifts. 
This is addressed by varying the number of magnetic field strengths used in the fits, 
fitting results obtained at  
$|\tilde n|\le \tilde n_{\rm max}$ with $\tilde n_{\rm max}=2,3,4,6,12$. 
The complexity of the fit form is also varied 
by either using the full form or by setting $\hat\delta=0$, 
$\hat\delta=\hat\gamma=0$ or $\hat\delta=\hat\gamma=\hat\beta=0$.
Further, either the smeared-sink or point-sink correlation functions are selected for  fitting. 
For $j=0$ states, 
a total of $4\times2\times 2=16$ different fits are considered for each bootstrap ensemble.
For the $j>0$ states,  coupled fits were performed to the magnetic field strength dependence of the 
$j_z=\pm j$  spin states and  $5\times 4\times2=40$ different fits are considered for each bootstrap ensemble. 
A large number of fits successfully described the results with an acceptable $\chi^2/d.o.f.$, although some did not.
The central values and uncertainties in the extracted parameters are evaluated from the distribution of the results for the acceptable fits, and are taken as the 50$^{\rm th}$ quantile and the  17$^{\rm th}$--83$^{\rm rd}$ range of quantiles, respectively. As an additional check, we have used the Bayes information criterion for a given fit to assess those that are acceptable and find uncertainties that are consistent with those defined from the $\chi^2$. 

A further complication arises from the  Landau-level nature of the eigenstates and the sub-optimal
projection of the interpolating operators onto them. 
In every case, correlation functions that have single exponential behavior 
over a significant range of time-slices  are found. However, it is clear that these states do not correspond to
the lowest Landau level. 
Expanding the field dependence in the non-relativistic limit, 
the magnetic moment contribution cancels in the spin-averaged energy shift, 
but a linear contribution survives with a coefficient determined by $n_L$,
\begin{equation}
a \left[ E_{h;j_z}({\bf B}) +  E_{h;-j_z}({\bf B})\right] = {a^2 Q_h | e\ {\bf B}| \over 2 a M_h} (2n_L+1)+ {\cal O}(|{\bf B}|^2)\,,
\end{equation}
where the masses are precisely known from previous studies \cite{Beane:2012vq}.
From examining the small field shifts for charged states, it is found that $n_L\ne0$ in all cases. 
Thus, the interpolating operators overlap strongly onto  excited states of the system and presumably  will 
relax to the ground state only at  large Euclidean times. 
Because of this, $n_L$ is treated as a fit parameter
and the fits themselves are used to determine which Landau levels  are dominating the various correlation functions. 
In the limit of vanishing lattice spacing, and neglecting structure effects, the allowed values of $n_L$ are 
integers and a somewhat complicated approach to fitting is required.
Two alternate procedures are considered.
In the first approach, the lowest magnetic field strengths are used to determine the linear term in the field-strength dependence, 
which is used to identify the integer value of $n_L$ that is most consistent with the numerical results.
This value is held fixed and then used in further fits utilizing the full form of Eq.~(\ref{eq:Bdependlatt}). 
In the second approach, $n_L$ is first treated as a real-valued fit parameter and the full fits are performed. 
Then, after considering  the different fits, the integer $\hat n_L$ closest to the mean 
of  successful fits is chosen and held fixed in the final set of fits.\footnote{
The extractions of magnetic moments and the $np \to d \gamma$ cross-section 
are independent of these complications as the Landau-level contributions cancel in the energy differences between spin states.
}
 An additional systematic is 
assessed by combining sets of fits (varying fit forms, data ranges, and types of correlation functions) 
with $\hat n_L\to \hat n_L\pm 1$ into the full suite of fits 
(for charged, $j>0$ states, a total of 120 different fits are considered for each bootstrap ensemble).
Both choices of  Landau-level procedures lead to consistent results after these systematic uncertainties are taken into account.

A related systematic uncertainty that is considered
is the potential ambiguity in identifying the Landau level associated with the plateau at each magnetic field strength.
In the fitting forms, it is implicit that the energies of a nucleon or nucleus  result from 
a single Landau level for the range of magnetic fields that are considered.  
This is expected to be the case at small magnetic fields, but might not be valid at larger magnetic fields.
To explore this issue, fit forms with different $\hat n_L$ for different magnetic fields are considered.
Keeping only the results from the lowest four  magnetic fields, and allowing different values of $\hat n_L$  
for each magnetic field does not result in acceptable fits except when the $\hat n_L$'s are all the same.
This leads to confidence in the assumption that the same Landau state is providing the energies that are dominating the fits.


\subsection{The Magnetic Properties of Nucleons and Nuclei}

\subsubsection{The Neutron}
\label{sec:n}

\begin{figure}[!t]
  \centering
  \includegraphics[width=0.7\columnwidth]{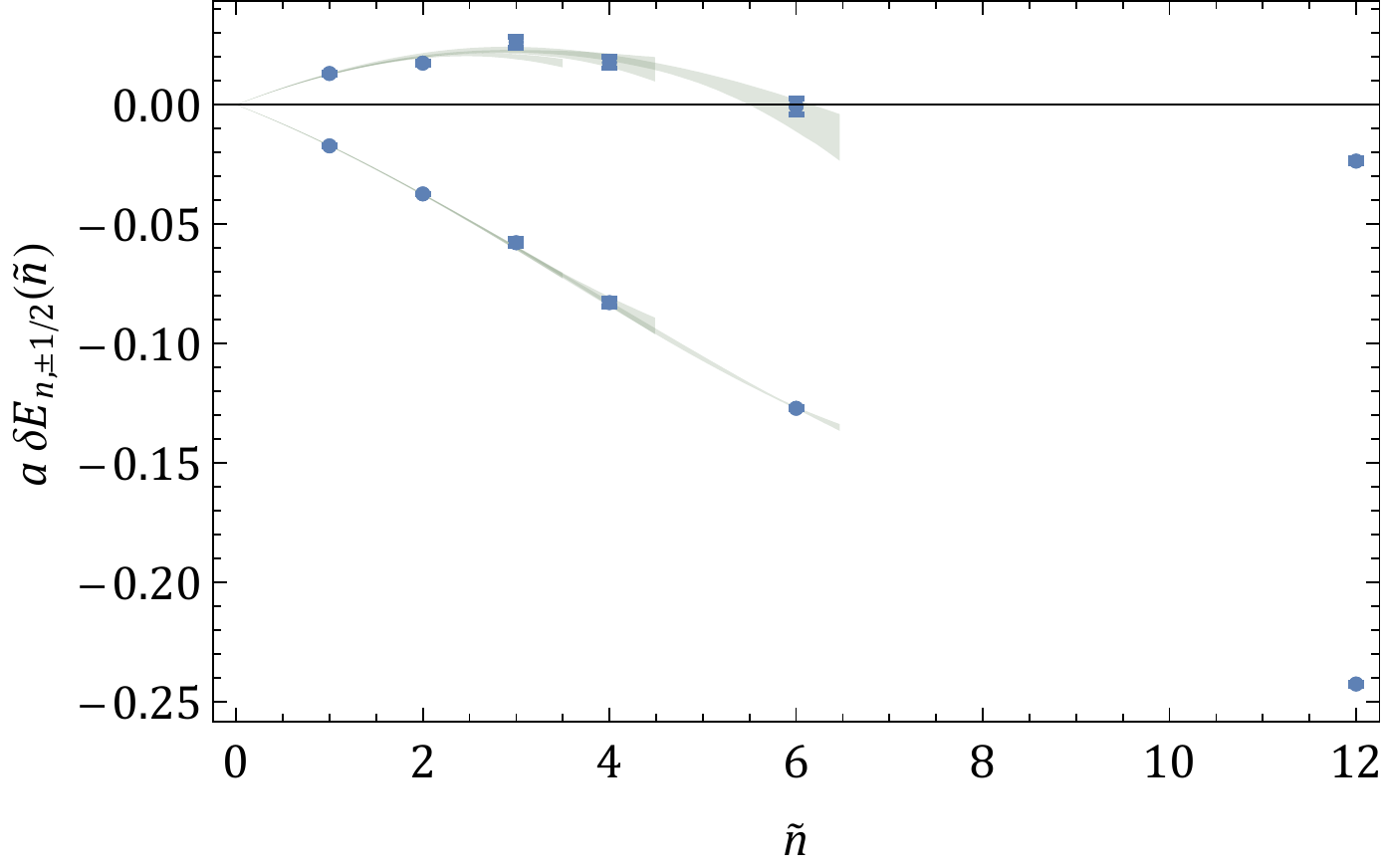}\
  \includegraphics[width=0.6\columnwidth]{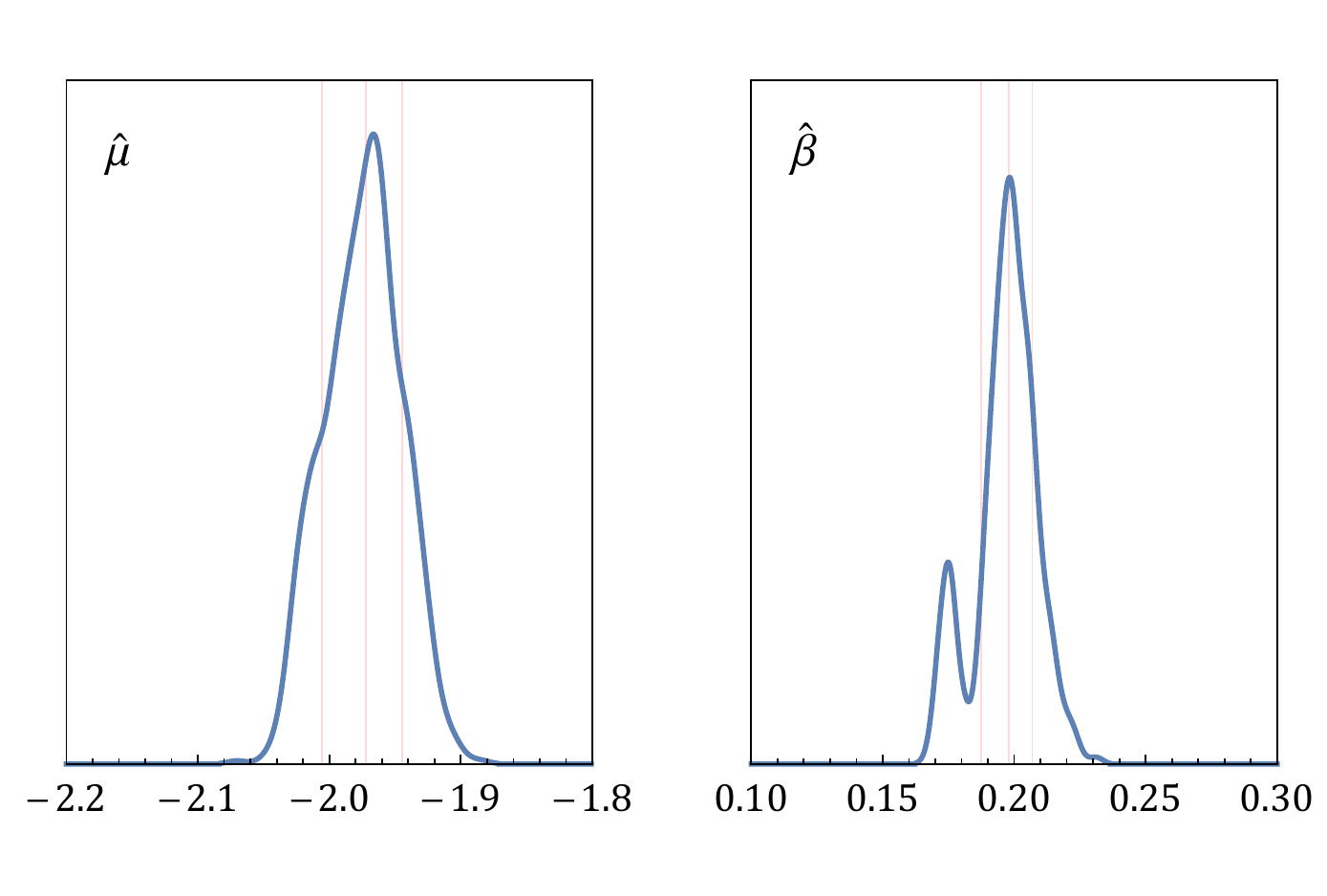}\
  \
  \caption{ 
  Results for the energy shifts of a spin-up (upper points) and a
    spin-down (lower points) neutron in a background magnetic field.
    The central 68$^{\rm th}$ quantile of successful fits is shown as the shaded bands. 
    Different overlapping bands are shown for fits over the different  ranges of $\tilde n$. 
    The lower panel shows the probability-density functions for the relevant fit parameters 
    $\hat \mu$ and $\hat \beta$, with the  vertical lines indicating the central value and uncertainties. 
  }
  \label{fig:dEneutspinstates}
\end{figure}

The neutron correlation functions  and their ratios 
for each spin component and magnetic field strength used in this analysis, 
along with the associated fits, are shown in Figs.~\ref{fig:corrs} and \ref{fig:corrratio1},
and the energy shifts extracted from these functions are presented  in Table~\ref{tab:energies}. 
Figure~\ref{fig:dEneutspinstates} 
shows the energy shifts of a spin-up and spin-down neutron as a function of the background magnetic field strength.
The two spin states behave quite differently in the presence of the magnetic field.  The energy shift of the spin-down
state (negatively shifted as the neutron magnetic moment is negative) 
responds almost linearly to the magnetic field, 
even out to $| e {\bf B}|\sim 0.71~{\rm GeV}^2$ ($\tilde n=12$),
while the response of the spin-up state exhibits significant
nonlinearities even for modest magnetic fields. 
The behavior of
the spin-up state is reminiscent of the lower level in a two-state system with
an avoided level crossing.  
Given the expected tower of   QCD
excitations of the nucleon, the observed behavior of the spin-up state
is consistent with the magnetic field inducing mixing between the
spin-up neutron and higher-lying states.  
Such mixing is expected from quark-hadron duality, 
and LQCD calculations that also probe the response of excited states to 
the magnetic field could be used to investigate this further. 
This behavior implies that spin-dependent polarizabilities are highly correlated 
with spin-independent polarizabilities, 
and it will be interesting to learn if this pattern persists as the 
quark masses are brought closer to their physical values.

As discussed in the previous subsection, a large number of 
different fits involving varying  ranges of field strengths and with a variety of functional forms are performed 
in order
to the analyze the energy shifts and determine
the magnetic moment and  polarizability.
The 17$^{\rm th}$ and 83$^{\rm rd}$ quantile range of all successful fits to each data range are shown as the shaded regions in Fig.~\ref{fig:dEneutspinstates}. Separate (overlapping) regions are shown for each data range  $\tilde n\leq n_{\rm max}$ for which successful fits were found. Note that the figure shows the results from smeared-smeared correlation functions only, 
but the fits that are considered also include those involving the energies extracted from smeared-point correlators.
The figure  also shows the probability density functions (PDFs) generated from combining the central values of all successful fits (considering fits over the energy shifts extracted from each of the 
bootstrap ensembles) for the two relevant parameters, $\hat{\mu}$ and $\hat{\beta}$. For fits involving additional parameters, these are integrated over, while for linear fits just involving the magnetic moment, these are ignored in determination of the PDF for $\hat\beta$.
Analysis of the suite of fits that are detailed in the previous subsection 
yields a neutron magnetic moment and polarizability of
\begin{eqnarray}
\hat\mu_n & = & \mun \,, \\
  \hat\beta_n   & = &   \betan \,,
  \label{eq:neutPOL}
\end{eqnarray}
where the first uncertainties combine the statistical  and  systematic uncertainties from the extraction of the energy shifts, as well as the
systematic uncertainty from the fit to the magnetic field strength. 
The second uncertainty estimates the effects of discretization and finite volume effects; as discussed in the conclusion this is assessed to be a 3\% multiplicative uncertainty on magnetic moments and a 5\% multiplicative uncertainty on polarizabilities. 
The above results are presented in the natural dimensionless units, and 
the values of the magnetic moment and 
polarizability in physical units are subject to  additional 
uncertainties from the lattice scale-setting procedure, which 
are  discussed in the conclusion.  
The  magnetic polarizability and  magnetic moment of the neutron have been calculated
previously with LQCD over a range of light-quark masses~\cite{Lee:2005dq,Primer:2013pva}
albeit with large uncertainties. 
The calculated magnetic moment is consistent with previous calculations at similar quark masses, 
and the value of $\beta_n$ is also consistent with previous calculations~\cite{Primer:2013pva}.\footnote{ 
The authors of Ref.~\cite{Primer:2013pva} report difficulties in identifying ground states.
}

\subsubsection{The dineutron}
\label{sec:nn}

At these unphysical quark masses, the dineutron 
(in the $\si$ channel)
is a bound state, with a
binding energy of $B_{nn} = 16(5)~{\rm MeV}$~\cite{Beane:2012vq}.  
As it is electrically neutral, comprised
of two neutrons in the $^1S_0$ channel with positive parity, the dineutron provides the simplest
nuclear system with which to explore the effects of binding on magnetic properties.
This system is discussed before proceeding to states that are electrically charged 
and therefore complicated by the presence of Landau levels.
\begin{figure}[!t]
  \centering
  \includegraphics[width=0.7\columnwidth]{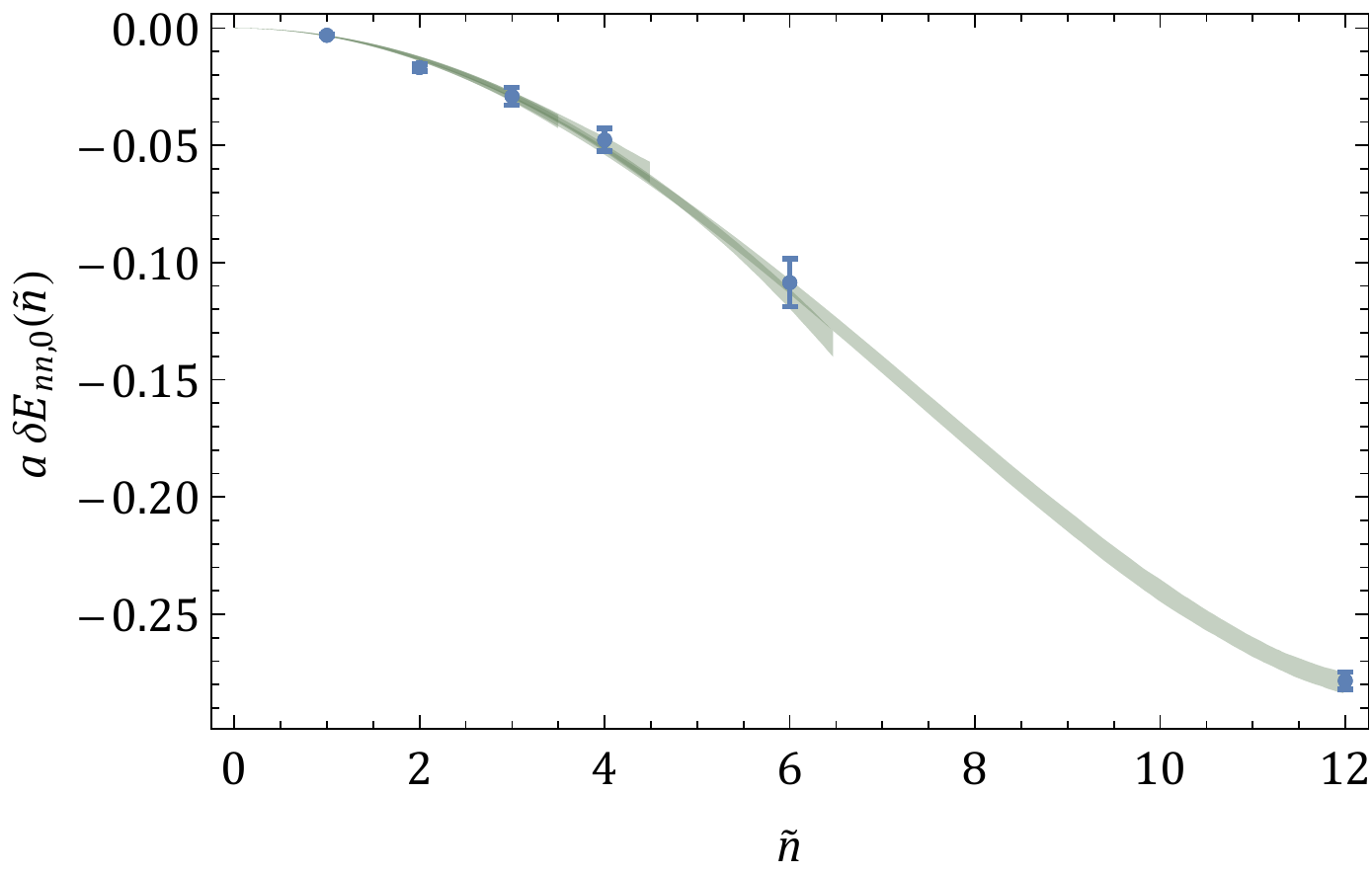}\
  \includegraphics[width=0.6\columnwidth]{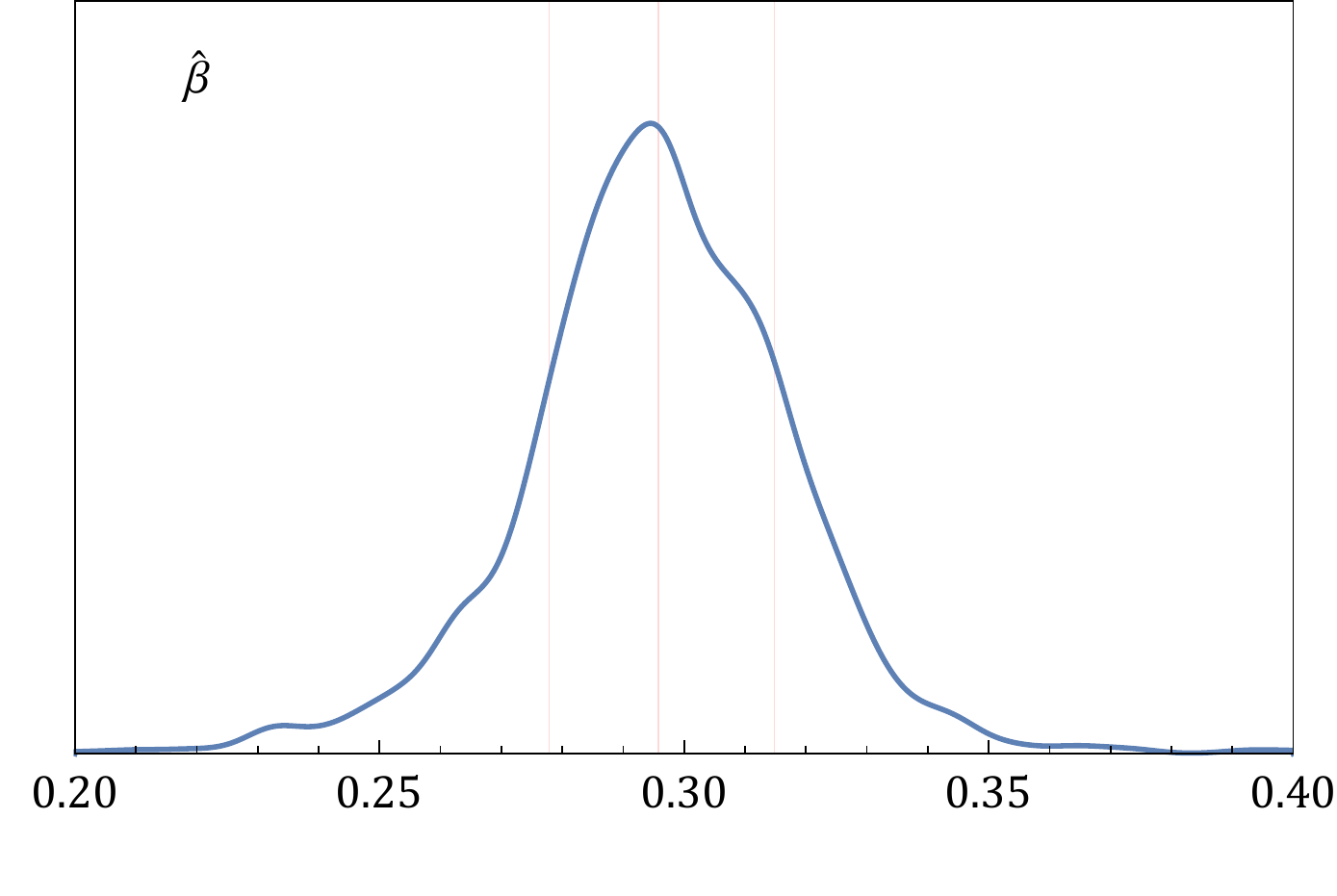}\
   \caption{ 
     The energy shifts of the dineutron as a function of the
    background magnetic field strength, $\tilde n$.   
The details of the figure are as in Fig.~\ref{fig:dEneutspinstates}.
The lower panel shows the PDF for the dineutron polarizability.
  }
  \label{fig:dEdineut}
\end{figure}

Figures~\ref{fig:corrs} and \ref{fig:corrratio1} show the dineutron correlation functions and the 
ratios of correlation functions for each field strength, along with  fits to the time dependence of the ratios. 
The energy shifts extracted from the  ratios  of correlation functions are given 
in Table~\ref{tab:energies} for each field strength, 
and Figure~\ref{fig:dEdineut} shows these shifts.
Combining all of the attempted fits to the energy shifts, as described in detail for the neutron,
yields a magnetic polarizability of
\begin{eqnarray}
  \hat\beta_{nn}   & = &   \betann  \,,
  \label{eq:dineutPOL}
\end{eqnarray}
where the uncertainties are as for the case of the neutron, 
and the result is presented in the dimensionless natural units of 
the system, defined in Eqs.~(\ref{eq:Bdependlatt}) and (\ref{eq:imlessvardef}).

\begin{figure}[!t]
  \centering
  \includegraphics[width=0.7\columnwidth]{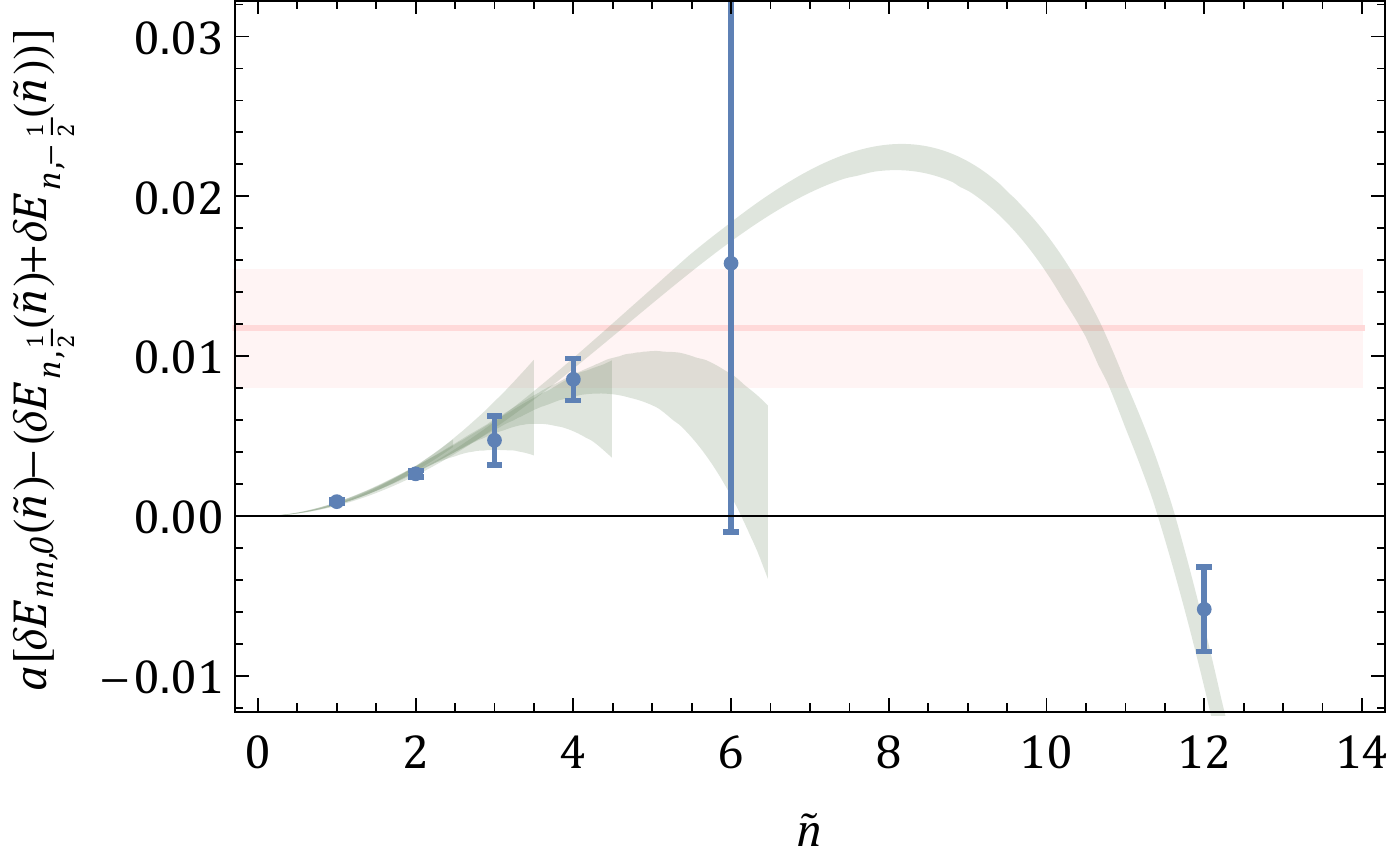}\\ \vspace*{5mm}
  \includegraphics[width=0.7\columnwidth]{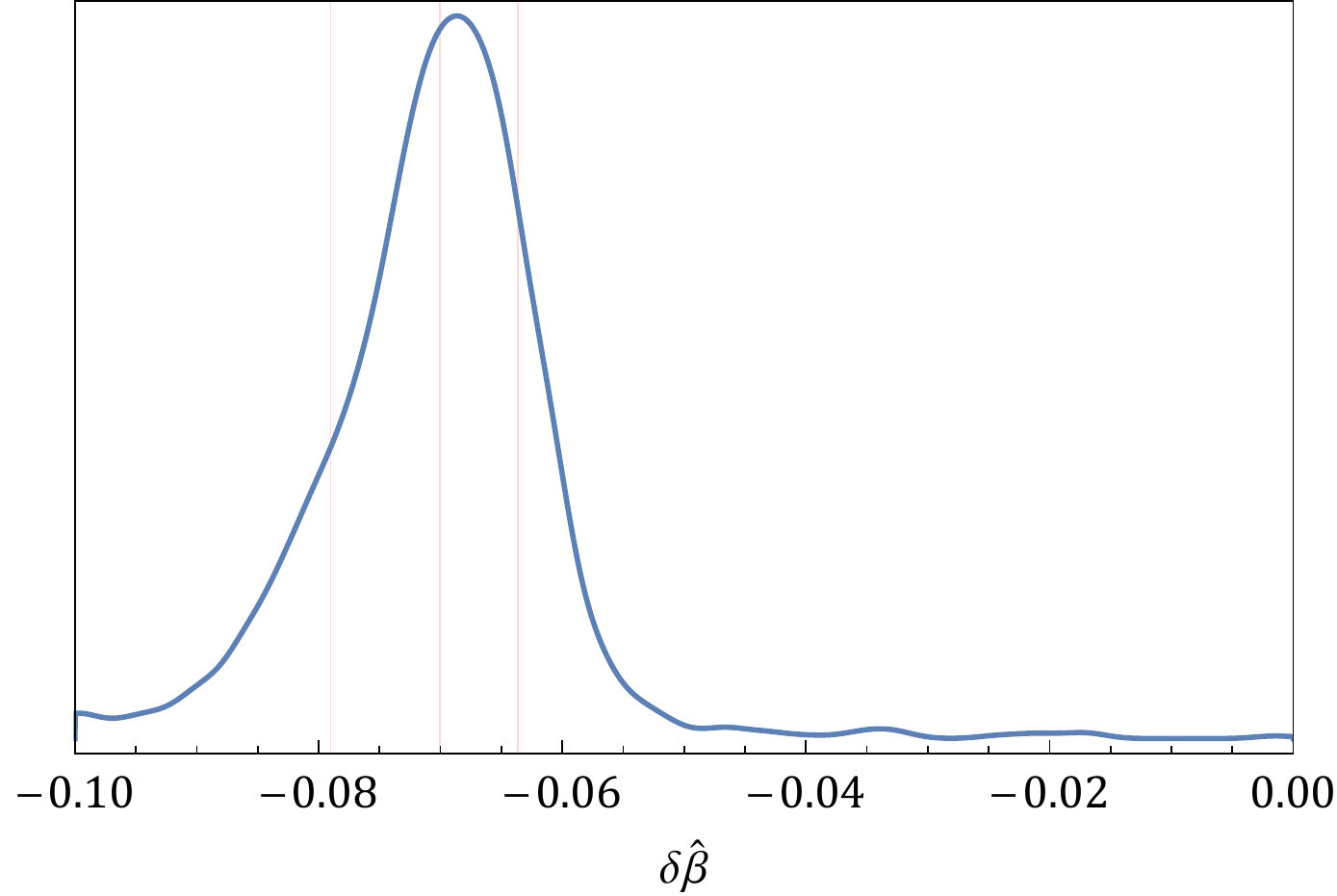}\
  \
  \caption{   Results for the difference between the energy shifts of the dineutron
    and a spin-up and spin-down neutron as a function of the
    background magnetic field strength.  
    The green shaded regions are the result of the suite of fits to the field strength dependence. 
    The horizontal red-shaded region shows the breakup threshold for the
    dineutron, above which the ground state of the system would be two
    neutrons in the continuum in the $\si$ channel.  }
  \label{fig:dEneutMnn}
\end{figure}

This polarizability is significantly smaller than twice the single neutron
polarizability with 
$  \delta \hat\beta_{nn} \ \equiv\ \hat\beta_{nn} - 2\hat\beta_{n}  \sim  \dbetann $. This difference can also be obtained from the ratio $\delta R_{nn,0}(t;{\bf B})$
in Eq.~(\ref{eq:ratcorrAminus}) that
probes the difference directly in a correlated manner. 
In the large time limit,
the exponential decay of this ratio is governed by the energy difference
$\delta E_{nn}({\bf B}) - \delta E_{n,\frac{1}{2}}({\bf B})- \delta E_{n,-\frac{1}{2}}({\bf B})$.
These ratios are displayed in Fig.~\ref{fig:deltaNNcorr} for the different field strengths
and the extracted energy shifts   are shown in Fig.~\ref{fig:dEneutMnn} as a function of the field strength.
In turn, $\delta \hat\beta_{nn}$ is the coefficient of the quadratic term in the field strength 
dependence of this energy difference. 
Analyzing these energy shifts using the same methods as above, leads to
\begin{eqnarray}
  \delta \hat\beta_{nn} \ \equiv\ \hat\beta_{nn} - 2\hat\beta_{n} & = &  \dbetannCORR \ .
\label{eq:dineutPOLDIFF}
\end{eqnarray} 

As discussed above, the neutron spin-down state is  magnetically rigid and remains undeformed
even at  large magnetic fields, 
while the spin-up state is  strongly deformed.  
For the dineutron, the overall energy is lowered in a
magnetic field, driven largely by the spin-down 
neutron.  As this also lowers the energy of the spin-up neutron, it
has a reduced mixing with other states and, therefore, becomes more
rigid. From Fig.~\ref{fig:dEneutMnn}, it is apparent that
the binding energy of the dineutron 
(the energy required to separate the spin-up and spin-down neutron) 
is reduced for $\tilde n\lsim 8$, 
but is larger for $\tilde n=12$, 
than at ${\bf B}=0$.  
If this behavior persists at the physical quark masses, 
it would indicate that
it is energetically disfavored for neutron matter (or near-neutron matter) in dense stellar objects 
to spontaneously generate a magnetic field through the formation of dineutron pairs. 
It is also interesting to note that at intermediate 
field strengths, the dineutron system is nearing a Feshbach resonance in which the binding energy is approaching zero and the scattering length is approaching infinity.

\subsubsection{The proton}
\label{sec:p}

The analysis of the proton in a magnetic field is
more complicated than that of the neutron and dineutron.  
 As discussed previously,  the interpolating operators used in this work project 
onto plane waves in all three spatial directions
rather than Landau levels which we expect to be closer to the eigenstates of the system, so the quality of the correlation functions for
charged systems is expected to be significantly worse than that for
electrically neutral systems.
This is indeed the case, as can be seen from the 
EMPs shown in the first row
of Fig.~\ref{fig:corrs};
in comparison  to the neutron, the proton correlation functions are of lower quality with plateaus setting in 
at later times and with significantly larger uncertainties. 
Further, the presence of 
Landau levels significantly complicates the spectrum of charged system and it 
is  clear that 
the plateaus that are evident do not correspond to the lowest Landau level, as discussed above.
The Landau level associated with the plateau is identified
through systematic analysis of the field-strength dependence, 
as discussed in Section \ref{sec:Bfits}.

\begin{figure}[!t]
  \centering
  \includegraphics[width=0.7\columnwidth]{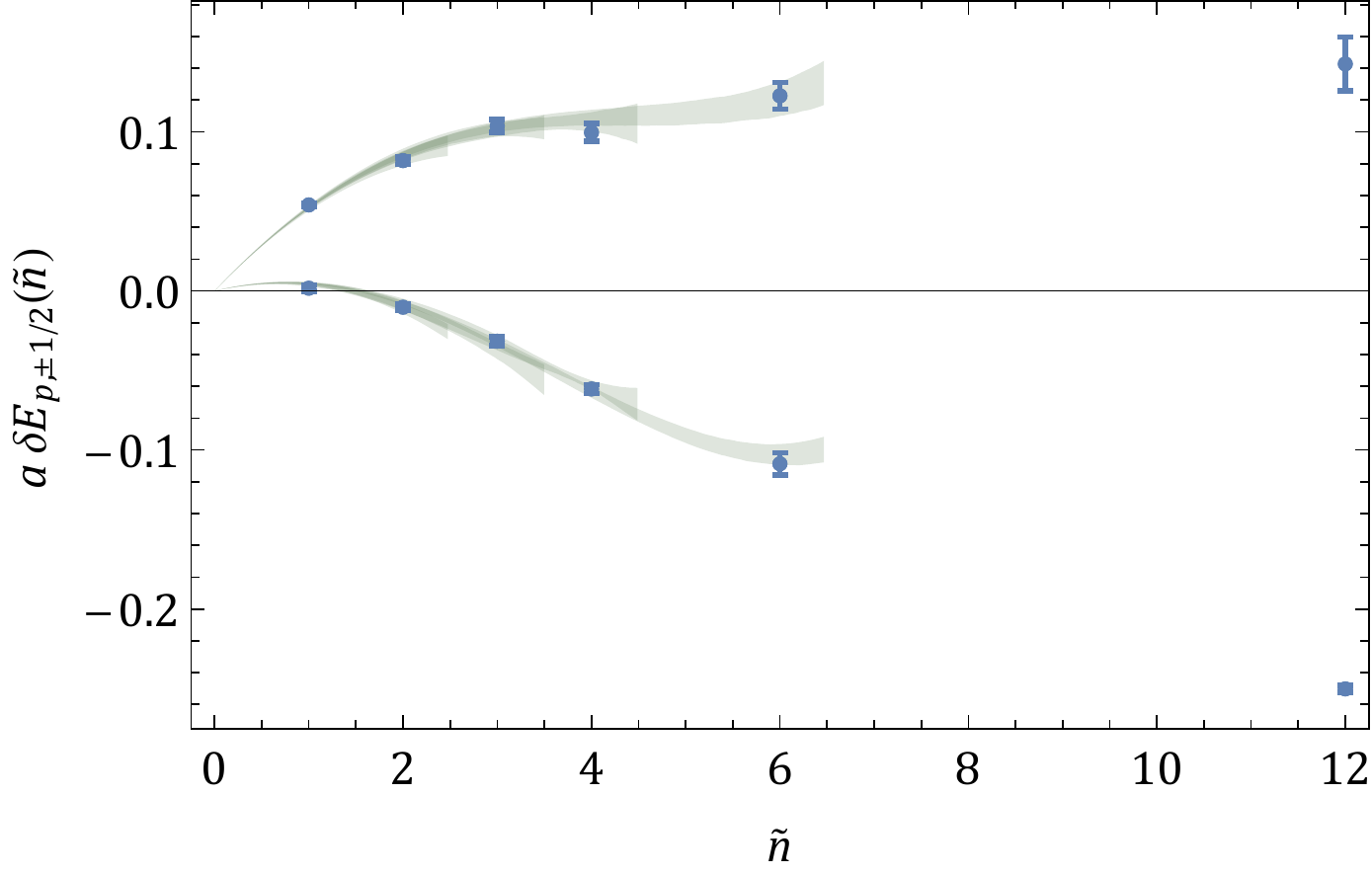}\
  \includegraphics[width=0.6\columnwidth]{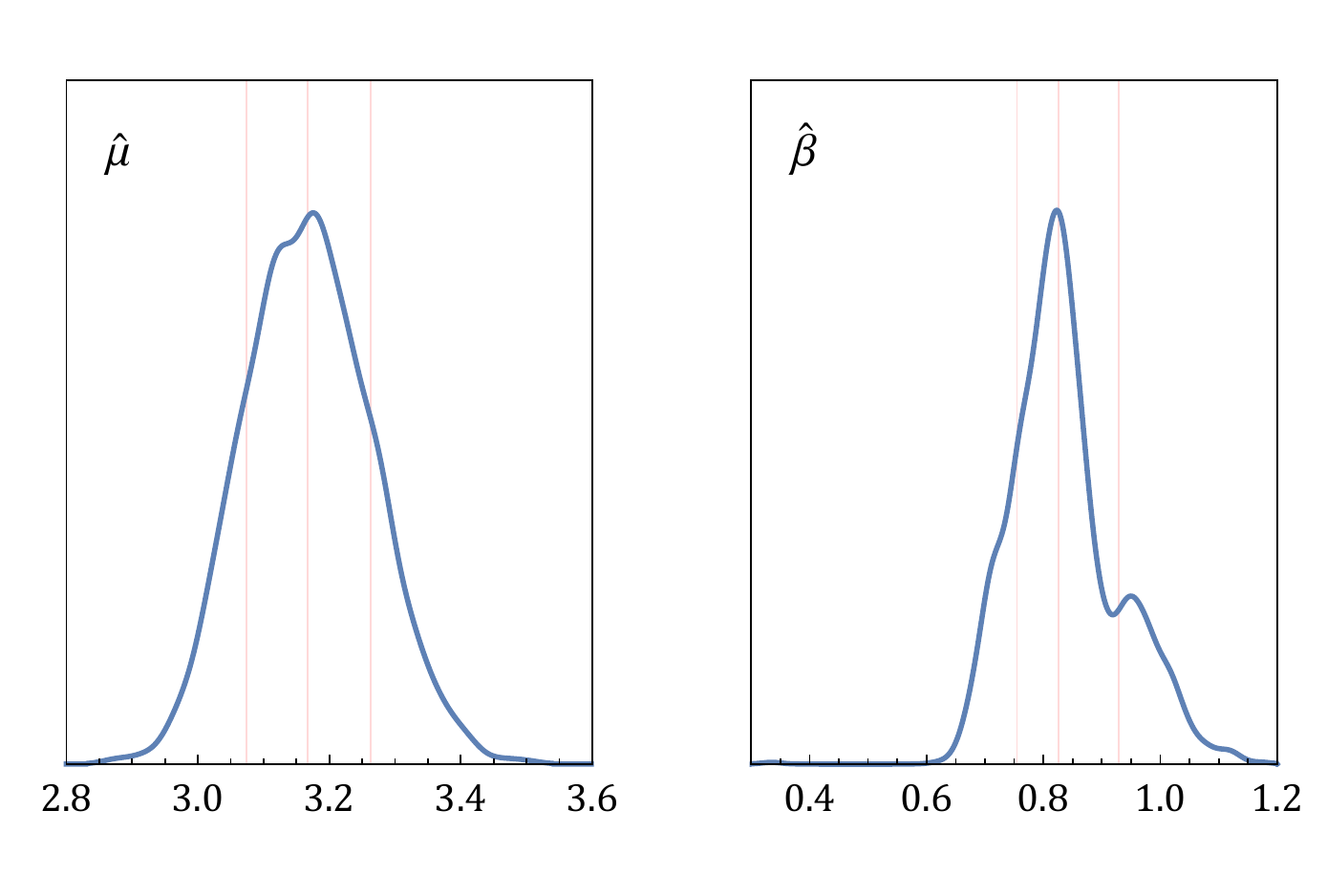}\
  \
  \caption{ 
    Results for the  energy shifts of a spin-up (lower points) and a
    spin-down (upper points) proton in a uniform background magnetic field. 
   The details of the figure are as in Fig.~\ref{fig:dEneutspinstates}. 
   The lower panel shows the PDFs for the fit parameters $\hat{\mu}$ and $\hat\beta
$.  }
  \label{fig:dEprotspinstates}
\end{figure}

In Fig.~\ref{fig:corrratio1}, the  ratio of correlation functions associated with each spin component and field strength,
and  associated fits, are shown.
The energy shifts resulting from these fits are given in Table~\ref{tab:energies} for each 
magnetic field strength and are shown in Fig.~\ref{fig:dEprotspinstates}.
The suite of fits that are performed  lead to a proton magnetic moment and polarizability of
\begin{eqnarray}
\hat\mu_p &=& \mup \,, \\
  \hat\beta_{p} & = &
  \betap
  \ ,
  \label{eq:betaMprot}
\end{eqnarray}
where the uncertainties are as discussed for the case of the neutron, and the results are presented in the
 dimensionless natural units 
defined in Eqs.~(\ref{eq:Bdependlatt}) and (\ref{eq:imlessvardef}).
The Landau level makes a contribution to the ${\cal O}\left(B^2\right)$ term that is suppressed by the mass of the proton, 
and its main contribution is to the term linear in $B$ where it can be well constrained by the coupled analysis of the two spin states. Consequently, 
the uncertainty in identifying the correct Landau level of the system 
does not lead to a particularly large uncertainty in the extracted value of its magnetic polarizability (although the neutron polarizability is considerably more precise).

The magnetic polarizability of the proton is found to be 
considerably larger than that of the neutron, $\hat{\beta}_p-\hat{\beta}_n=0.63(10)(4)$, indicating a significant
isovector component at this unphysical pion mass.  Currently, there are no
other LQCD calculations of the proton magnetic polarizability with
which to compare,
however, it can be compared
with the experimental value. As quoted previously, $\beta^{\rm phys}_p =
3.15(0.35)(0.2)(0.3) \times 10^{-4}~{\rm
  fm}^3$~\cite{Federspiel:1991yd,Zieger:1992jq,MacGibbon:1995in,Beringer:1900zz,Myers:2014ace,Lensky:2014efa}
  which corresponds to $\hat \beta^{\rm phys}_p = 0.116(13)(7)(11)$ in dimensionless units.
The physical value results from cancellations between pion-loop (chiral physics) 
and $\Delta$-pole contributions
that are both ${\cal O}(10 \times 10^{-4}~{\rm fm}^3)$. Since the pion-loop contribution
is strongly suppressed at heavy quark masses and the  $\Delta$-pole contribution  depends 
less strongly on mass, the size of the magnetic polarizability determined at the SU(3) point is in line with expectations.

\subsubsection{The diproton}
\label{sec:pp}

The diproton is in the same $\si$ isotriplet as the dineutron and, neglecting the electroweak interactions and the difference in mass between the up- and down-quarks, 
it would have the same properties as
the dineutron at zero magnetic field.  
However,  the presence of the background
magnetic field breaks  isospin symmetry through the 
light-quark electric charges, so the 
diproton magnetic properties are expected to be quite different from the dineutron, 
even neglecting the issue of Landau levels.

\begin{figure}[!ht]
  \centering
  \includegraphics[width=0.7\columnwidth]{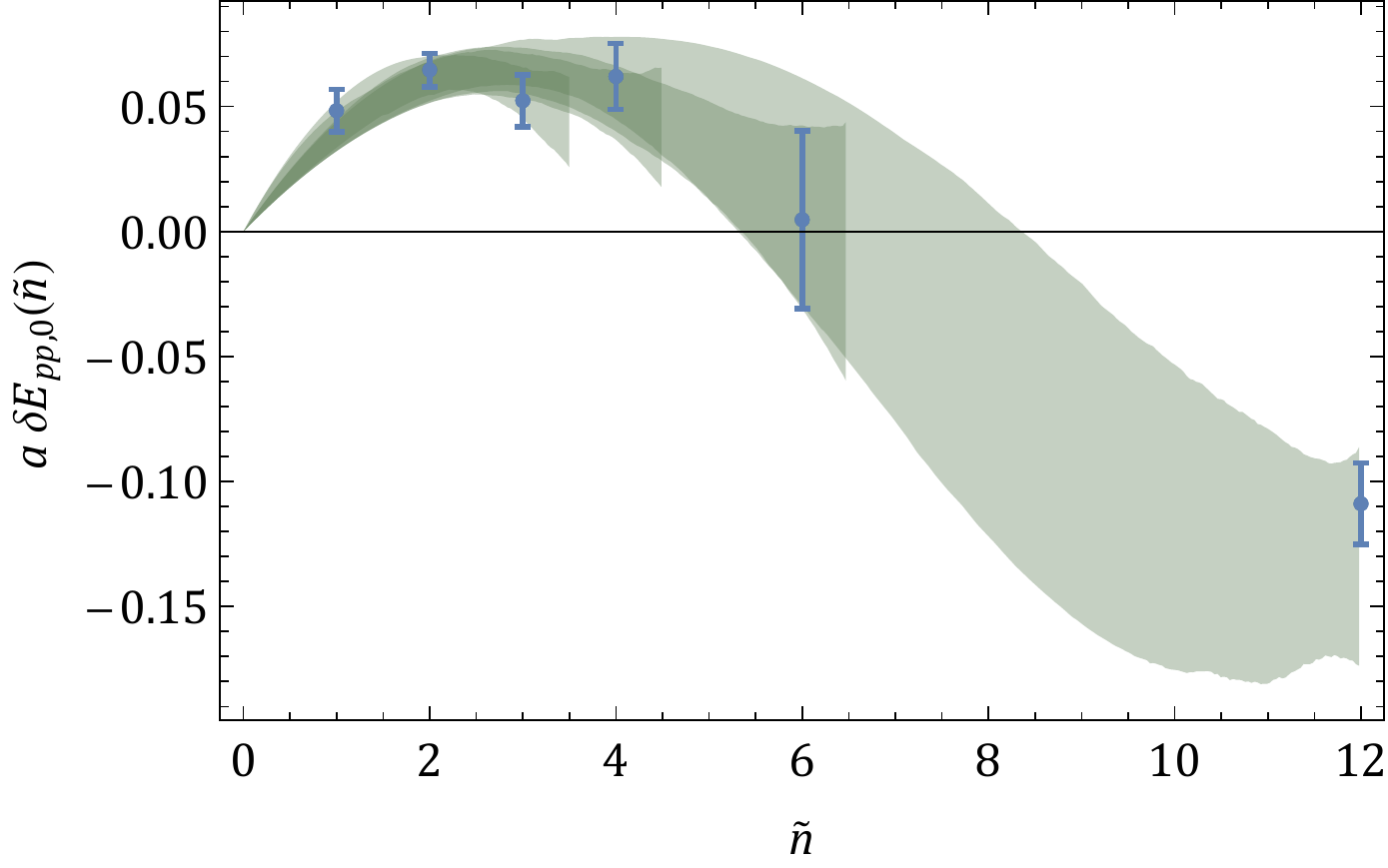}\
  \includegraphics[width=0.6\columnwidth]{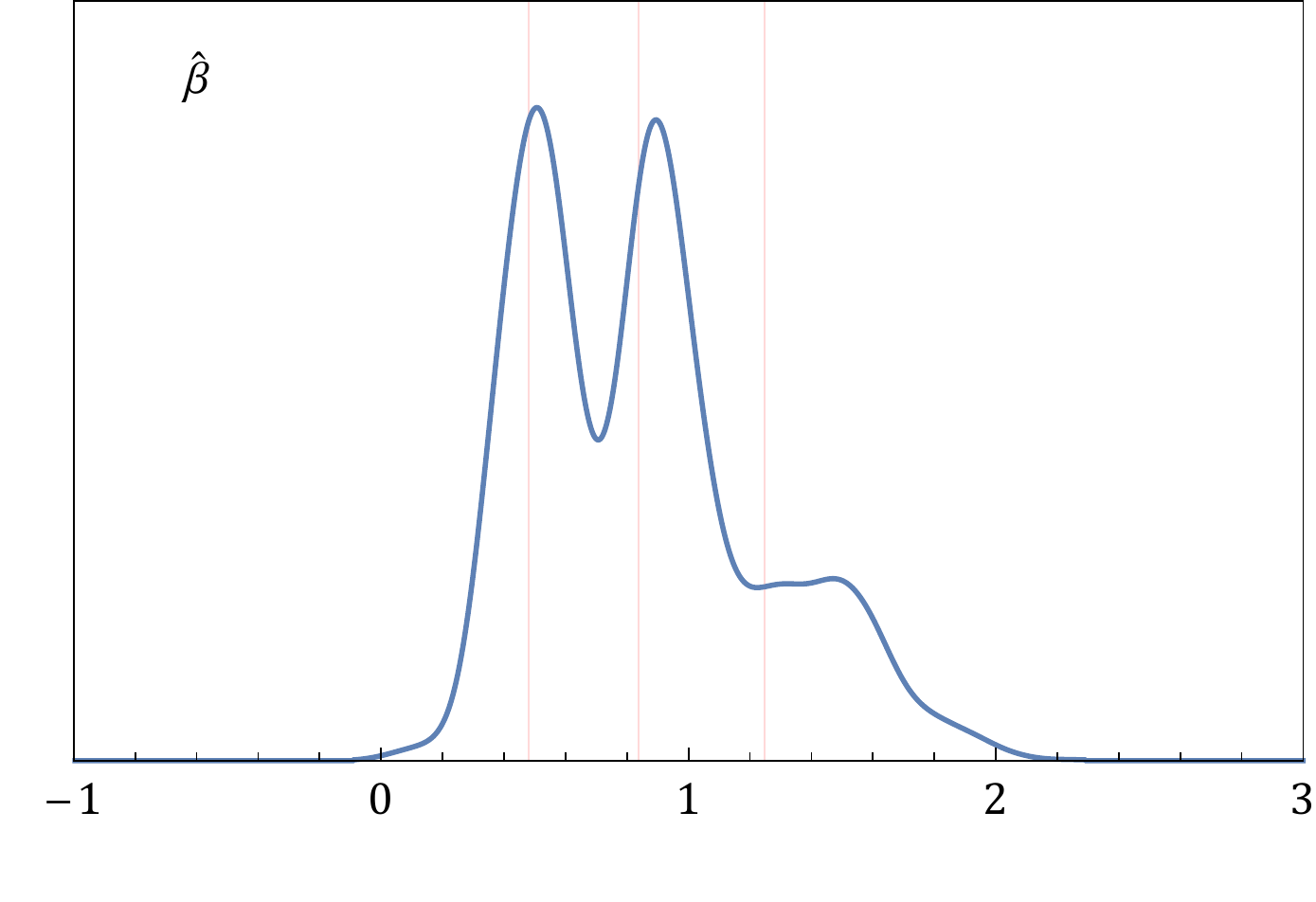}\
  \
  \caption{ 
    Results for the energy shifts of the diproton as a function of the
    background magnetic field strength. 
    The details are as those in Fig.~\protect{\ref{fig:dEdineut}}. 
     The lower panel shows the PDFs for the fit parameter $\hat\beta$. }
  \label{fig:dEdiprot}
\end{figure}

\begin{figure}[!ht]
  \centering
  \includegraphics[width=0.7\columnwidth]{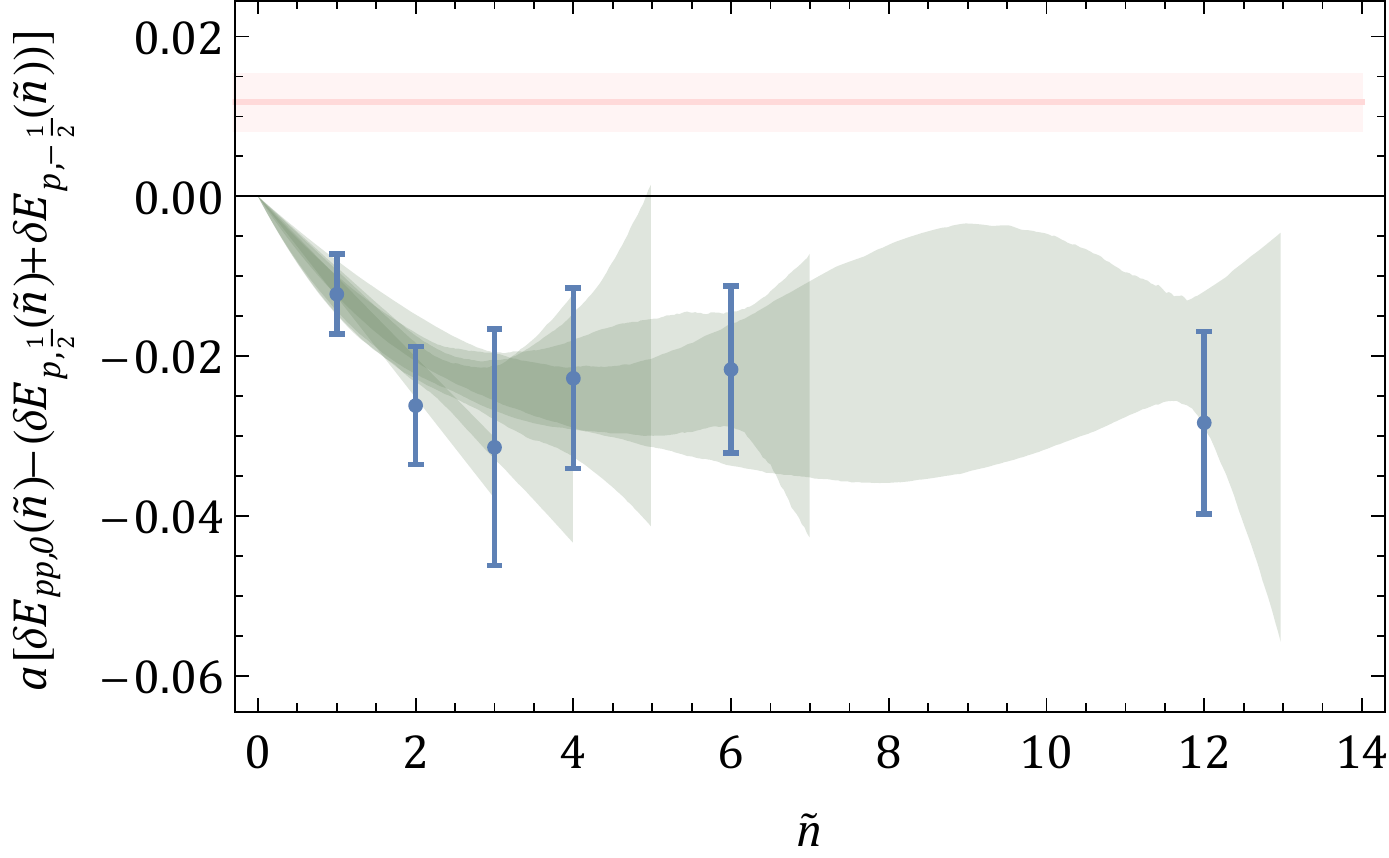}
  \caption{   Results for the difference between the energy shifts of the diproton
    and a spin-up and spin-down proton as a function of the background
    magnetic field strength. 
    The red-shaded region corresponds to the breakup threshold, above which the ground state of the system would be two
    protons in the continuum in the $\si$ channel.  }
  \label{fig:dEprotMpp}
\end{figure}

Extracting energy differences from fits to the ratios of correlation functions shown in Fig.~\ref{fig:corrratio1}
leads to the results shown in Fig.~\ref{fig:dEdiprot}.
Fitting the energy shifts, as discussed previously, allows for an extraction of the diproton 
polarizability of
\begin{eqnarray}
  \hat\beta_{pp} & = & \betapp \ .
  \label{eq:betapp}
\end{eqnarray}
As in the
case of the dineutron, the correlated ratios of the diproton and the spin-up and spin-down
proton correlation functions directly determines the difference of energy splittings.
Figure~\ref{fig:deltaNNcorr} shows these ratios, leading to the energy shifts shown 
in Fig.~\ref{fig:dEprotMpp}. 
 The figure also shows the envelopes of the 
ensemble of acceptable fits that were performed using polynomials 
of up to quartic order. 

It is clear from Fig.~\ref{fig:dEprotMpp} that the magnetic field strengthens the binding of the diproton by a significant amount
that rapidly increases until $\tilde n\sim 3$ and then remains constant for larger field strengths.
This behavior is interesting in the context of the suggestion that at the physical quark masses, the diproton can overcome the Coulomb
repulsion and form a bound state~\cite{Allor:2006zy} in a strong
enough magnetic field, although this argument requires the system to be near unitarity.
However, the form of the difference is more complicated in this case than for the dineutron
because the contributions of Landau levels in the diproton and spin averaged protons 
may be different. The difference in magnetic polarizabilities  is therefore estimated in the naive way,
giving
\begin{eqnarray}
	 \delta  \hat\beta_{pp} & = &
  \dbetapp
  \ ,
\end{eqnarray}
where the uncertainties of the diproton and proton polarizabilities are combined in quadrature.

\subsubsection{The deuteron : $j_z=\pm 1$}
\label{sec:deutjzpmone}
The deuteron
is a bound state in the positive parity  $\siii$-$\diii$ coupled channels.  
In a
background magnetic field, while the $j_z=\pm 1$ states remain
isolated in the $\siii$-$\diii$ coupled channels (in infinite volume), 
the $j_z=0$ state
mixes with the positive parity $\si$ isotriplet $np$ channel.  
Here, the focus  is
on the $j_z=\pm1$ states which are used to and extract the magnetic moment and a combination 
of the scalar and tensor polarizabilities. The $j_z=I_z=0$ coupled states are addressed in the following subsection.

\begin{figure}[!ht]
  \centering
  \includegraphics[width=0.7\columnwidth]{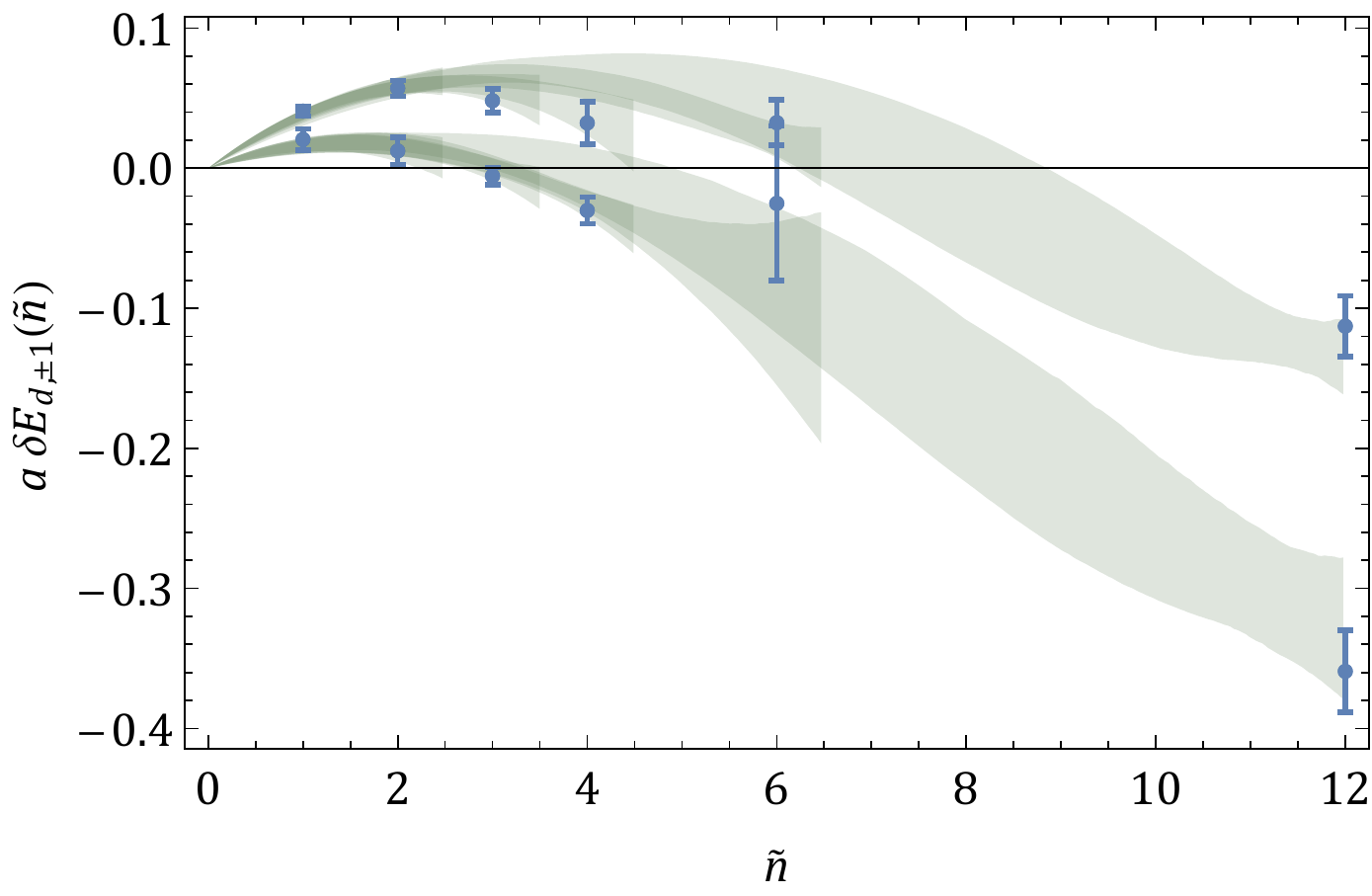}\ \
  \includegraphics[width=0.6\columnwidth]{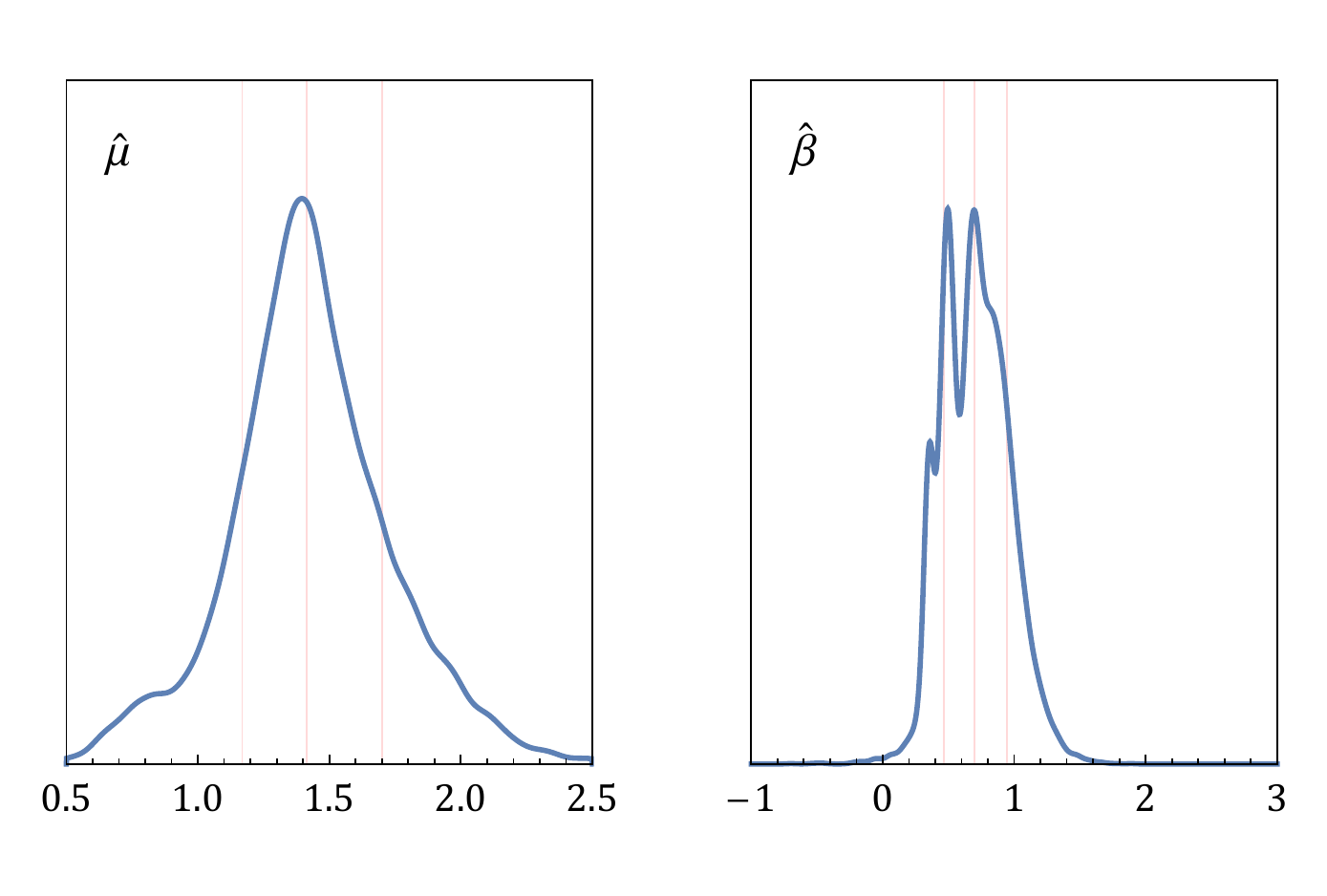}\
  \caption{
    Results for the energy shifts of the deuteron in the $j_z=\pm 1$
    states as a function of the background magnetic field strength 
    (the lower points correspond to the $j_z=+1$ state).
   The details of the figure are as in Fig.~\protect{\ref{fig:dEdineut}}. 
   The lower panel shows the PDFs for the fit parameters $\hat{\mu}$ and $\hat\beta
   $. }
  \label{fig:dEdeut}
\end{figure}

Figure~\ref{fig:corrs} shows  the effective masses 
resulting from the $j_z=\pm1$ deuteron correlation functions
and Fig.~\ref{fig:corrratio2} shows the  ratios of these correlation functions, 
along with  fits to their time dependence.
The energy shifts extracted from these ratios are shown  in Fig.~\ref{fig:dEdeut}.
Analysis of the field strength dependence through a suite of coupled fits to the two spin states, as discussed above,
leads to a magnetic moment and polarizability  of
\begin{eqnarray}
 \hat\mu_d &=& \mud\,, \\
  \hat\beta^{(M0)}_{d}  +  {1\over 3} \hat\beta^{(M2)}_{d} 
  & = &
  \betadpm
  \ .
\end{eqnarray}
As the deuteron has $j=1$, both the scalar and tensor
polarizabilities contribute to the quadratic dependence on the
magnetic field strength, as presented in Eq.~(\ref{eq:Eshift}).\footnote{
With further 
analysis, the ${\cal O}(|e {\bf B}|^2)$ shifts in the $j_z=0$ $np$ coupled system should
determine an orthogonal combination of the  scalar and tensor polarizabilities, 
$ \hat\beta^{(M0)}_{d}  -  {2\over 3} \hat\beta^{(M2)}_{d} $ as given in Eq.~(\ref{eq:Eshift}),
but this extraction is not pursued in the present study.
}

\begin{figure}[!ht]
	\centering
	\includegraphics[width=0.7\columnwidth]{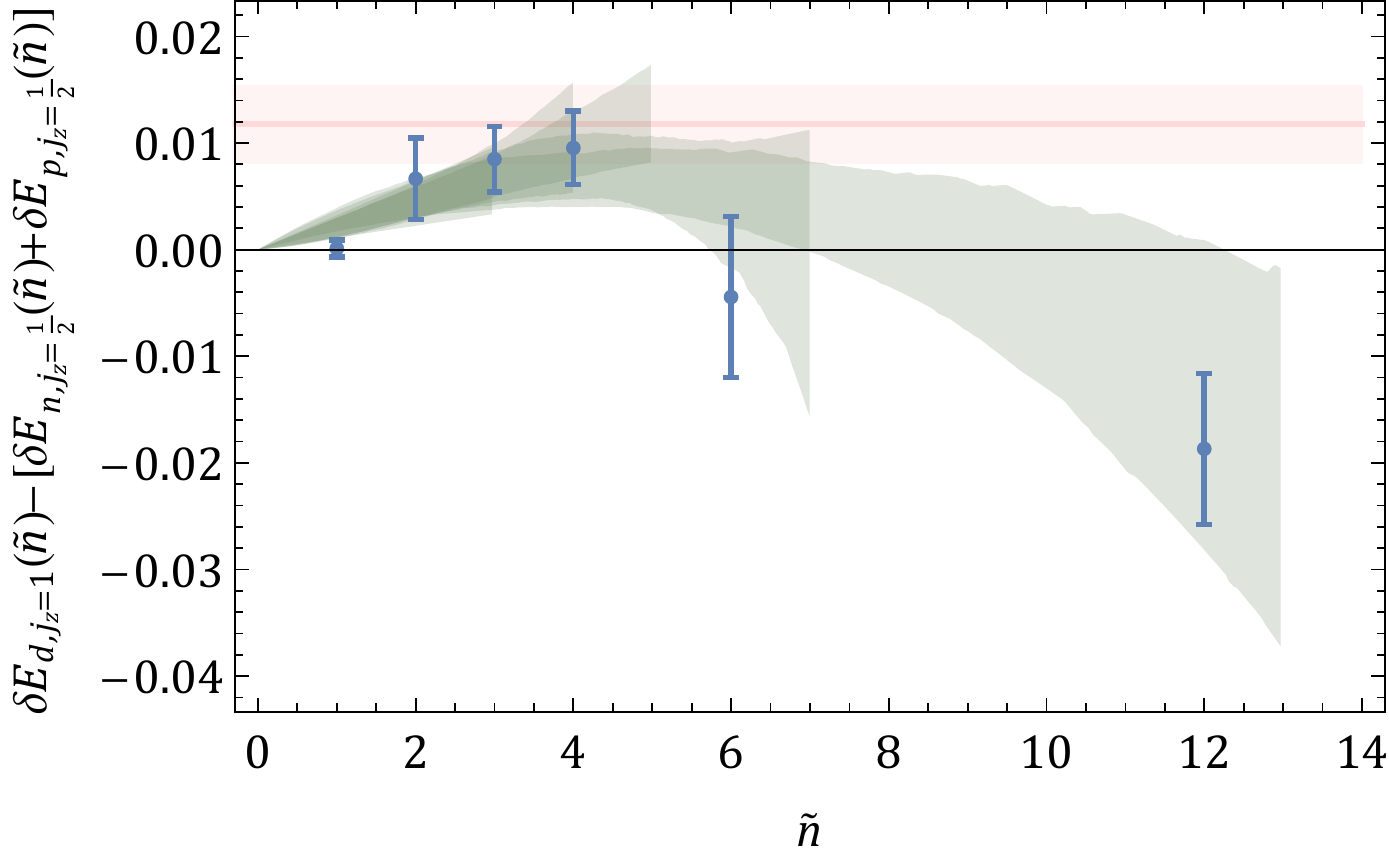}
	\caption{ 
	  Results for the difference between the energy shifts of the $j_z= + 1$ spin states of the deuteron 
		and that of a spin-up neutron and spin-up proton as a function of the background
		magnetic field strength. 
		The red-shaded horizontal band corresponds to the breakup threshold, above 
		which the ground state of the system would be a proton and a neutron in the  $\si$ continuum. The green shaded regions correspond to the envelopes of the 
		fits discussed in the main text. }
	\label{fig:dEdeutMnMp}
\end{figure}

The sum of the proton and neutron magnetic polarizabilities at this pion mass is
$\hat \beta_p+\hat\beta_n\sim 1.02\left({\tiny \begin{array}{c}
	+0.10 \\ -0.07
	\end{array}}\right)(0.05)$, 
so the deuteron
in the $j_z=\pm 1$ states is somewhat more magnetically rigid
than the sum of its constituents.  While they cannot be separated from
this result alone, the nuclear forces and gauge-invariant electromagnetic  two-nucleon
operators are responsible for this difference. 
If the difference persists at the physical quark masses, 
this would suggest that the extraction of the neutron polarizability from experiments on the deuteron is problematic.
Figure \ref{fig:dEdeutMnMp} shows 
the splitting between the $j_z=\pm 1$ spin states of the deuteron and the breakup 
threshold as a function of the field strength. As in the case of the dineutron, 
the magnetic field pushes the  $j_z=\pm 1$ spin states of the deuteron towards threshold
and at $\tilde n\sim 5$, the deuteron becomes potentially unbound before rebinding at larger field strengths. The figure also shows the envelopes of the 
ensemble of acceptable fits that we perform using polynomials 
of up to quartic order. As for the case of the diproton, the 
presence of Landau levels that may differ between the deuteron and proton complicates the analysis of the
field strength dependence and we do not report 
a value of $\delta\beta_d$.
%

\subsubsection{$^3$He}
\label{sec:hethree}

At the physical quark masses, $^3$He can be thought of, to a large degree, as
two protons spin-paired in the $\si$ channel and a single unpaired
$S$-wave neutron.  
The ground state is positive parity with spin-half and  is an isospin partner with the ground state of
the triton, $^3$H.  A naive shell-model prediction is that the
magnetic moment of the ground state of $^3$He is that of the neutron
(with the spin-paired protons not contributing) and that the magnetic
moment of the ground state of the triton is that of the proton (with
the spin-paired neutrons not contributing).  The experimental
values of both magnetic moments deviate only slightly from these naive
predictions. Recent calculations have shown that this feature persists even
at heavier quark masses~\cite{Beane:2014ora}, in particular, at the
pion mass employed in the present analysis.

\begin{figure}[!ht]
  \centering
  \includegraphics[width=0.7\columnwidth]{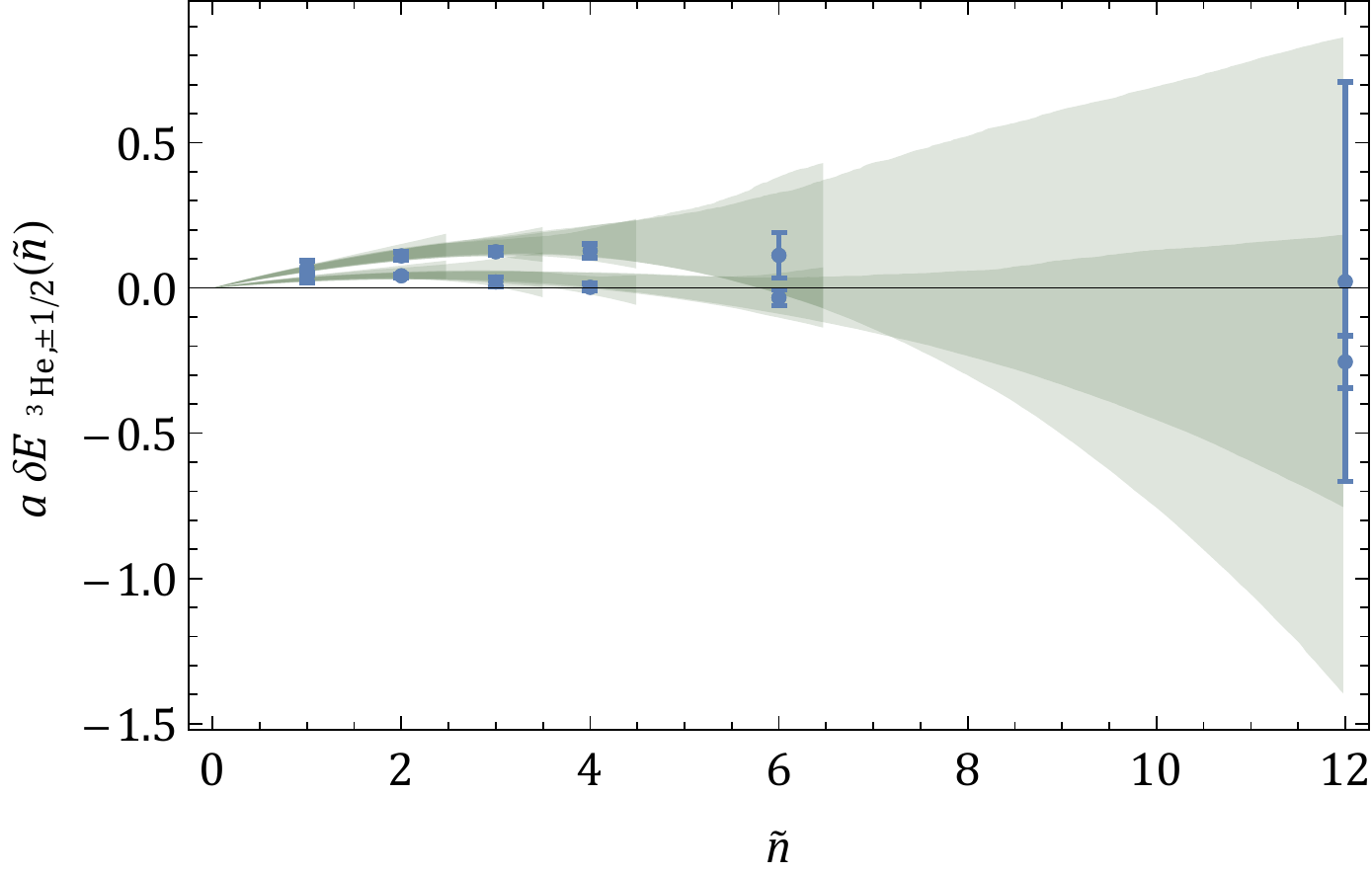}\
  \includegraphics[width=0.6\columnwidth]{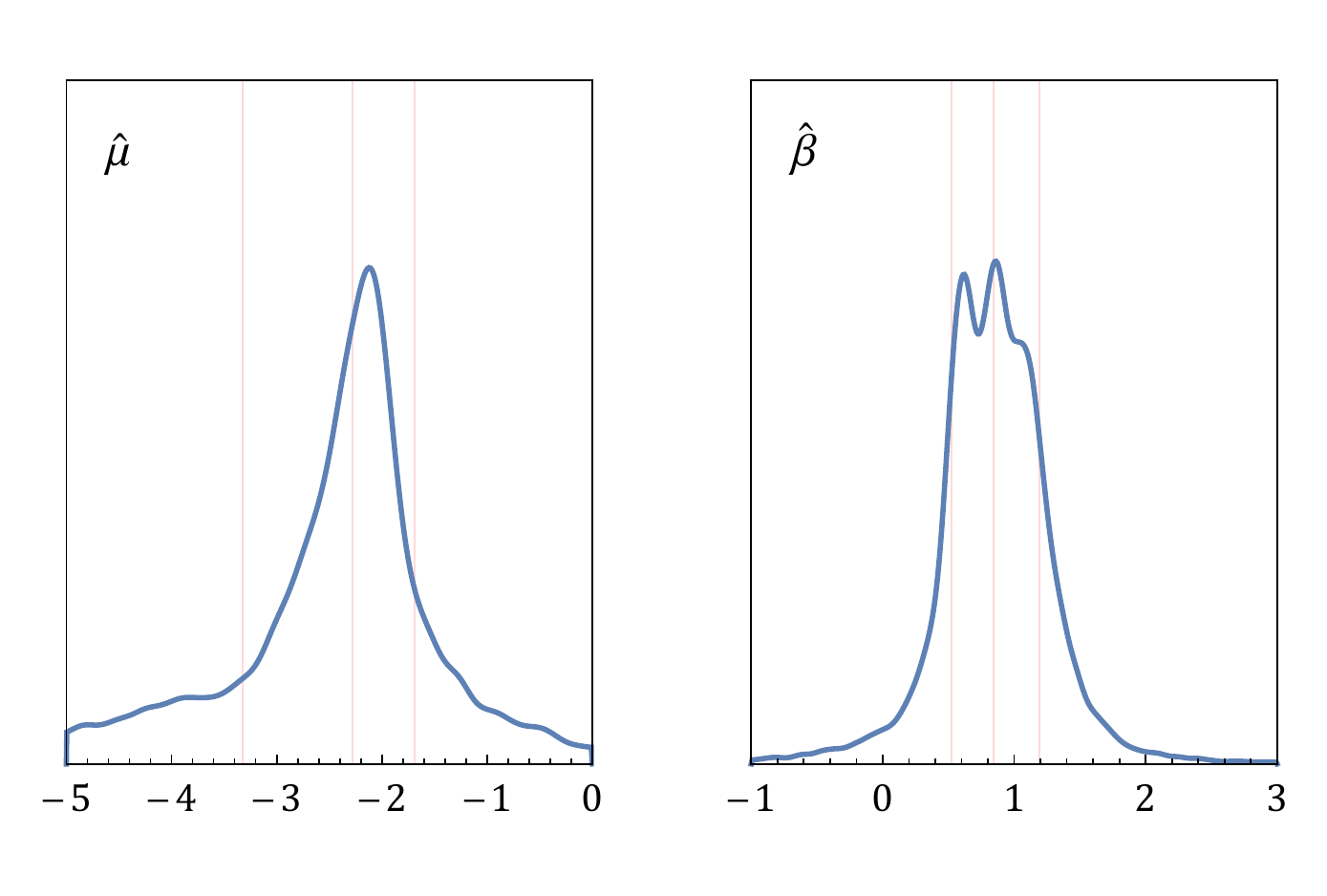}\
  \
  \caption{   Results for the energy shifts of $^3$He as a function of the
    background magnetic field strength along with the fit envelopes.  
    The details of the figure are as in Fig.~\protect{\ref{fig:dEdineut}}. 
The lower energy points correspond to the $j_z=-{1\over 2}$ state, while the upper points correspond to  $j_z=+{1\over 2}$. The lower panel shows the PDFs for the fit parameters $\hat{\mu}$ and $\hat\beta
$. 
}
  \label{fig:dEHe3}
\end{figure}

The EMPs obtained  
from the $^3$He correlation functions in the background magnetic fields
are shown in Fig.~\ref{fig:corrs} and the ratios of correlation functions for 
each spin state are shown in Fig.~\ref{fig:corrratio2},
along with fits to their time dependence. 
The quality of these  ratios is inferior to those obtained in the 
one-nucleon and two-nucleon sectors, but strong signals are still evident. The energies that
are extracted from these ratios, are shown in
Fig.~\ref{fig:dEHe3}.
Analysis of the field strength dependence of the two spin states
allows the magnetic responses to be determined,
leading
to a magnetic moment and polarizability of $^3$He of
\begin{eqnarray}
  \hat\mu_{^3 {\rm He}} & = &
\muHethree  \,, \\
  \hat\beta_{^3 {\rm He}} & = &
  \betaHethree
  \ ,
\end{eqnarray}
in natural dimensionless units. Given that the magnetic moment favors the naive 
shell model expectation, 
$ \mu_{^3 {\rm He}} =  \mu_n$, one might naively expect  the polarizability 
of $^3$He to arise from the polarizability of the diproton combined with that 
of the neutron. 
Within the uncertainties, that expectation is found to hold. 

The uncertainties in the magnetic polarizabilities of  $^3$He  
are sufficiently large that statistically significant deviations
from the contributions from the one-body contribution are not 
obtained, and hence we have no meaningful constraint on the MEC
contributions.

\subsubsection{The triton}
\label{sec:hthree}

\begin{figure}[!ht]
  \centering
  \includegraphics[width=0.7\columnwidth]{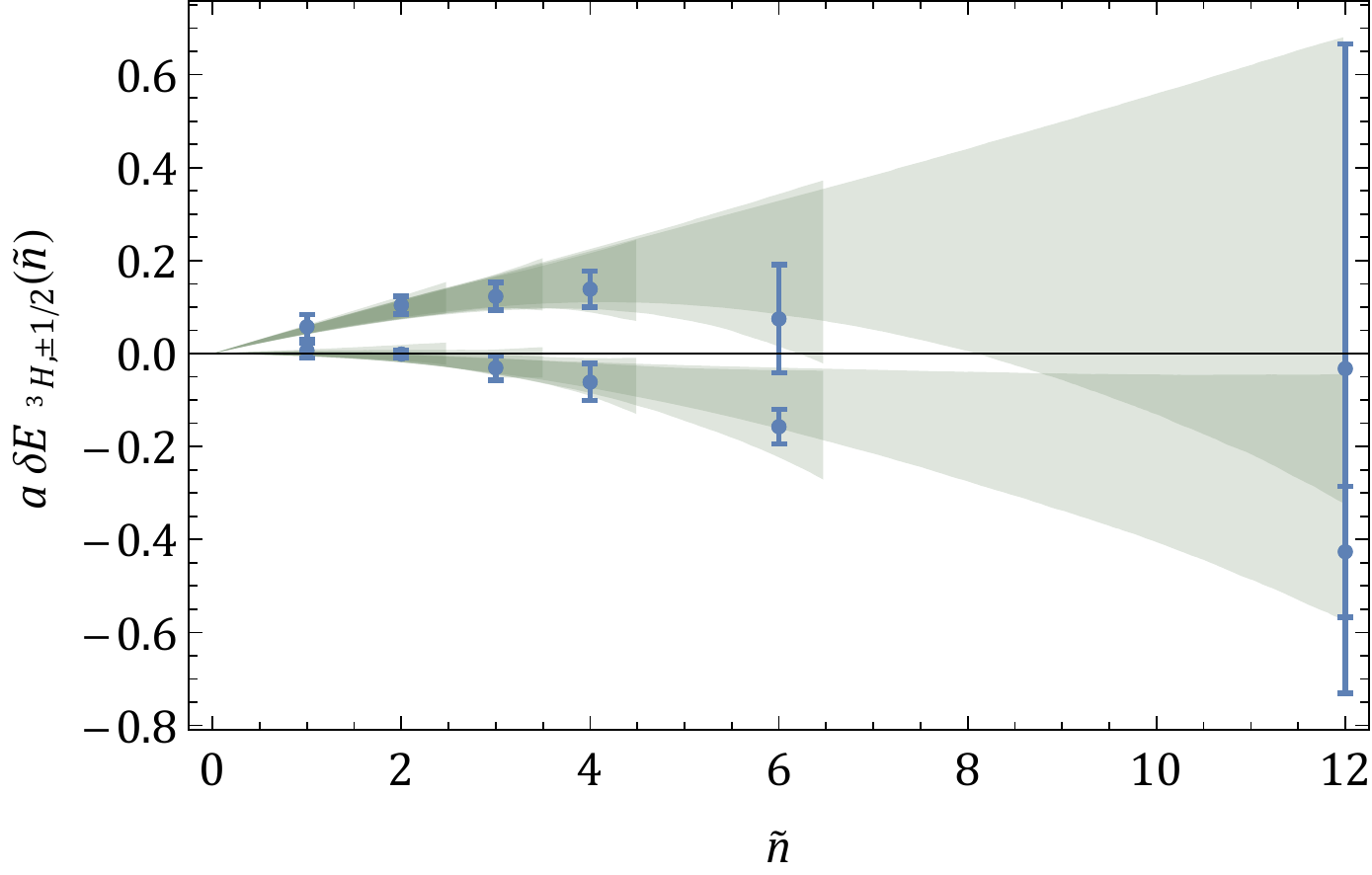}\
  \includegraphics[width=0.6\columnwidth]{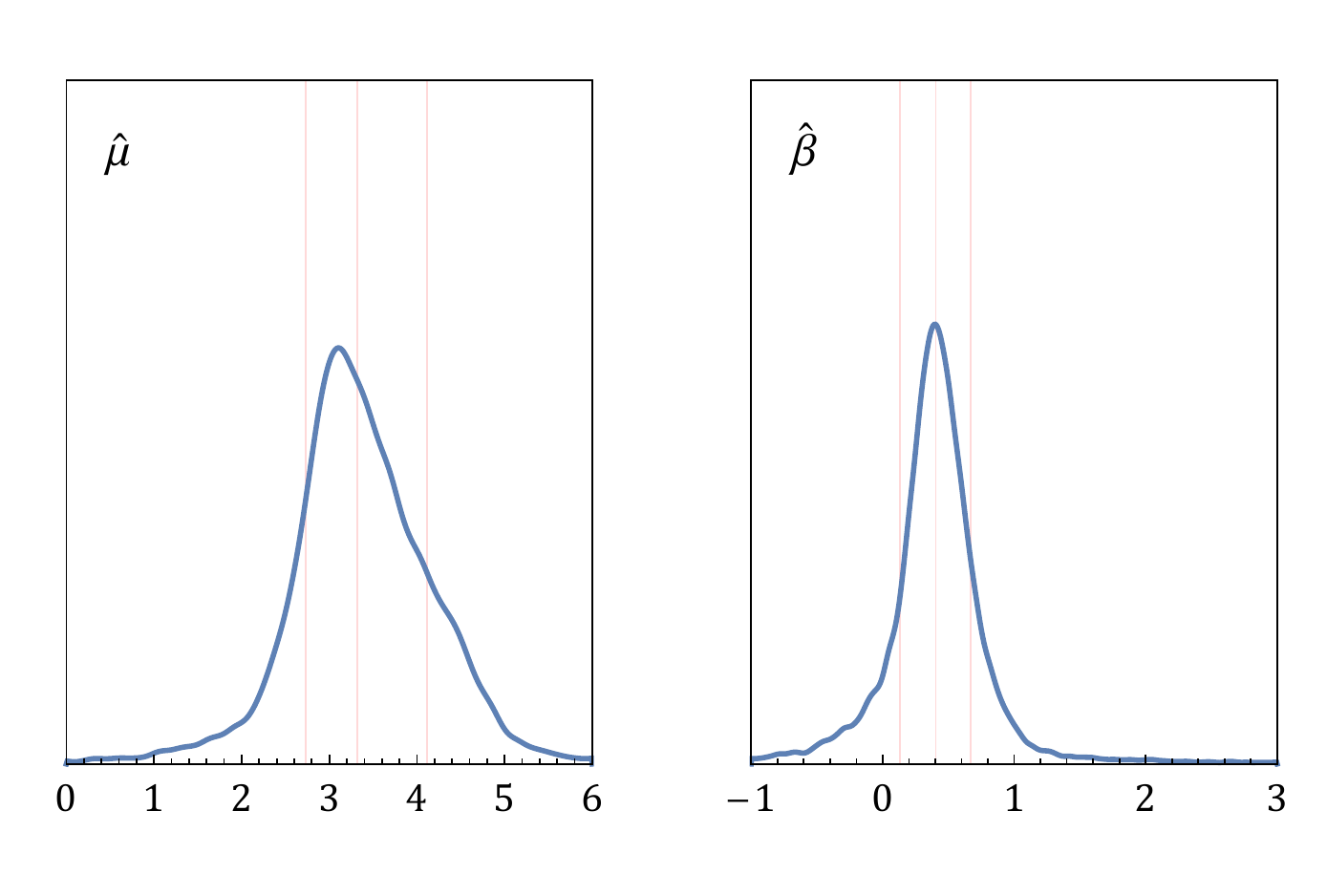}\
  \
  \caption{   Results for the energy shifts of $^3$H as a function of the
    background magnetic field strength, along with envelopes of fits.  
    The details of the figure are the same as in Fig.~\protect{\ref{fig:dEdineut}}.  
    The lower energy points correspond to the $j_z=+{1\over 2}$ state, while the upper points correspond to  $j_z=-{1\over 2}$. The lower panel shows the PDFs for the fit parameters $\hat{\mu}$ and $\hat\beta
    $.  
    }
  \label{fig:dEH3}
\end{figure}

As in the case of $^3$He, the  ratios of the triton correlation functions are 
significantly less well-defined than those in the one-body and two-body sectors.   
The energy shifts extracted from the
correlation functions are shown in Fig.~\ref{fig:dEH3}.
Fits to the magnetic field strength dependence of the energies of the two spin states
enable an extraction of the magnetic moment and polarizability of the triton of
\begin{eqnarray}
  \hat\mu_{^3 {\rm H}} & = & \mutriton\,, \\
  \hat\beta_{^3 {\rm H}} & = &
  \betatriton
  \ .
\end{eqnarray}
The value of the triton polarizability is considerably smaller than the naive 
expectation of the sum of the polarizability of the dineutron and of the proton, 
$\beta_p+\beta_{nn}=1.12\left({\tiny \begin{array}{c}
	+0.11 \\ -0.07
	\end{array}}\right)$
and this difference could potentially be used to provide a constraint on two- and three-nucleon electromagnetic
interactions.

\subsubsection{$^4$He}
\label{sec:hefour}

The $^4$He nucleus has the quantum numbers  of two protons and two neutrons in a spin-zero, even-parity configuration.  
The energy of the ground state
has been determined at unphysical quark masses in previous LQCD
calculations~\cite{Yamazaki:2009ua,Yamazaki:2012hi,Beane:2012vq,Yamazaki:2013rna},
and at this pion mass it is bound by $B_{^4{\rm He}} =
107(24)~{\rm MeV}$~\cite{Beane:2012vq}.  
While it has no magnetic moment, it can be polarized by electromagnetic
fields. 

\begin{figure}[!t]
  \centering
  \includegraphics[width=0.7\columnwidth]{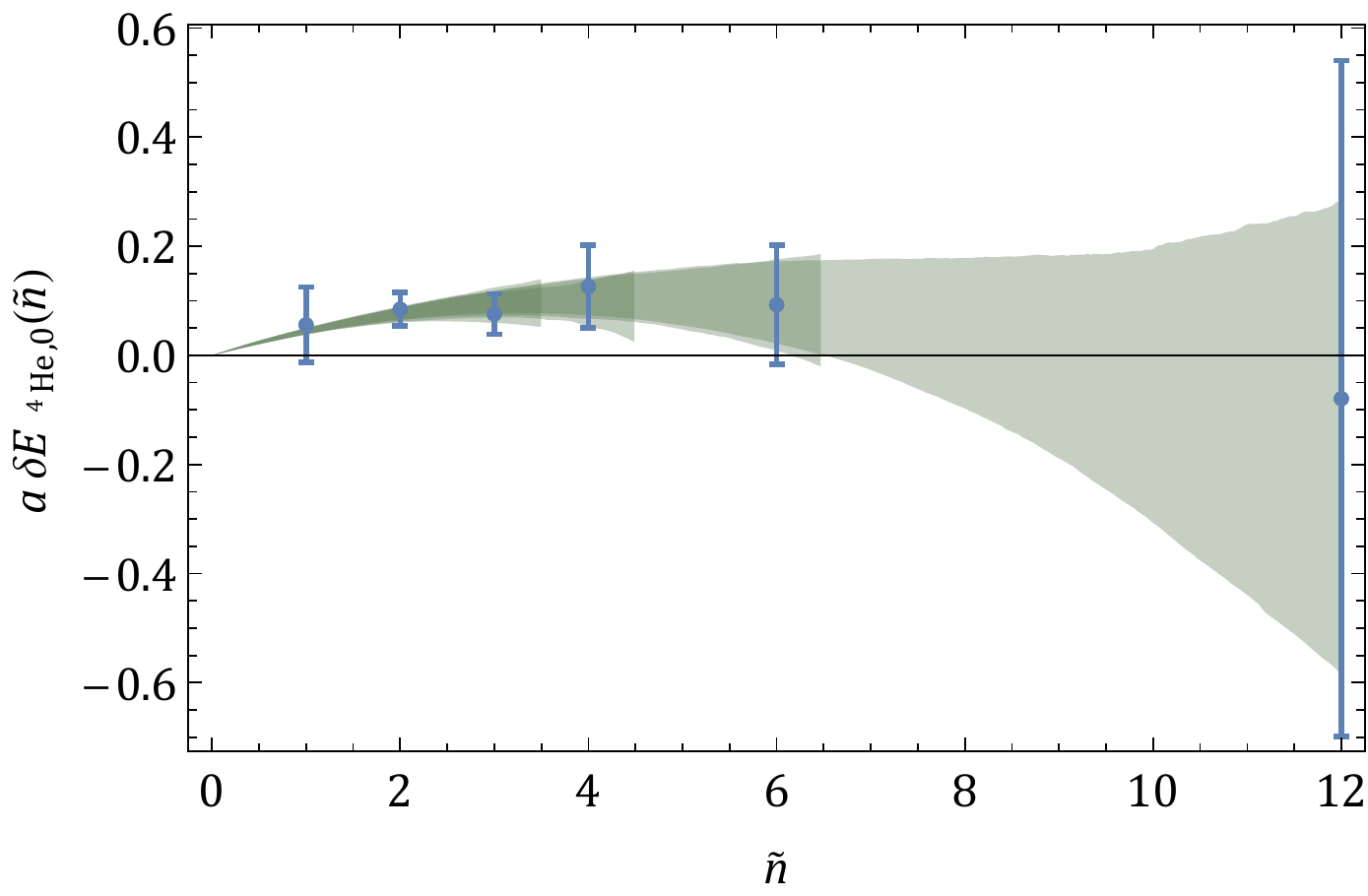}\
  \includegraphics[width=0.6\columnwidth]{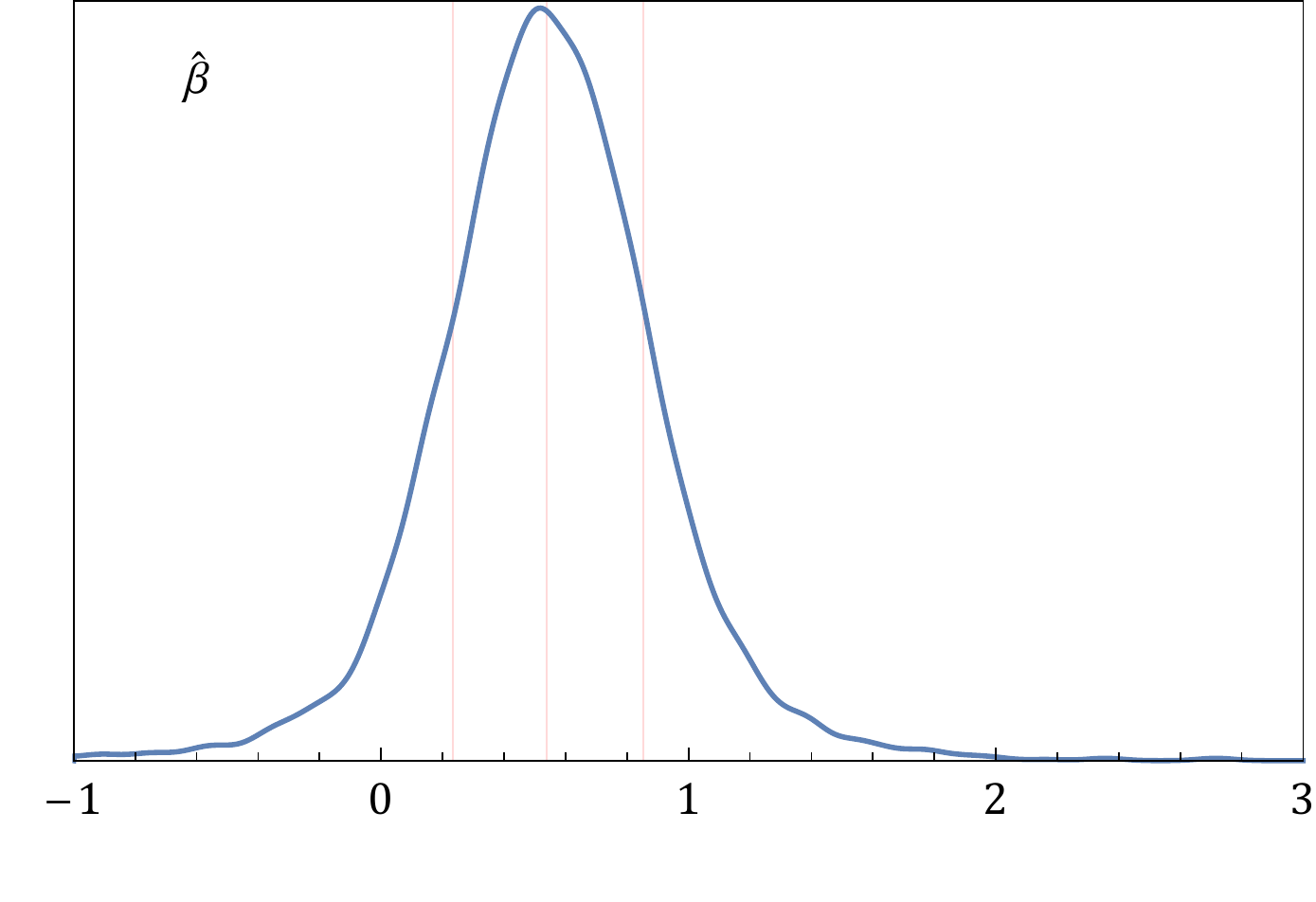}\
  \
  \caption{   Results for the energy shifts of $^4$He as a function of the
    background magnetic field strength. The details of the figure are as in Fig.~\protect{\ref{fig:dEdineut}}. The lower panel shows the PDF for the fit parameters  $\hat\beta$. }
  \label{fig:dEHe4}
\end{figure}

The EMPs obtained from  $^4$He correlation functions in the background magnetic fields
are shown in Fig.~\ref{fig:corrs},
and the  ratios of correlation functions are shown in Fig.~\ref{fig:corrratio2},
 along with fits to their time dependence. 
The energy shifts extracted from fits to these ratios are given in Table~\ref{tab:energies} 
and are shown in Fig.~\ref{fig:dEHe4}.
Analysis of the magnetic field strength dependence of the $^4$He energies
enables an extraction of the magnetic polarizability, giving
\begin{eqnarray}
  \hat\beta_{^4 {\rm He}} & = &
  \betaHefour
  \ .
\end{eqnarray}
%

\subsection{The $j_z=I_z=0$  $np$ states and the $np\to d\gamma$ transition}
\label{sec:deutjzzero}

\begin{figure}[!ht]
  \centering
  \includegraphics[width=0.7\columnwidth]{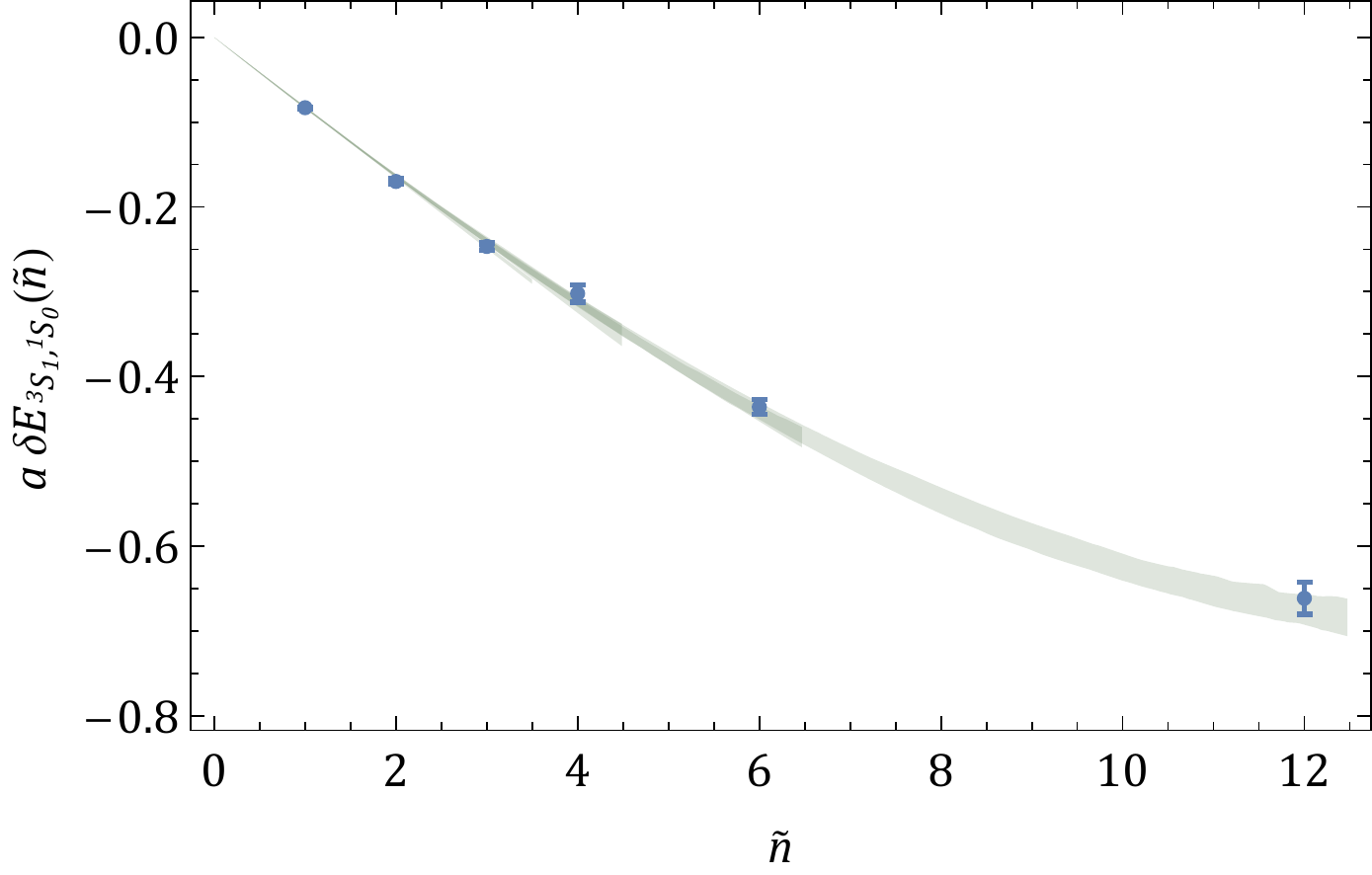}\\ \vspace*{5mm}
  \includegraphics[width=0.7\columnwidth]{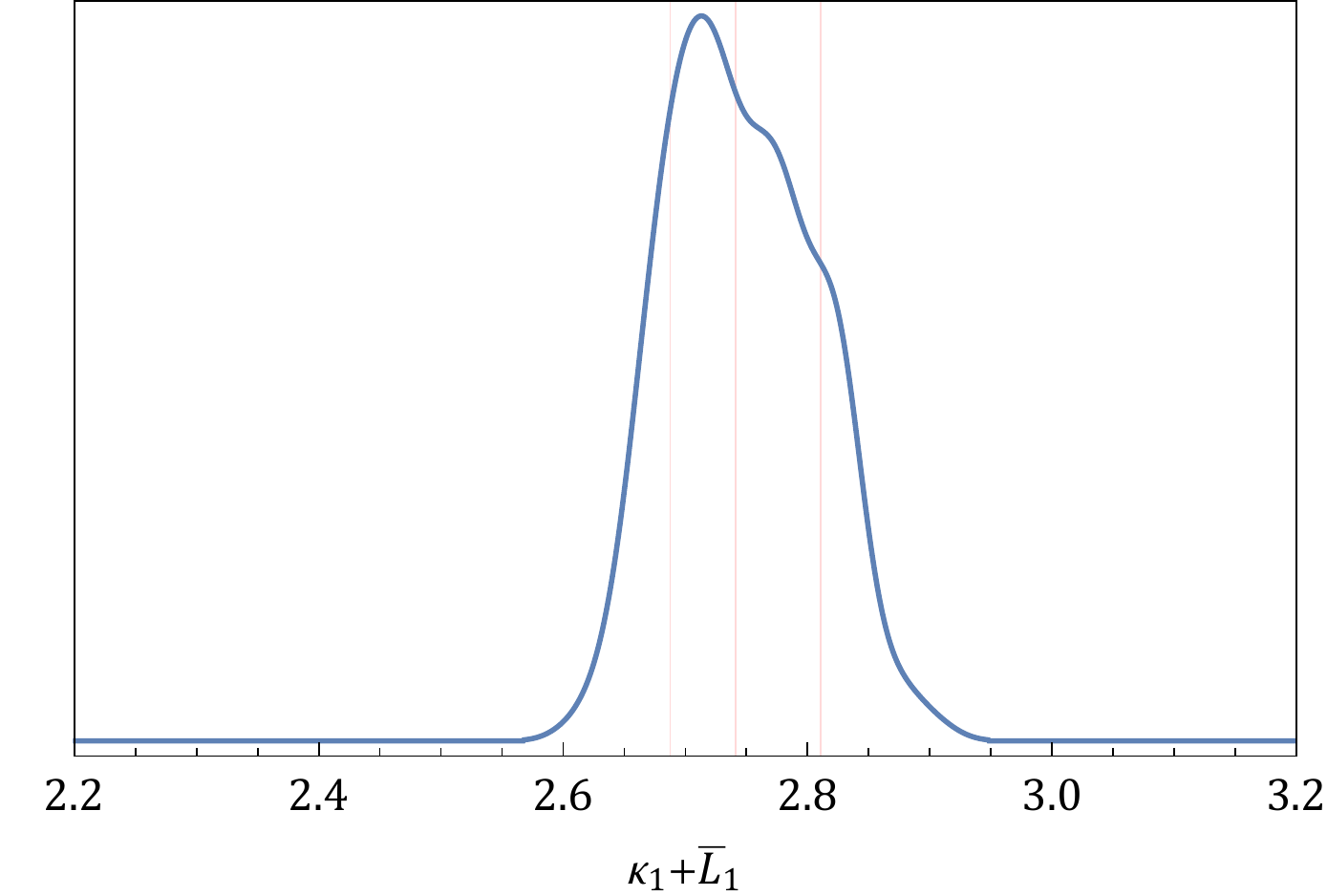}\
  \
  \caption{ 
    Results for the differences in energy shifts between the two $j_z=I_z=0$
  $np$ energy eigenstates as a function of the background
    magnetic field strength.  
    The details of the figure are the same as in Fig.~\protect{\ref{fig:dEdineut}}.  
The lower panel shows  the PDF for the coefficient of the linear field dependence, $\kappa_1 + \overline{L}_1$.  }
  \label{fig:dEdeutjz0}
\end{figure}

The   two energy eigenvalues of the  coupled $j_z=I_z=0$   $np$ channels
 are shown in
Fig.~\ref{fig:corrJzeq0} for each magnetic field strength.
In order to extract the $\siii$-$\si$
mixing, and hence the short-distance two-nucleon  (MEC) 
contribution to $np\rightarrow d\gamma$, 
the  ratios of correlation functions
$R_{\siii,\si}(t;{\bf B})$ in Eq.~(\ref{eq:R3s11s0}) are constructed,
and shown in Fig.~\ref{fig:corrratioJzeq0},
 along with fits to the time dependence. 
The energy shifts extracted from these ratios are shown in Fig.~\ref{fig:dEdeutjz0},
along with 
the envelope of the ensemble of successful fits, from which the linear coefficient is found to be
\begin{eqnarray}
  \kappa_1 + \overline{L}_1 
  & = &  
  \kappaplusLonebar
  \ \ \ ,
\end{eqnarray}
where the result is presented in  dimensionless units determined by the natural nuclear magneton
at this pion mass. Fits of up to  quadratic order are considered in this analysis. 

\begin{figure}[!ht]
  \centering
  \includegraphics[width=0.7\columnwidth]{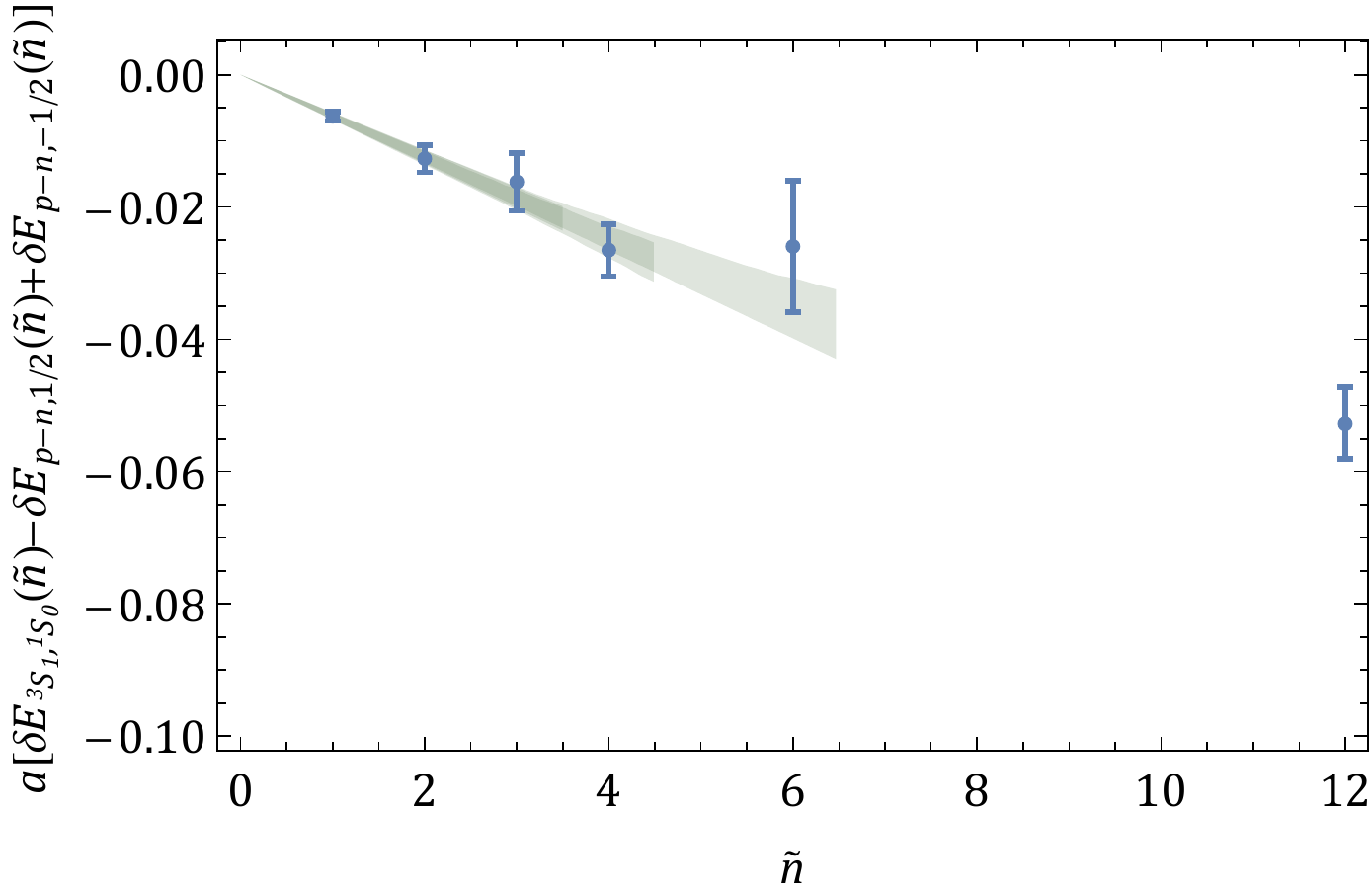}\
  \includegraphics[width=0.6\columnwidth]{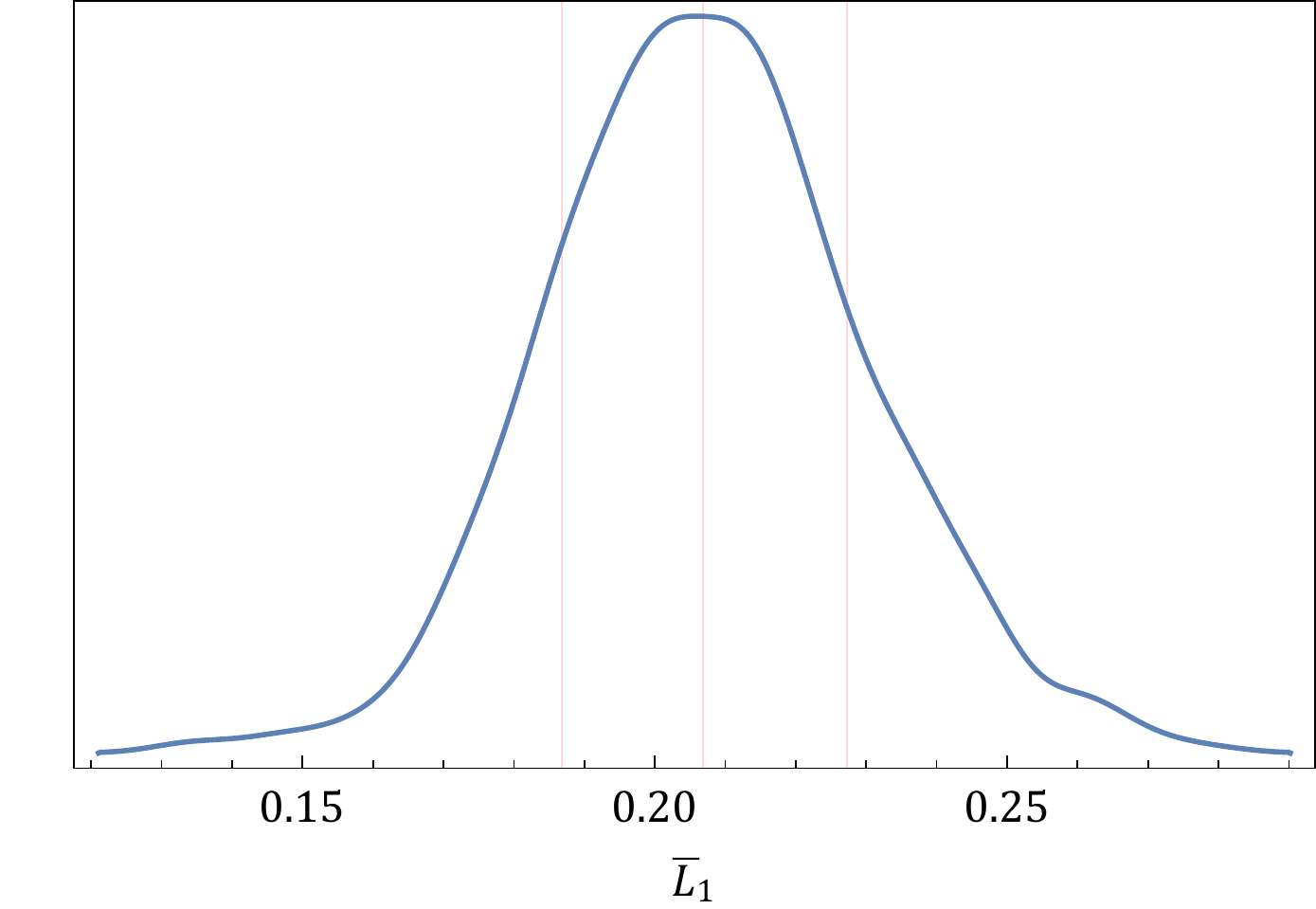}\
  \
  \caption{ 
  The correlated difference between the differences in energy shifts between the two $np$
    $j_z=0$ energy eigenstates and 
    those of  the isovector nucleon (see Eq. (\ref{eq:deltaR3s11s0})) as a function of the background
    magnetic field strength.      The details of the figure are the same as in Fig.~\protect{\ref{fig:dEdineut}}.  
    }
  \label{fig:dEdeutjz0Shift}
\end{figure}

To further isolate the short-distance two-nucleon contribution, 
the ratios $\delta R_{\siii,\si}(t;{\bf B}) $, defined in Eq.~(\ref{eq:deltaR3s11s0}), 
are formed. By design, the energy shifts extracted from these ratios (see Fig.~\ref{fig:corrratioratioJzeq0}) 
have the form $2 |e {\bf B}| \overline L_1 /M + {\cal O}(|e {\bf B}|^2)$.
These shifts are shown in Fig.~\ref{fig:dEdeutjz0Shift},
and performing polynomial fits to the dependence on the magnetic field strength leads to 
\begin{eqnarray}
  \overline{L}_1 = 
  \Lonebar 
  \ \ \ .
  \label{eq:Liconstraint}
\end{eqnarray}

Given the isovector magnetic moment and the short-distance two-nucleon contribution, 
the cross-section for the 
process $np\to d \gamma$  can be determined near threshold at $m_\pi\sim\pionmass$. 
Even though the LQCD calculations determines these parameters from
mixing between bound states,
the 
EFT($\pislash$)
 framework is valid for low-energy scattering states and 
can be immediately applied.
It is conventional to use a multipole expansion to define the 
cross-section for the radiative-capture process $np\rightarrow d\gamma$ at low 
energies~\cite{Chen:1999tn,Chen:1999bg,Rupak:1999rk},
\begin{eqnarray}
\sigma(np\rightarrow d\gamma) & = & 
{e^2(\gamma_0^2+|{\bf p}|^2)^3\over M_N^4 \gamma_0^3 |{\bf p}|}
\left[\ 
|\tilde X_{M1}|^2  + |\tilde X_{E1}|^2  + ...
\ \right]
\ ,
\label{eq:npdgsigma}
\end{eqnarray}
where 
$\tilde X_{M1}$ is the $M1$ amplitude and $\tilde X_{E1}$ is the $E1$ amplitude
for the process,
$\gamma_0$ is the binding momentum of the deuteron and ${\bf p}$ is the momentum of each incoming nucleon in the center-of-mass frame.
The ellipsis denotes higher-order amplitudes, suppressed by powers of the photon momentum.
Following Refs.~\cite{Beane:2000fi,Detmold:2004qn}, it is straightforward to show that the amplitudes at NLO,
with the dibaryon parameterization of Eq.~(\ref{eq:traniL}),
are~
\begin{eqnarray}
\tilde X_{E1} & = & -{1\over\sqrt{1-\gamma_0 r_3} }
{ |{\bf p}| M_N \gamma_0^2 \over ( |{\bf p}| ^2+\gamma_0^2)^2}
\nonumber\\
\tilde X_{M1} & = & 
{Z_d\over  -{1\over a_1} + {1\over 2} r_1 |{\bf p}|^2 - i |{\bf p}|}
 \left[\ {\kappa_1 \gamma_0^2\over \gamma_0^2 +|{\bf p}|^2}\left( \gamma_0 - {1\over a_1} + {1\over 2} r_1 |{\bf p}|^2 \right)
+ {\gamma_0^2\over 2} l_1
\right]
\ ,
\label{eq:npdgMone}
\end{eqnarray}
where the quantities appearing in this expression are defined in Section~\ref{sec:Jz0corrs}. Near threshold, the $E1$ amplitude is sub-leading and will be ignored.
Inserting the extracted values for $\kappa_1$, $\overline{L}_1$, the binding energy from 
Ref.~\cite{Beane:2012vq}, and the scattering lengths and effective ranges from Ref.~\cite{Beane:2013br},
leads to 
a radiative capture cross-section at the SU(3) symmetric point of
\begin{eqnarray}
\left.\sigma({np\to d \gamma})\right|_{m_\pi\sim \pionmass}= 17\left({\tiny\begin{array}{c}
+101 \\ -16
\end{array}}\right) ~{\rm mb}
\ \ ,
\label{eq:xsec800}
\end{eqnarray}
for an incident neutron speed of $v=2200$ m/s, accurate up to NLO in EFT($\pislash$). Because of the non-linear nature of the dependence of the cross-section on the inputs, the distribution is extremely non-Gaussian; the  central value is reported as the 50$^{\rm th}$ quantile and the uncertainty bounds as the 17$^{\rm th}$ and 83$^{\rm rd}$ quantiles of the full distribution.
Improving on this uncertainty requires significantly better determinations of the scattering parameters and the binding momentum. At the physical point,  the cross-section is known to be 
$\sigma ({np\to d \gamma}) = 334.2(0.5)~{\rm mb}$ ~\cite{Cox1965497} at this relative velocity, which is significantly larger. 
The short-distance two-body contribution in the calculated cross-section (Eq.~(\ref{eq:xsec800})) is about 10\%, just as in the phenomenological determinations.
Accounting for the significantly different phase space available at the SU(3) point, 
and the greatly different scattering parameters in both channels.
The discrepancy in the cross-section is unsurprising.
In Ref.~\cite{Beane:2015yha}, this result is combined with an analogous result 
at $m_\pi\sim 450$ MeV to extrapolate to the physical point and
postdict a cross-section of 
$\sigma^{\rm lqcd} ({np\to d \gamma}) = 332.4\left(^{+5.4}_{-4.7}\right)~{\rm mb}$.

\section{Summary}
\label{sec:end}
The magnetic moments and magnetic polarizabilities of the lightest few nuclei have been
calculated at a pion mass of $m_\pi\sim\pionmass$ using 
LQCD in the presence of background magnetic fields.  
In addition, by considering the 
mixing of two-nucleon states with $j_z=I_z=0$, the $\overline{L}_1$
counterterm of EFT($\pislash$) 
that governs short-distance two-nucleon contributions to the 
radiative-capture process $np \to d \gamma$ 
has been determined. 
This has then been used to predict the near threshold capture cross-section 
at this pion mass.
The success of the calculations presented in this work, and in Refs.~\cite{Beane:2014ora,Beane:2015yha},
demonstrate the feasibility of studying the structure of nuclei
directly from QCD and open the way to a variety of additional QCD 
calculations of the structure and interactions of light nuclei.

The LQCD calculations have been performed at a single lattice
spacing and in one lattice volume, and the lack of continuum and
infinite-volume extrapolations introduces systematic uncertainties
into the results.  
The effects of the FV used in this work 
on the binding energies of the light nuclei
have been explicitly quantified in previous works~\cite{Beane:2012vq}
and found to be small.
It is expected that such effects  in the  moments and polarizabilities are of comparable size. 
An additional uncertainty  of $e^{-\gamma L}$ is assigned to the extracted values of nuclear moments and polarizabilities and for simplicity, we conservatively use $\gamma =\gamma_d \sim 190$ MeV for the binding momentum, leading to a $\sim 3\%$ uncertainty. 
For the single nucleons, the expected volume effects of 
${\cal O}(e^{-m_\pi L})$ are negligibly small.
Calculations with multiple lattice spacings have not been performed and this
systematic uncertainty remains to be quantified.  
The electromagnetic contributions to the action are perturbatively
improved as they are included as a background field in the link
variables. 
Therefore, the lattice-spacing artifacts are expected to be small, 
appearing at ${\cal O}(\Lambda_{\rm QCD}^2a^2, \alpha^2 \Lambda_{\rm QCD} a)\sim 3$\% for
$\Lambda_{\rm QCD}=300$ MeV.  
To account for these effects in dimensionless quantities, 
an overall  multiplicative systematic uncertainty of 3\% 
is assigned to the extracted magnetic  moments and $\overline{L}_1$, and an uncertainty of 5\% is assessed on all of the polarizabilities,
where more complicated effects that compound these uncertainties may be possible. 
For nuclei, these contributions are small compared to the other systematic uncertainties.
The main results are presented in terms of dimensionless quantities, but in Table \ref{tab:summary}, we also convert the polarizabilities to physical units using the lattice spacing $a=0.110(1)$; since the units of polarizabilities are fm$^3$, the scale-setting uncertainty corresponds to an additional 3\% uncertainty that is added in quadrature.
Unfortunately, the calculations of the individual polarizabilities are incomplete because of
the omission of the disconnected contributions (the coupling of the external field to the sea quarks), 
however empirical evidence~\cite{Freeman:2014kka,Deka:2013zha,Green:2015wqa}
suggests that the omitted contributions will lead to only  small
modifications that lie within the current uncertainties. 
Confirming this expectation is left to future work.
We stress that the magnetic moments and the  $M1$ transition amplitude for $np\to d\gamma$ are not afflicted by the absence of disconnected diagrams (and nor are  isovector differences such as $\beta_p-\beta_n$ at the SU(3) point).

\begin{figure}[!th]
  \centering
\includegraphics[width=0.72\columnwidth]{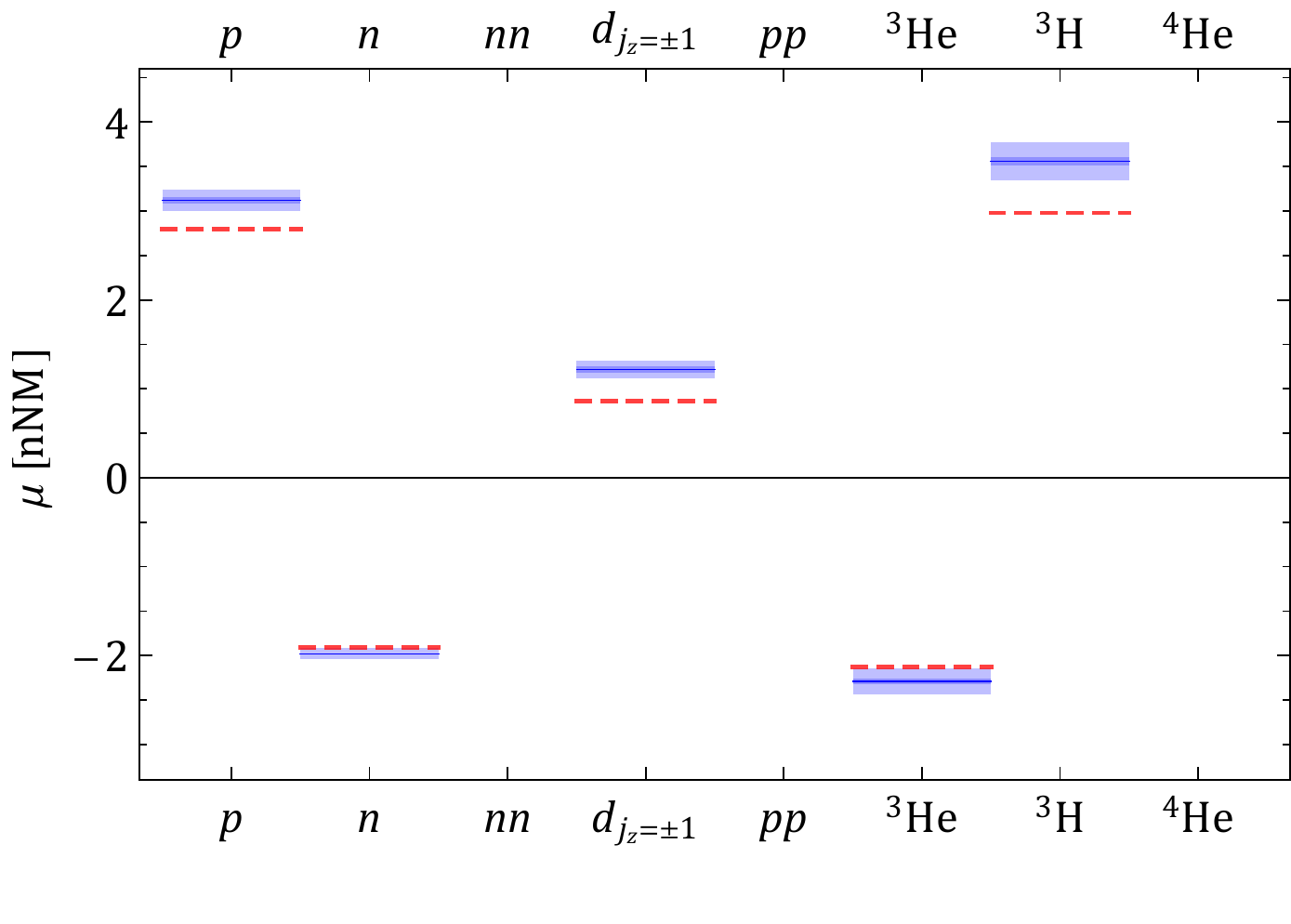}
  \caption{ 
  A summary of the magnetic moments of the nucleons and
    light nuclei calculated with LQCD at SU(3) symmetric quark masses corresponding to a pion mass of
    $m_\pi\sim\pionmass$. 
    The results are presented in units of  natural nuclear magnetons.
    The red dashed lines correspond to the experimental magnetic moments.
}
  \label{fig:summaryMU}
\end{figure}

\begin{figure}[!th]
  \centering
  \includegraphics[width=0.7\columnwidth]{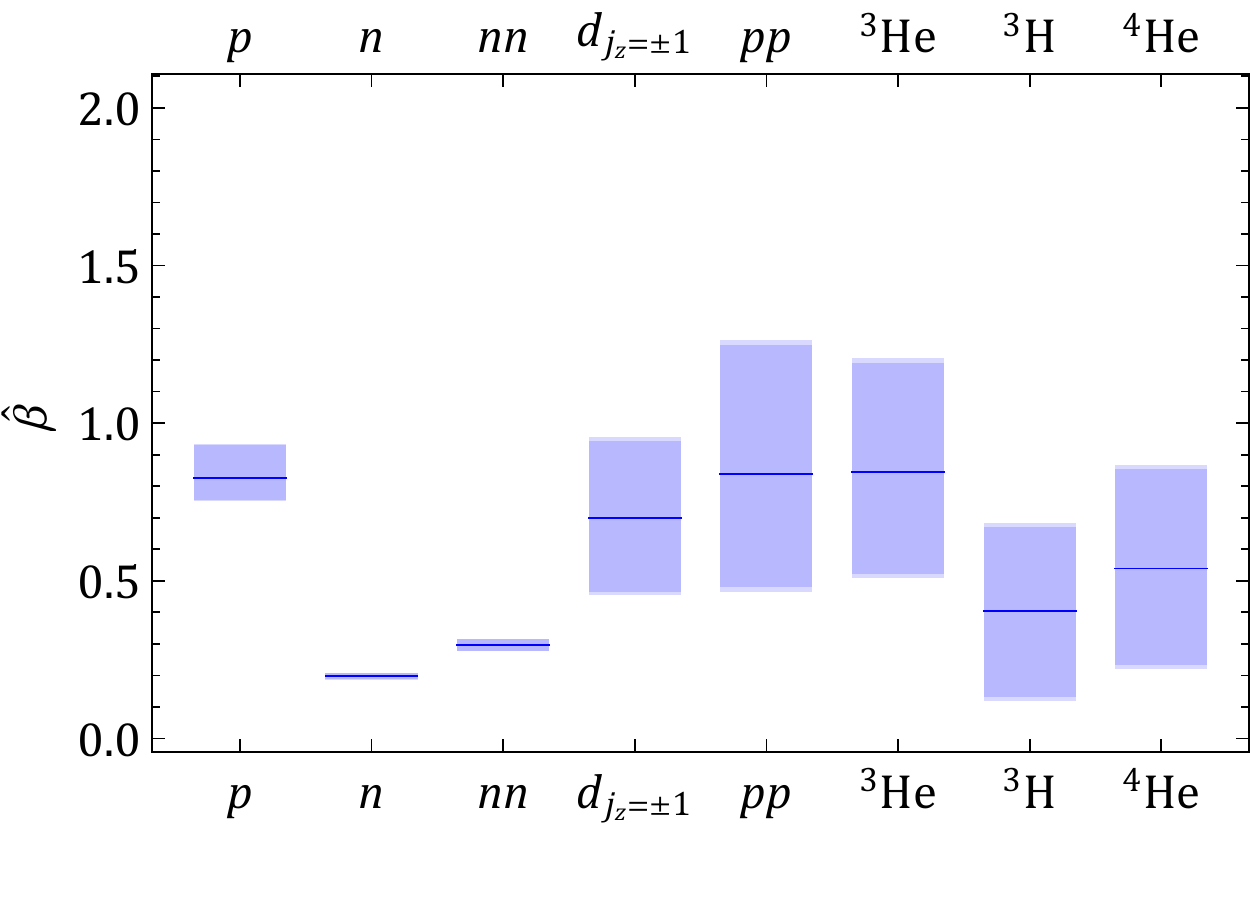}\ \\ 
  \vspace*{5mm}
  \includegraphics[width=0.72\columnwidth]{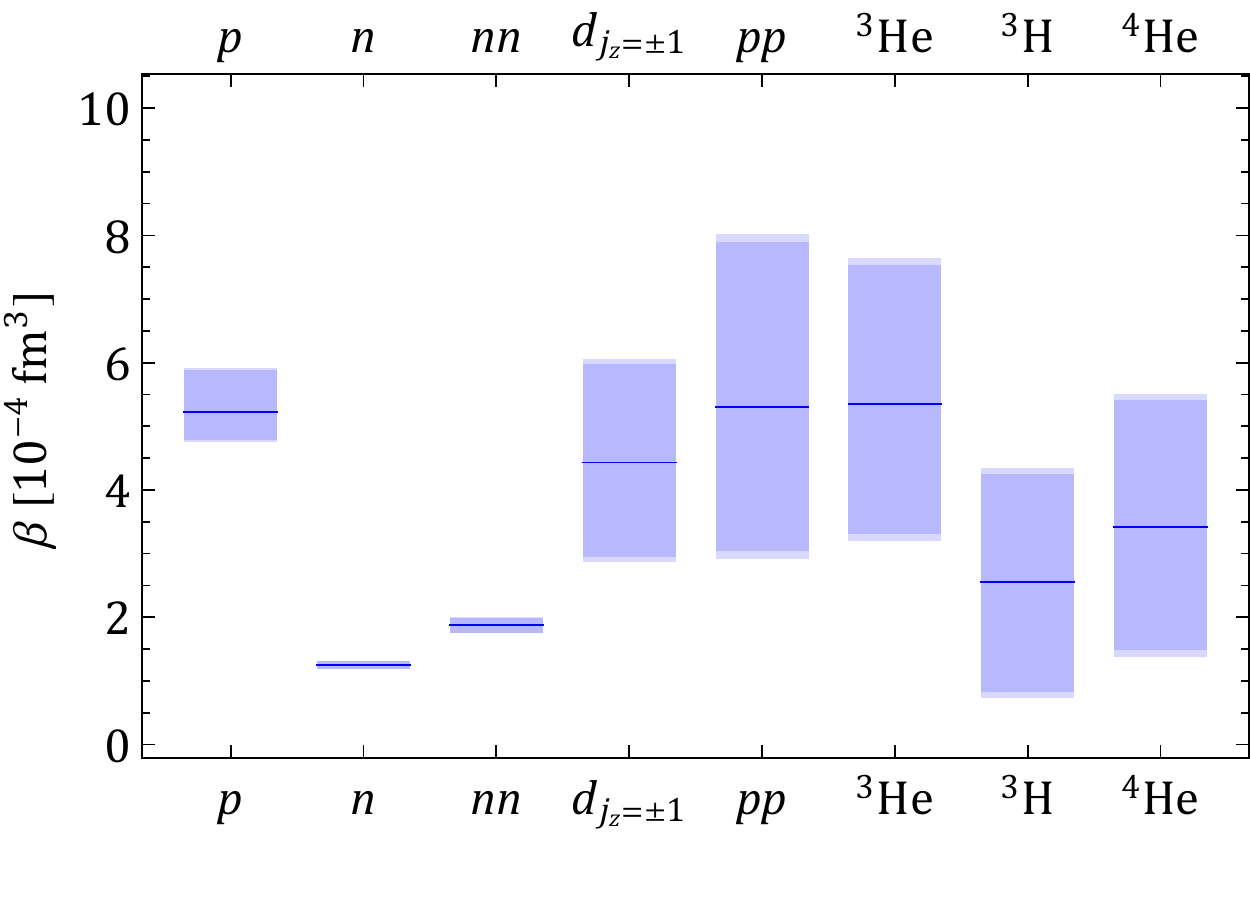}
  \caption{ 
  A summary of the magnetic polarizabilities of the nucleons and
    light nuclei calculated with LQCD at a pion mass of
    $m_\pi\sim \pionmass$. 
    The upper panel presents the dimensionless quantity
    $\hat\beta=M_N^2(M_\Delta - M_N)\beta/e^2$ obtained from the fits with the inner shaded region 
    representing the total uncertainty arising from statistical and fitting systematic uncertainties. 
    The outer shaded region assesses additional systematic uncertainties from discretization effects 
    and FV effects, combined in quadrature and applied multiplicatively.
    The lower panel presents the
    polarizabilities in physical units; in this case, the outer shaded region also includes the effect of 
    the scale setting uncertainty.
     }
  \label{fig:summaryBETA}
\end{figure}

The magnetic moments and polarizabilities that  have been determined in this work and in Ref.~\cite{Beane:2014ora} are summarized in
Tables~\ref{tab:musummary} and \ref{tab:summary}
and are shown in Figs.~\ref{fig:summaryMU} and \ref{fig:summaryBETA}  (the magnetic moments calculated from spin splittings in Ref.~\cite{Beane:2014ora} are the most precise determinations).
The electrically neutral systems are found to be
by far the most precise because the electrically charged systems  are defined by Landau levels, 
which have less than ideal overlap with the interpolating operators used to form the  correlation functions.
\begin{table}[!ht]
  \begin{ruledtabular}
    \begin{tabular}{ c c  }
      State & $\mu [{\rm nNM}]$ \\
      \hline
      $n$ & $-1.981(05)(18)$  \\
      $p$ & $+3.119(33)(64)$ \\
      $d (j_z=\pm1)$ & $+1.218(38)(87)$  \\
      $^3$He & $-2.29(03)(12)$  \\
      $^3$H & $+3.56(05)(18)$  \\
    \end{tabular}
  \end{ruledtabular}
  \caption{
    The results of our previous calculations of the nucleon and nuclear 
    magnetic moments~\protect\cite{Beane:2014ora}  from spin splittings 
    at a pion mass of $m_\pi\sim \pionmass$.
    The first uncertainty is statistical while the second is the complete systematic.
    As discussed in the text, these values are more precise than those determined from the more complex analysis required to extract the polarizabilities.
  }
  \label{tab:musummary}
\end{table}
\begin{table}[!ht]
  \begin{ruledtabular}
    \begin{tabular}{ ccc }
      State & $\hat{\beta}=M_N^2(M_\Delta-M_N)/e^2 \times \beta$ & $\beta\ [10^{-4} {\rm fm}^3 ]$ \\
      \hline
      $n$ & $\betan$ & $\betanPhysNoUnit$ \\
      $p$ & $\betap$ & $\betapPhysNoUnit$ \\
      $nn$ & $\betann$ & $\betannPhysNoUnit$ \\
      $pp$ & $\betapp$ & $\betappPhysNoUnit$ \\
      $d(j_z=\pm 1)$ & $\betadpm$ & $\betadpmPhysNoUnit$ \\
      $^3$He & $\betaHethree$  & $\betaHethreePhysNoUnit$ \\
      $^3$H & $\betatriton$ & $\betatritonPhysNoUnit$ \\
      $^4$He & $\betaHefour$  & $\betaHefourPhysNoUnit$ 
    \end{tabular}
  \end{ruledtabular}
  \caption{
    The magnetic polarizabilities calculated in this work at a pion mass of $m_\pi\sim\pionmass$.
    An additional 5\% uncertainty is associated with each polarizability as an estimate of 
    discretization and finite volume effects.
    For the polarizabilities  presented in physical units, an additional scale setting 
    systematic uncertainty (3\%)  is  included in quadrature in the second uncertainty.
  }
  \label{tab:summary}
\end{table}

These results, while not obtained at the physical values of the light-quark
masses, are interesting in their own right and suggest important
features of these systems.  First, our
calculations are sufficiently precise to determine that the strong
interactions between neutrons are such that when placed into a
magnetic field, the two-neutron system is more magnetically rigid than
the sum of the individual neutrons.  This is consistent with
expectations at the physical quark masses based upon phenomenological
nuclear interactions \cite{sanjayPC}, and indicates that it is energetically
disfavored for a neutron star to lower its energy by spontaneously
generating a large magnetic field.   Second, a large
isovector component to the nucleon magnetic polarizability is found.
The proton polarizability is found to be considerably
larger than that in nature while
the neutron polarizability is consistent with the phenomenological value, but much
more precise. 
Third, analysis of the $j_z=I_z=0$ $np$ system leads to a precise extraction
of the coefficient, $\overline{L}_1$, of the short-distance two-body magnetic current operator connecting the $\siii$ and
$\si$ states 
in the context of EFT($\pislash$).  
This operator provides an important
contribution to the $np\rightarrow
d\gamma$ capture cross-section near threshold, which is a critical input for calculations of 
the production of elements in big-bang nucleosynthesis and in other
environments as is discussed further in Ref~\cite{Beane:2015yha}.

These calculations are the first of their kind and are the initial steps in a
comprehensive program to determine the electromagnetic
properties of the light nuclei as well as the response of nuclei to electroweak
currents.
The next steps will include calculations of axial matrix elements in the various light nuclei, 
as these are of significant phenomenological interest in neutrino-nucleus scattering experiments.
Further, this points the way to calculating matrix elements of 
interactions required  in laboratory searches for dark matter 
and other potential beyond the Standard Model scenarios 
which involve nuclear matrix elements of a variety of currents.  
Calculations at smaller lattice spacings and 
in other volumes,  as well as for lighter quark masses where direct connection to
experiment can be made are important to this program. 
Finally it is important to include the presently omitted
couplings of sea quarks to the background fields.
Calculations addressing these goals are planned for the  future.


\
\begin{acknowledgments}
We would like to thank Zohreh Davoudi, Harald Grie{\ss}hammer, Daekyoung Kang, David B. Kaplan 
and Sanjay Reddy for several interesting discussions.
  Calculations were performed using computational resources provided
  by the Extreme Science and Engineering Discovery Environment
  (XSEDE), which is supported by National Science Foundation grant
  number OCI-1053575, NERSC (supported by U.S. Department of
  Energy Grant Number DE-AC02-05CH11231),
  and by the USQCD
  collaboration.  This research used resources of the Oak Ridge Leadership 
  Computing Facility at the Oak Ridge National Laboratory, which is supported 
  by the Office of Science of the U.S. Department of Energy under Contract 
  No. DE-AC05-00OR22725. 
  The PRACE Research Infrastructure resources Curie based in France at the Tr\`es Grand Centre de Calcul and MareNostrum-III based in Spain at the Barcelona Supercomputing Center were also used.
  Parts of the calculations used the Chroma software
  suite~\cite{Edwards:2004sx}.  SRB was partially supported by NSF
  continuing grant PHY1206498.  WD was supported by the
  U.S. Department of Energy Early Career Research Award DE-SC0010495.
  KO was supported by the U.S. Department of Energy through Grant
  Number DE- FG02-04ER41302 and through Grant Number DE-AC05-06OR23177
  under which JSA operates the Thomas Jefferson National Accelerator
  Facility.  The work of AP was supported by the contract
  FIS2011-24154 from MEC (Spain) and FEDER. MJS was supported 
  by DOE grant No.~DE-FG02-00ER41132.  BCT was supported in part by a
  joint City College of New York-RIKEN/Brookhaven Research Center
  fellowship, a grant from the Professional Staff Congress of the
  CUNY, and by the U.S. National Science Foundation, under Grant
  No. PHY12-05778.
\end{acknowledgments}
%


\appendix

\section{Charged-Particle Correlation Functions: Source Location and Gauge Origin}  
\label{app:CPcors}

\subsection{General discussion}

In addition to a uniform magnetic field,  the Abelian gauge links in 
Eq.~(\ref{eq:bkdgfield}) lead to two further gauge-invariant quantities that are
 finite volume artifacts. 
These quantities are the Wilson loops $W_1(x_2)$ and $W_2(x_1)$ appearing in 
Eq.~(\ref{eq:wilsonloops}), which express the non-vanishing holonomies of the background gauge field. 
A major consequence of these non-vanishing holonomies is the breaking of discrete translational invariance down to a smaller subgroup, which is referred to as the magnetic translation group,  see~Ref.~\cite{AlHashimi:2008hr}. 
The size of this subgroup depends on the magnetic-flux quantum,  $\tilde{n}$,  as
\begin{equation}
W_1 \left(x_2 + \frac{j}{3 q_q \tilde{n}} L\right) =  W_1(x_2), 
\quad  \text{and}  \quad
W_2 \left(x_1 + \frac{j}{3 q_q \tilde{n}} L \right) =  W_2(x_1)
\ \ ,
\end{equation}
for $j = 0, 1, \ldots, 3 |q_q |\tilde{n} -1$. 
Consequently lattice translational invariance for down-type quarks coupled  to $U_\mu^{(Q)} (x)$, 
for example,  is reduced to
$\mathbb{Z}_{\tilde{n}} \times \mathbb{Z}_{\tilde{n}} \times \mathbb{Z}_{\frac{L}{a}}$. 
This is to be contrasted with the infinite-volume case, 
where full translational invariance in a uniform magnetic field is maintained due to gauge invariance. 
On a torus,  gauge invariance is more restrictive due to the additional specification of Wilson loops, 
and translational invariance is consequently reduced.

For charged-particle correlation functions, 
the reduced translational invariance can lead to subtle effects. 
For example,  consider electromagnetic gauge links that are unity at the origin, 
$U_\mu^{(Q)} (0)= 1$, 
such as those in  Eq.~(\ref{eq:bkdgfield}), 
and  the hadronic source to be at the spatial position
$\bm{x}_i$. 
The electromagnetic gauge links can be altered so that they become unity at the source location, 
however, 
this cannot be achieved by a gauge transformation because the Wilson loops would be modified 
by\footnote{
The electromagnetic gauge links are accordingly modified in the form
$U_\mu^{(Q)} (x) \to U_\mu^{(Q)} \left[ ( x - x_i) \mod \frac{L}{a} \right]$. 
The new links  $U_\mu^{(Q)} \left[ ( x - x_i) \mod \frac{L}{a} \right]$
are related to $U_\mu^{(Q)} ( x - x_i)$
through a gauge transformation, 
however,  such a transformation alters the functional dependence of charged-particle correlation functions. 
Because the Wilson loops are gauge invariant, 
there is no difference, 
for example, 
between 
$W_1 \left[ ( x_2 - x_{i,2}) \mod \frac{L}{a} \right]$
and
$W_1(x_2 - x_{i,2}) = W_1 (x_2) W^\dagger_1(x_{i,2})$, 
which appears above. 
}
\begin{equation}
W_1(x_2) \to W_1( x_2) W^\dagger_1( x_{i,2}), 
\quad
\text{and}
\quad
W_2(x_1) \to W_2(x_1) W^\dagger_2( x_{i,1})
\label{eq:shift}
\ \ .
\end{equation}
Similarly, 
translational invariance cannot be used to relocate the hadronic source to the origin without altering the correlation function. 
Consequently charged-particle correlation functions depend on the the origin of the gauge potential, 
and the location of the source. 
Even when 
$\bm{x}_i$
is related to the origin of the gauge potential by a discrete magnetic translation, 
which corresponds to the special case where the required translation is equivalent to a gauge transformation, 
the charged-particle correlation function will not be identical due to gauge dependence.\footnote{
The definition of the charged particle two-point function could be altered 
to include an electromagnetic gauge link between the source and sink locations. 
The resulting correlation functions would be gauge invariant;
but,
the cost is the introduction of a path for the gauge link. 
Because magnetic flux threads closed loops that are oriented transverse to the magnetic field, 
the resulting correlation function is then path dependent. 
}
As the degree to which lattice translational invariance is reduced depends on the strength of the magnetic field, 
steps are required to ameliorate this situation. 

One way to deal with the problem is to fix the background field entirely including the holonomies. 
This approach can be implemented by randomly choosing a source location before
including the $U_\mu^{(Q)}(x)$ links. 
The hadron source location must still be chosen relative to the gauge potential, 
and a convenient choice is to make them coincident. 
This method corresponds to implementing periodic-boundary conditions (BCs) on the quarks,\footnote{
Technically the quarks are only periodic (or twisted) up to a gauge transformation.
This is often referred to as a magnetic periodic BC. 
} 
and was employed in~\cite{Primer:2013pva}
to investigate magnetic properties of the nucleon. 
The choice of a coincident location, however,  is not required;  and,  other choices for the relative separation 
that are not related by a magnetic translation could be averaged over to mitigate FV effects. 

An alternative approach to reduce FV effects consists of varying the holonomies. 
One way to achieve this consists of introducing a constant shift in the gauge potential, 
which corresponds to implementing twisted-BCs on the quarks.\footnote{
Shifting the source and shifting the gauge potential are equivalent up to a gauge transformation, 
however, 
this leads to different expectations for charged particle correlation functions, 
as shown below.
} 
These are flavor dependent BCs
due to the difference in quark electric charges. 
Ultimately to remove the arbitrariness of this choice, 
all non-equivalent shifts should be averaged over. 
The resulting twist average removes the FV effect associated with translational invariance, 
and related proposals have recently been suggested more generally to reduce FV effects in 
other lattice QCD computations~\cite{Briceno:2013hya,Lehner:2015bga}.\footnote{
In the analogous case of time-varying gauge potentials which lead to electric fields, 
a variant of this procedure was carried out for the neutron~\cite{Engelhardt:2007ub}.
In that study, 
results at second order in the gauge potential, 
$A_\mu$,
were directly isolated by perturbatively expanding external field correlation functions, 
and enforcing temporal Dirichlet BCs. 
While twist averaging does not eliminate the FV effect in that case, 
it was shown that the neutron electric polarizability can nonetheless be separated from finite-size effects by efficacious shifts of the gauge potential. 
}

The present calculations were performed with the following approach to handle the lack of lattice translational invariance.  
The source locations were varied relative to the origin of the gauge potential, rather than varying the  gauge holonomies. 
Varying the source locations
allows for the restoration of 
lattice translational invariance in two equivalent ways:
averaging over all sources on a given configuration, and then performing the ensemble average; 
or 
performing the ensemble average with a fixed source location, and then averaging over all locations. 
In this work, a hybrid approach was chosen 
because of limited computational resources.
Each QCD gauge configuration was post-multiplied by the Abelian gauge links
$U_\mu^{(Q)}(x)$, 
and then a random offset was introduced. 
The random offset was the same for each value of the magnetic flux quantum, 
$\tilde{n}$, 
in order to maximize correlations between differing field strengths. 
On each configuration,  quark propagators were calculated using 
$48$
symmetrically distributed source locations. 
In addition to improving statistics, 
the source averaging partially restores lattice translational invariance on each configuration. 
Translational invariance is further improved in the ensemble average, 
because a random offset is chosen for each configuration.

While quark propagators are subject to finite-size effects owing to the reduction in translational invariance, 
hadronic correlation functions for neutral particles are less susceptible. 
Using the neutron as an example, 
the non-vanishing holonomies of the gauge field show up only in exponentially small FV corrections to the 
neutron energy~\cite{Tiburzi:2014zva}. 
These corrections arise primarily from charged pion fluctuations that wind around the torus. 
For charged particles, 
however, 
the reduction in translational invariance has a direct effect on the coordinate dependence of their correlation functions. 
In turn, 
the overlap of a given hadronic interpolating field onto Landau levels  depends on the hadron source location and gauge holonomy
in a field-dependent way. 
To make this discussion more concrete, 
the simplified case of a point-like charged particle subject to the same method utilized in the present LQCD calculation is explored.

\subsection{Expectations for a Point-Like Charged Particle}

To illustrate the dependence of charged particle correlation functions on the source location and gauge holonomy, 
a point-like particle propagating in an external magnetic field on a torus is considered. 
In the point-like approximation, 
the spin-projected hadron propagator can be derived following the arguments 
in~\cite{Tiburzi:2012ks,Tiburzi:2014zva}.
Various ways  to reduce finite-size effects 
by shifting the gauge potential 
and/or 
shifting the source location are considered. 

In practice, 
the point-like approximation is valid only when the typical hadronic size 
cannot be resolved in a given Landau level. 
For example, 
higher-lying Landau levels have a larger (rms) radius,
$\propto \sqrt{(n_L + \frac{1}{2})/|Q_h eB|}$; 
hence, 
the details of the hadronic state will be less relevant compared with lower Landau levels. 
By contrast, 
the lowest Landau level is the most sensitive to hadron structure, 
and the most likely to be dynamically altered away from the point-like result. 
With magnetic fields that are not arbitrarily weak, 
more dependence on hadronic structure can be expected, and 
the point-like particle case may thus only provide a guide.
Further study is needed to address the dynamical Landau levels of bound states, 
and to design better interpolating operators for hadrons in magnetic fields.

Employing a uniform shift of the gauge potential transverse to the magnetic field direction, 
the gauge links are  modified to
\begin{eqnarray}
U^{(Q)}_1(x) 
&=& 
 \exp [ - i q_q \theta_1 ]
\times
\left\{ \begin{array}{lcr}  
1 &  {\rm for} 
 &  x_1 \ne  L -a \\
 \exp\left[-i q_q \tilde n {6\pi  x_2  \over L} \right]
 & {\rm for}& x_1 = L-a
 \end{array}\right. ,
   \nonumber\\
 U^{(Q)}_2(x) & = & 
 \exp [ - i q_q \theta_2 ]
 \,
  \exp\left[i q_q \tilde n {6\pi  a \, x_1  \over L^2} \right]\,,
  \nonumber\\
  U^{(Q)}_{3,4}(x)& = & 1
  \ .
  \label{eq:gen_bkdgfield}
\end{eqnarray}
The propagator for a structureless charged particle then has the form
\begin{eqnarray}
C_{h, j_z} (x, x_i; \bm{\theta}, \bm{B})
=
\sum_{\bm{\nu} }
e^{ - i Q_h \bm{\theta} \cdot \, \bm{\nu}_\perp}
[W^\dagger_2(x_1)]^{\nu_2}\ 
C^{(\infty)}_{h, j_z} (x + \bm{\nu} L, x_i; \bm{B})
\ \ \ ,
\label{eq:FVprop}
\end{eqnarray}
where $x$ and $x_i$ are the spacetime locations of the sink and source relative to the origin, 
respectively, and $\bm{\nu}$ is a triplet of integers .
Images transverse to the magnetic field direction, 
$\bm{\nu}_\perp = ( \nu_1, \nu_2)$, 
pick up phases arising from the constant shift of the gauge potential. 
Notice the Euclidean time direction is treated as infinite in extent. 
The propagator 
$C^{(\infty)}_{h, j_z} (x,x_i;\bm{B})$
is that for a charged particle in infinite volume.
Ignoring discretization effects, 
the continuum form of the infinite volume propagator is employed, 
which in coordinate space is~\cite{Tiburzi:2014zva} 
\begin{eqnarray}
C^{(\infty)}_{h, j_z} (x, x_i; \bm{B})
&=&
\mathfrak{W}^\dagger (x,x_i)
\mathfrak{C}^{(\infty)}_{h, j_z} (x -x_i; \bm{B})
,\end{eqnarray}
and consists of two parts. 
There is a spacetime translationally invariant part
\begin{eqnarray}
\mathfrak{C}^{(\infty)}_{h, j_z} (z; \bm{B})
&=&
\int_0^\infty \frac{ds}{(4 \pi s)^2}
\frac{Q_h e B s}{\sinh (Q_h e B s)}
\exp
\left\{
- 
s \tilde{E}^2_{h,j_z} 
-
\frac{Q_h e B}{4 \tanh ( Q_h e B s)}
[z_1^2 + z_2^2]
- 
\frac{1}{4s} [z_3^2+z_4^2]
\right\}
,
\notag \\
\end{eqnarray}
which contains the hadron energy 
$\tilde{E}_{h, j_z}$
appearing in
Eq.~(\ref{eq:Eshift}),
however, 
the tilde denotes that it excludes the contribution to the energy from the $n_L$-th Landau level.
Instead, 
contributions from all Landau levels are contained in this propagator~\cite{Schwinger:1951nm}.
The remaining part
$\mathfrak{W}^\dagger (x,x_i)$ 
is not translationally invariant, and accordingly depends on the gauge. 
It can be written as a Wilson line evaluated on the straight-line path between 
$x_i$ and $x$
\begin{equation}
\mathfrak{W} (x,x_i)
=
\exp \left[  i Q_h e \int_{x_i}^x dz_\mu A_\mu (z) \right]
= 
\exp \left[ - \frac{i}{2}  Q_h e B (x_1 - x_{i,1}) (x_2 + x_{i,2}) \right]
\ \ \ .
\end{equation}

In writing the FV propagator in Eq.~(\ref{eq:FVprop}), the Wilson loops have been implicitly modified 
to include the hadron's electric charge, $Q_h e$, instead of the quark charge, $q_q e$,
and the charged hadron propagator $C_{h, j_z} (x, x_i; \bm{\theta}, \bm{B})$
satisfies the following BCs in the directions transverse to the magnetic field:
\begin{eqnarray}
C_{h, j_z} (x + L \hat{x}_1, x_i; \bm{\theta}, \bm{B})
&=&
e^{ i Q_h \theta_1}
C_{h, j_z} (x, x_i; \bm{\theta}, \bm{B})
,\notag \\
C_{h, j_z} (x + L \hat{x}_2, x_i; \bm{\theta}, \bm{B})
&=&
e^{ i Q_h \theta_2}
W_2(x_1)
C_{h, j_z} (x, x_i; \bm{\theta}, \bm{B})
\label{eq:bdy}
\ \ .
\end{eqnarray}
The first is a twisted-BC, 
while the second emerges as a magnetic-twisted-BC. 
In the actual computation,  
the quark propagators are calculated with periodic BCs, 
however, 
the implementation of the external field with links that are periodic up to a gauge transformation, 
and on configurations with a gauge shift leads to the BCs in Eq.~(\ref{eq:bdy}).  
Because of the gauge shift, 
the gauge potential no longer vanishes at the origin, 
e.g.,  the links $U_2^{(Q)} (x)$ are unity when  $x_1 = \theta_2 L^2  / 6 \pi a \tilde{n}$.  
Finally, notice the asymmetric appearance of the holonomy  $W(x_1)$ in  Eq.~(\ref{eq:FVprop}). 
The Wilson loop  $W(x_2)$ does not appear explicitly in the sum over the winding number $\nu_1$, 
and the charged-particle propagator is twisted in the  $x_1$-direction rather than magnetic twisted, 
see Eq.~(\ref{eq:bdy}).  The effect of this Wilson loop,  however,  is contained implicitly in the sum over winding number 
$\nu_1$ because of the  $x_2$-coordinate  dependence of the Wilson line $\mathfrak{W}(x,x_i)$.  
This asymmetry in the charged-particle propagator and the BCs that result from it  are directly related to the 
asymmetric choice of gauge.

Given the form of the propagator in 
Eq.~\eqref{eq:FVprop}, 
a natural question to ask is whether shifting both the gauge potential and the source location is superfluous. 
To answer this question, 
one can express the propagator in terms of the source-sink separation, 
$\Delta \bm{x} = \bm{x} - \bm{x}_i$, 
and attempt to absorb the remaining dependence on the source location into a redefinition of the twist angles 
$\bm{\theta}$.
Due to the breaking of translational invariance, 
this is not possible. 
By virtue of the Wilson line
$\mathfrak{W} (x, x_i)$, 
the correlation function retains explicit dependence on 
$x_{i,2}$, 
which is measured relative to the origin.  
The origin has no significance for gauge-invariant quantities, 
however, 
in terms of the gauge links, 
the origin can readily be discerned, 
see
Eq.~(\ref{eq:gen_bkdgfield}). 
As a consequence, 
gauge variant quantities, 
such as the charged-particle propagator, 
can depend on positions relative to the origin.

Four scenarios are considered, denoted by $\Gamma$:
i) periodic BCs with coincident origins of the gauge potential and source for the correlation functions ($\Gamma=0$),
ii) shifting the gauge potential ($\Gamma={\bm\theta}$),
iii) shifting the source location ($\Gamma=X$),
and 
iv) varying the shift in the gauge potential and shifting the source location ($\Gamma={\bm\theta} X$).

\subsubsection{Periodic BCs with coincident origins $(\Gamma=0)$}

Choosing the origin of the gauge potential to coincide with that of the source
and not including a uniform gauge field 
corresponds to specifying
$\bm{x}_i = \bm{0}$, 
and
$\bm{\theta} = \bm{0}$. 
The latter leads to periodic quarks (up to a gauge transformation). 
This method was chosen in~\cite{Primer:2013pva}.
Additionally in that study, 
as in this work too, 
the spatial sink location is summed over, 
which projects the correlator onto vanishing three-momentum. 
Three-momentum states do not have definite energy eigenvalues, 
however,
and one expects correlation functions to contain multiple low-lying Landau levels. 
For a point-like particle on a continuous torus, 
consider the spatially-integrated correlation function,
\begin{equation}
C_0(t)
=
\int_0^L d\bm{x}
\, \,
C_{h,j_z}(x,0;\bm{0},\bm{B})
\ \ \ .
\end{equation}
Carrying out the three-momentum projection gives,
\begin{equation}
C_0(t)
=
\frac{1}{2}
\int_0^\infty 
\frac{ds}{\sqrt{2 \pi s}}
e^{- \frac{1}{2} s \tilde{E}_{h, jz}^2 - \frac{1}{4s} t^2}
\int_{0}^L dx_2
\sum_{\nu_2  = - \infty}^\infty
\langle x_2, s |  \nu_2 L, 0 \rangle
\ \ ,
\end{equation}
which has been written in terms of the quantum-mechanical propagator for the simple harmonic oscillator
\begin{eqnarray}
\langle x', t' | x, t \rangle
&=& 
\theta(t'-t)
\sqrt{\frac{Q_h e B } {2 \pi \sinh[ Q_h e B ( t' - t)]}}
\notag \\
&&
\times
\exp 
\left\{ - \frac{Q_h e B}{2 \sinh [ Q_h e B (t' - t)]}
\left[
(x^{\prime 2} + x^2) \cosh [ Q_h e B ( t' - t)]
-  2 x' x
\right]
\right\}
\ ,
\end{eqnarray}
where  $t$ and $t'$ are Euclidean times. 
In terms of Landau levels,  the correlation function can be written as
\begin{equation}
C_0(t)
=
\sum_{n_L=0}^\infty Z_{n_L}^{(0)}\  \frac{e^{ - E_{h, j_z} t}}{2 E_{h, j_z}}
\ \ ,
\end{equation}
where the energies 
$E_{h, j_z}$
include the Landau energy, 
and are given in 
Eq.~(\ref{eq:Eshift}). 
The dimensionless coefficients 
$Z_{n_L}^{(0)}$
are given by
\begin{equation}
Z_{n_L}^{(0)} = \int_{0}^L dx_2 \ \psi^*_{n_L}(x_2)
\sum_{\nu_2 = -\infty}^\infty
\psi_{n_L} (\nu_2 L)
\ \ ,
\end{equation}
and are written in terms of the coordinate wave functions, 
which have the standard form in terms of Hermite polynomials
\begin{equation}
\psi_{n_L}(x)
= 
\frac{1}{\sqrt{2^{n_L}\,  n_L! \sqrt{\frac{\pi}{|Q_h e B|}}}}
H_{n_L} \left( \sqrt{|Q_h e B|} x \right)
e^{ - \frac{1}{2} |Q_h e B| x^2 }
\ \ .
\end{equation}
Notice that contributions from odd-parity Landau levels are absent due to the sum over winding number. 
The coefficients $Z_{n_L}^{(0)}$
are not positive; spectral positivity is not maintained due to the lack of translational invariance. 
For quantized values of the magnetic field, 
these coefficients depend on the flux quantum $\tilde{n}$, but not on the size $L$. 
This dependence on  $\tilde{n}$,  however, is exponentially suppressed.

\subsubsection{Shifting the gauge potential $(\Gamma={\theta})$}

While shifting the gauge potential has not been pursued as a means to reduce FV effects, 
it is instructive to discuss briefly point-like expectations for this method. 
Averaging over all possible shifts of the gauge potential is equivalent to averaging over quarks with randomly twisted boundary 
conditions. 
As a result, 
the twist-averaged propagator in Eq.~\eqref{eq:FVprop}
receives contributions only from zero winding numbers
$\bm{\nu}_\perp = \bm{0}$. 
While this is a desirable feature, 
there is no further simplification of the charged particle correlation function. 
The twists can be utilized, 
however,
to form the infinite volume propagator via constructing the Fourier transformation in blocks%
~\cite{Lehner:2015bga}. 
The lack of lattice translational invariance means that the magnetic periodic images, rather than periodic images,
must be summed over. 
In effect, this provides access to 
$C^{(\infty)}_{h,j_z} (x, x_i; \bm{B})$, 
for 
$\vec{x}_\perp$ 
over the whole transverse plane. 
The resulting zero-momentum correlation function has the form
\begin{equation}
C_\theta(t)
=
\frac{1}{2} \int_0^\infty  ds
\frac{
e^{ - \frac{1}{2} s \tilde{E}^2_{h,j_z} - \frac{1}{4s} t^2}
}
{\sqrt{2 \pi s}}
\int_{-\infty}^\infty 
dx_2
\, 
\langle x_2, s | x_{i,2}, 0 \rangle
\ \ .
\end{equation}
In terms of Landau levels, 
the expected behavior is thus
\begin{equation}
C_\theta(t)
=
\sum_{{n_L}=0}^\infty
Z_{n_L}^{(\theta)}
\
\frac{e^{ - E_{h, j_z} t}}{2 E_{h, j_z}}
\ \ ,
\end{equation}
where the coefficients are given by
\begin{equation}
Z_{n_L}^{(\theta)}
=
\int_{-\infty}^\infty 
dx_2
\,
\psi^*_{n_L}(x_2)
\,
\psi_{n_L} (x_{i,2})
\ \ ,
\end{equation}
in which there is no remaining dependence on 
$L$.
When the source is located at the origin, 
there is no dependence on the magnetic flux quantum, 
$\tilde{n}$. 
Notice that there are no contributions from odd-parity Landau levels, 
and further that spectral positivity does not emerge.

\subsubsection{Shifting the source location $(\Gamma=X)$}

As described previously, source-to-all propagators are calculated in the present LQCD  calculation, 
with sources  randomly located relative to the gauge origin. 
A constant shift of the gauge potential is not implemented, and thus 
$\bm{\theta} = \bm{0}$, 
but with an approximate average over 
$\bm{x}_i$. 
This leads  to the following expression for the correlation function of a point-like charged particle,
\begin{equation}
C_{X} (t)
=
\frac{1}{2}
\int_0^\infty ds
\frac{
e^{ - \frac{1}{2} s \tilde{E}^2_{h,j_z} - \frac{1}{4s} t^2}
}
{\sqrt{2 \pi s}}
\int_0^L
dx_2 
\int_0^L
\frac{dx_{i,2}}{L}
\
\big\langle x_2 , s \, \big| x_{i,2}, 0 \big\rangle
\ \ .
\end{equation}
Performing the integration over source location 
$x_{i,1}$
restricted the winding number expansion to the sector with 
$\nu_2 = 0$. 
Using the spectral decomposition for the quantum-mechanical harmonic-oscillator propagator 
gives,
\begin{equation}
C_{X} (t)
=
\sum_{{n_L}=0}^\infty
Z^{(X)}_{n_L}
\,
\frac{e^{-E_{h,j_z} t}
}{2 E_{h,j_z}}
\ \ ,
\end{equation}
where 
$E_{h,j_z}$
includes the energy of the $n_L$-th Landau level.
The corresponding spectral weights are
\begin{equation}
Z^{(X)}_{n_L}
=
\frac{1}{L}
\left|
\int_{0}^{L} dx
\, 
\psi_{n_L}(x)
\right|^2
\ \ ,
\end{equation}
and give the probability of finding the charged particle in the 
${n_L}$-th Landau level. 
Positivity of these weights arises due to the symmetric treatment of the source and sink locations. 
Landau levels of both parities contribute to the correlation function. 
When evaluated for quantized magnetic fields, 
the weights only depend on the flux quantum 
$\tilde{n}$. 
Ratios of coefficients, 
however,  
depend on 
$\tilde{n}$
through exponentially small terms.

\subsubsection{Varying the gauge shift and source location $(\Gamma={\theta} X)$}

Despite computational requirements, 
it remains worthwhile to consider averaging over the shift of the gauge potential and the source location. 
The former can be used to construct magnetic-periodic images and build the infinite-volume propagator in blocks. 
The result of this procedure leads to 
\begin{equation}
C_{\theta X} (t)
=
\sum_{{n_L}=0}^\infty
Z^{(\theta X)}_{n_L}
\,
\frac{e^{-E_{h,j_z} t}
}{2 E_{h,j_z}}
\ \  ,
\end{equation}
where $E_{h,j_z}$ includes the energy of the ${n_L}$-th Landau level, 
and the corresponding spectral weights are
\begin{equation}
Z^{(\theta X)}_{n_L}
=
\frac{1}{L}
\left|
\int_{-\infty}^{\infty} dx
\, 
\psi_{n_L}(x)
\right|^2
\ \ .
\end{equation}
This procedure  eliminates all 
$L$ dependence,  and $\tilde{n}$
dependence is completely absent in ratios of coefficients. 
Furthermore, 
the procedure excludes contributions from odd-parity Landau levels, 
and maintains spectral positivity.

\subsection{Results for a point-like particle}

\begin{figure}[!ht]
\includegraphics[width=0.6\textwidth]{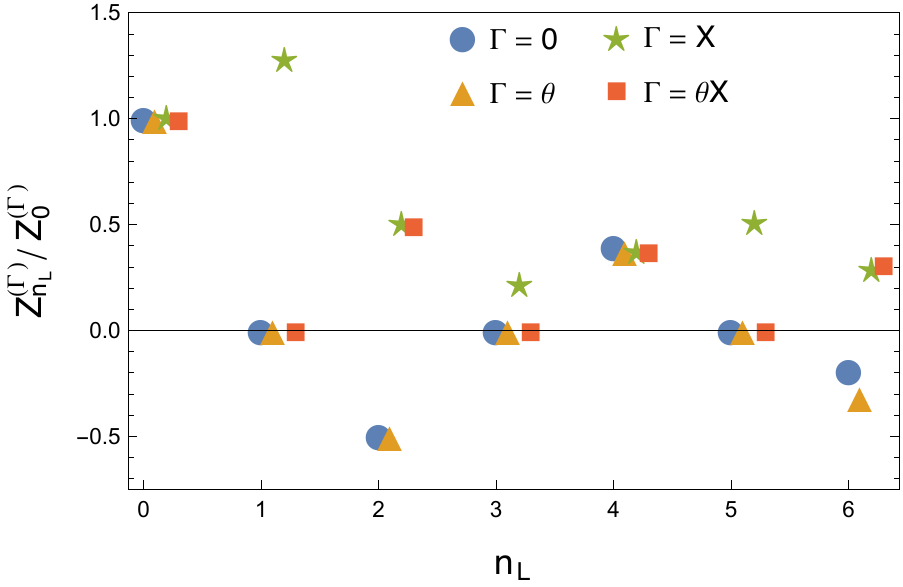}
\caption{
Contributions to the correlation functions of a point-like charged particle from the first few Landau levels compared with the  
contribution from the lowest Landau level.
Four such coefficient ratios are considered, corresponding to different ways of dealing with the lack of lattice translational invariance,
as described in the text. 
These are labeled by  $\Gamma = 0$, $\theta$, $X$, and $\theta X$, 
which correspond to:  coincident origin,  twist-averaged,  source-averaged,  and  twist and source-averaged, 
respectively.  Values have been slightly displaced in $n_L$ to allow different ratios to be discernible. 
For quantized magnetic fields, all ratios are independent of the lattice size  $L$, 
and ratios are either independent of the flux quantum  $\tilde{n}$,  or depend on it only through exponentially small terms. 
}
\label{f:coeffplot}
\end{figure}

Having determined the charged-particle correlation functions for four different 
methods of dealing with the lack of translational invariance, 
the relative contributions of the lowest-lying Landau levels are compared in Fig.~\ref{f:coeffplot}.
In plotting  the coefficient ratios,  the $n_L$-dependence of the hadron energies 
$E_{h,j_z}$ appearing in the correlation functions has been ignored.  
These energy denominators will lead to smaller coefficients, 
but only for higher-lying Landau levels. 

\bibliography{bib_nukepols}
\end{document}